\documentclass[useAMS,usenatbib]{mn2e}
\usepackage{epsfig,psfrag,array}
\usepackage{amsmath,amsfonts,amssymb}
\usepackage{natbib}
\usepackage{flafter}
\usepackage{graphicx}
\usepackage{mathptmx}
\usepackage{dcolumn}
\usepackage{subfig}
\usepackage{journals}
\newcolumntype{d}{D{.}{.}{-1}}

\graphicspath{{./}}
\newcounter{todo}
\renewcommand\thetodo{\Alph{todo}}
\def\todo#1{\addtocounter{todo}{1}[[\thetodo: #1]]\strut\vadjust{%
\kern-\dp\strutbox{\vtop to \dp\strutbox{\baselineskip\dp\strutbox\vss\rlap{%
\hskip\hsize\ \rm{$\leftarrow$\thetodo}}\null}}}}
\def\note#1{\strut\vadjust{\kern-\dp\strutbox{\vtop to \dp\strutbox{%
\baselineskip\dp\strutbox\vss\rlap{\hskip\hsize\ {\tiny\rm #1}}\null}}}}
\title[Bayesian analysis of weak-gravitational-lensing and Sunyaev--Zel'dovich data for six galaxy clusters]
{Bayesian analysis of weak-gravitational-lensing and Sunyaev--Zel'dovich data for six galaxy clusters\thanks{We request that any reference to this paper cites ``AMI Consortium: Hurley-Walker et al. 2011"}}

\label{firstpage}

\author[Hurley-Walker et~al.]
{AMI Consortium:
Natasha~Hurley-Walker$\thanks{Issuing author; E-mail: nhw@icrar.org}^{1,2}$,
Sarah~Bridle$^{3}$, Eduardo~S.~Cypriano$^{4}$, \newauthor
Matthew~L.~Davies$^{1}$, Thomas~Erben$^{5}$, Farhan Feroz$^{1}$, Thomas M. O. Franzen$^{6}$,\newauthor
Keith~Grainge$^{1,7}$, Michael~P.~Hobson$^{1}$, Anthony~Lasenby$^{1,7}$, P.~J.~Marshall$^{8}$, \newauthor
Malak~Olamaie$^{1}$, Guy~Pooley$^{1}$, Carmen~Rodr\'{i}guez-Gonz\'{a}lvez$^{1}$, \newauthor
Richard~D.~E.~Saunders$^{1,7}$, Anna~M.~M.~Scaife$^{9}$, Michel~P.~Schammel$^{1}$, Paul~F.~Scott$^{1}$, \newauthor
Timothy Shimwell$^{1}$, David~Titterington$^{1}$, Elizabeth~Waldram$^{1}$, and Jonathan~T.~L.~Zwart$^{10}$\\
 $^1$ Astrophysics Group, Cavendish Laboratory,
      19 J.~J.~Thomson Avenue, Cambridge CB3 0HE \\
 $^2$ International Centre for Radio Astronomy Research, Curtin Institute of Radio Astronomy, 1 Turner Avenue, Technology Park, Bentley, WA 6845, Australia \\
 $^3$ Department of Physics and Astronomy, University College London, London WC1E 6BT\\
 $^4$ Departamento de Astronomia, Instituto de Astronomia Geof\`{i}sica e
Ci\^{e}ncias Atmosf\`{e}ricas da Universidade de S\~{a}o Paulo \\ Rua do Mat\~{a}o,
1226, 05508-900 S\~ao Paulo, Brazil \\
 $^5$ Argelander-Institut f\"{u}r Astronomie, University of Bonn, Auf demo H\'{u}gel 71, D-53121, Germany \\
 $^6$ CSIRO Astronomy \& Space Science, Australia Telescope National Facility, PO Box 76, Epping, NSW 1710, Australia \\
 $^7$ Kavli Institute for Cosmology Cambridge,
      Madingley Road, Cambridge CB3 0HA\\
 $^8$ Oxford Astrophysics, Department of Physics, Denys Wilkinson Building, Keble Road, Oxford, OX1 3RH\\
 $^9$ Dublin Institute for Advanced Studies, 31 Fitzwilliam Place,
      Dublin 2, Ireland \\
 $^{10}$ Columbia Astrophysics Laboratory, Columbia University, 550 West
      120th Street, New York, NY 10027, USA}

\date{Accepted ---; received ---; in original form \today}
\pagerange{\pageref{firstpage}--\pageref{lastpage}}
\pubyear{2011}

\begin{document}
\maketitle

\begin{abstract}
We present an analysis of observations made with the Arcminute Microkelvin Imager (AMI) and the
Canada-France-Hawaii Telescope (CFHT) of six galaxy clusters in a redshift range of 0.16--0.41. The cluster
gas is modelled using the Sunyaev--Zel'dovich (SZ) data provided by AMI,
while the total mass is modelled using the lensing data from the CFHT. In this paper, we:
i) find very good agreement between SZ measurements (assuming large-scale virialisation and a gas-fraction prior)
and lensing measurements of the total cluster masses out to $r_{200}$;
ii) perform the first multiple-component weak-lensing analysis of A115;
iii) confirm the unusual separation between the gas and mass components in A1914;
iv) jointly analyse the SZ and lensing data for the relaxed cluster A611, confirming our use of a simulation-derived mass-temperature
relation for parameterizing measurements of the SZ effect.
\end{abstract}

\begin{keywords}

   cosmology: observations - cosmic microwave background - radiation mechanisms: non-thermal 
 galaxies:clusters - Sunyaev-Zel'dovich - weak gravitational lensing
 - galaxies:clusters:individual (Abell~115, Abell~611, Abell~851, Abell~1914, Abell~2111, and Abell~2259)

\end{keywords}
\section{Introduction}
We employ two independent methods in a pilot study to investigate mass distributions
of six galaxy clusters selected to cover a range of redshifts and merging states:
\begin{enumerate}
\item{Weak gravitational lensing (see e.g. \citealt{2001PhR...340..291B} for a review),
in which images of background objects
are distorted by a mass lying along the line-of-sight, and can be used to directly probe
the cluster mass distribution.}
\item{The Sunyaev--Zel'dovich effect (SZ; see e.g. \citealt{SZ/Bir99} and \citealt{carlstrom02} for reviews),
the inverse-Compton scattering of the Cosmic Microwave Background (CMB) by the hot cluster gas, which
effectively measures the gas pressure. The dark matter content is normally assessed by combining the SZ information with
an X-ray measurement of gas temperature and using the assumption of hydrostatic equilibrium;
but neither the X-ray temperature nor this assumption are necessary if we assume the cluster is
virialised and we incorporate a sensible prior on the ratio of gas to dark matter
\citep{malak-param2010}.}
\end{enumerate}
SZ and weak lensing have a natural
complementarity, as they both have the potential to measure the distributed outskirts of clusters, with no strong bias towards
concentrations of gas or mass. Also possible is the determination of the gas fraction of a given cluster by
calculating the gas mass from the SZ and the total mass from lensing.
It is expected that thousands of galaxy clusters will be detected by new SZ surveys
performed by the Atacama Cosmology Telescope (ACT: \citealt{ACT11}),
the South Pole Telescope (SPT: \citealt{SPTGalaxyClusters}) and \textit{Planck} \citep{PlanckMission}; the last has produced the first all-sky SZ catalogue, the first
release of which is available in \cite{PlanckESZ}. Upcoming large-area multi-wavelength optical surveys such as the
the Dark Energy Survey \citep{DES2005WP}
and eventually the Large Synoptic Survey Telescope survey \citep{LSSTOverview} will improve photometric redshift measurements of galaxies
and release lensing-quality data over tens of thousands of square degrees of sky.
Combining SZ and lensing measurements of very
large samples of galaxy clusters may allow us to model their internal physics well enough for
cosmological applications \citep{HoekstraLensingReview}. 

The Arcminute Microkelvin Imager (AMI) is a radio interferometer that has made
observations of hundreds of known galaxy clusters to measure their gas masses and structures via the SZ.
A limitation of the large-area SZ survey instruments is their inability to resolve the morphology of cluster gas,
so with these instruments it is difficult to examine the cluster dynamical state. Interferometric arrays such as AMI 
and the Combined Array for Research in Millimeter-wave Astronomy \citep{CARMA}
allow the examination of structures on scales
between the high resolution of X-ray instruments and the somewhat lower resolution of the SZ survey instruments,
as well as covering the northern sky.

This study concerns a small selection of clusters of known X-ray structure, observed by AMI,
for which publicly-available CFHT data were accessible at the time the AMI observations were made.
As a pilot study for a future SZ-lensing comparison with a larger sample, we examine here six clusters with:
a redshift range of 0.16--0.41, a range of masses and varying degrees of merging activity. Given the depth of the optical
observations (see Section~\ref{sec:GL}), clusters at higher redshift would be more difficult to observe
as the field galaxy selection would likely be contaminated with foreground galaxies, and with only two optical bands
we would be unable to easily reduce this contamination. Below a redshift of about 0.08, AMI starts to resolve out
the cluster gas; this sample of clusters should not be affected by either of these issues, allowing us to
examine the agreement of lensing and SZ mass measurements and the effect of the cluster dynamical states.
AMI is limited to observing above Declination $20^{\circ}$ and is effectively
limited to measuring clusters of internal gas temperatures $\gtrapprox2$\,keV. Clusters
of lower temperature may be detectable but would require very high (unphysical) electron densities to produce
detectable SZ flux.

A data reduction pipeline
was developed to extract a weak-lensing catalogue from background galaxies in the CFHT fields. This allowed
the measurement of the total matter distributions of the galaxy clusters, which could then be directly compared
with gas measurements from the SZ. The SZ observations and data reduction are described in Section~\ref{sec:SZ},
and weak lensing in Section~\ref{sec:GL}. The Bayesian analysis, including sampling parameters
and priors, is described in Section~\ref{sec:modelling}.
We present notes on each cluster and its available data in Section~\ref{sec:results},
discuss the ramifications of these results in Section~\ref{sec:discussion}
and outline our conclusions in Section~\ref{sec:conclusions}.

Throughout, we assume a `concordance' $\Lambda$CDM cosmology, with $\Omega_{\mathrm{m,0}}=0.3$,
$\Omega_{\Lambda,0}=0.7$, and $H_{0}=70$km\,s$^{-1}$Mpc$^{-1}$ (and thus $h=H_{0}/100=0.7$).
%Relevant parameters are given in terms of $h$ so may be converted easily to a different value.
All coordinates are J2000 epoch and all optical magnitudes use the AB system.
\section{SZ observations}\label{sec:SZ}
AMI is a dual set of aperture-synthesis arrays located at the Mullard Radio
Astronomy Observatory, Lord's Bridge, Cambridge, UK. The AMI Small Array (SA) consists
of ten 3.7-m-diameter equatorially-mounted dishes with a baseline range of
$\simeq 5$--20\,m, while the AMI Large Array (LA) has eight 12.8-m-diameter dishes
with a baseline range of $\simeq 20$--100\,m. Both arrays observe I\,+\,Q polarisation flux densities in the band 12--18\,GHz, each with
system temperatures of about 25~K.

The back ends are analogue Fourier
transform spectrometers, from which the complex signals in each of eight
channels of 750-MHz bandwidth are synthesised, and the signals in the synthesised channels
are correlated at the $\simeq 10$ per cent level. In practice, the
two lowest-frequency channels are generally not used due to interference and
a poor correlator response in this frequency range. Further details are given in
\cite{2008MNRAS.391.1545Z}.

Details of the observations for these six clusters, made in 2008--2009, are shown in Table~\ref{tab:obslist}.
The SA observed a single pointing
while the LA used a (61+19)-point raster mode with 4\,arcmin\,spacing.
Phase calibrators were chosen from the Jodrell Bank VLA Survey
(\citealt{1992MNRAS.254..655P, 1998MNRAS.293..257B,
1998MNRAS.300..790W}) on the basis of proximity ($\leq5^\circ$) and flux density ($\geq1$\,Jy at 15\,GHz).

\begin{table*}
\begin{center}
\begin{tabular}{lcccccccc}
\hline
Cluster & RA & Dec & $z$ &       SA observing & SA map noise / & LA observing& LA 19-pt map noise /& LA 61-pt map noise /\tabularnewline
    & (J2000) & (J2000) & & time / hrs & $\mu$Jy\,beam$^{-1}$ & time / hrs & $\mu$Jy\,beam$^{-1}$ & $\mu$Jy\,beam$^{-1}$\tabularnewline
\hline
A115 & 00 55 59.50 & +26 19 14.0 & 0.197 & 67 & 114 & 19 & 101 & 237\tabularnewline
A1914 & 14 26 02.15 & +37 50 05.8 & 0.171 & 47 & 115 & 22 & 111 & 291\tabularnewline
A2111 & 15 39 44.08 & +34 24 56.2 & 0.229 & 43 & 90 & 22 & 99 & 268\tabularnewline
A2259 & 17 20 10.60 & +27 40 08.4 & 0.164 & 52 & 100 & 18 & 113 & 248 \tabularnewline
A611 & 08 00 59.40 & +36 03 01.0 & 0.288 & 76 & 85 & 23 & 76 & 208 \tabularnewline
A851 & 09 43 07.08 & +46 59 51.0 & 0.410 & 77 & 75 & 28 & 70 & 188 \tabularnewline
\hline
\end{tabular}
\caption{Details of the AMI observations of the six clusters.
\label{tab:obslist}}
\end{center}
\end{table*}
The AMI data reduction was performed using our in-house reduction software \textsc{reduce}.
This is used to apply path-delay corrections, to flag interference,
shadowing and hardware errors, to apply phase and amplitude calibrations and to
Fourier transform the correlated data to synthesise the frequency channels,
before writing output $uv$-\textsc{FITS} files suitable for imaging in the Astronomical
Image Processing System (\textsc{aips}; \citealt{aips}).

Flux calibration was performed using short observations of 3C48 and 3C286 near
the beginning and end of each run, with assumed I\,+\,Q flux densities for these sources in
the AMI channels consistent with \citet{1977A+A....61...99B} (see Table~\ref{tab:Fluxes-of-3C286}).
As \citeauthor{1977A+A....61...99B} measure I and AMI measures I\,+\,Q, these flux densities
include corrections for the polarization of the sources.
After phase calibration, the phase of AMI over one hour is generally
stable to $5^{\circ}$ for channels 4--7, and to $10^{\circ}$ for channels 3 and 8.
The system temperatures of each AMI antenna are
continously monitored using a modulated noise signal injected at each
antenna; this is used to continuously correct the amplitude
scale in a frequency-independent way. The overall consistency of the
flux-density scale is estimated to be better than five~per~cent.

\begin{table}
\centering
\begin{tabular}{cccccc}\hline
 Channel & $\nu$/ GHz & $S(\mathrm{3C286})$/Jy & $S(\mathrm{3C48})$/Jy &
 FWHM$_{\mathrm{LA}}$/ arcmin \\ \hline
 3 & 13.9 & 3.74 & 1.89 & 6.08\\
 4 & 14.6 & 3.60 & 1.78 & 5.89\\
 5 & 15.3 & 3.47 & 1.68 & 5.70\\
 6 & 16.1 & 3.35 & 1.60 & 5.53\\
 7 & 16.9 & 3.24 & 1.52 & 5.39\\
 8 & 17.6 & 3.14 & 1.45 & 5.25\\ \hline
\end{tabular}
\caption{Assumed I\,+\,Q flux densities of 3C286 and 3C48 over the commonly-used AMI band, and the AMI
LA primary beam Full-Width-Half-Maxima (FWHM) for each channel.  \label{tab:Fluxes-of-3C286}}
\end{table}
\def\subcaption#1{\hbox to 7cm{\hfill #1 \hfill}}
\begin{figure*}
\centerline{\includegraphics[width=7cm,clip=]{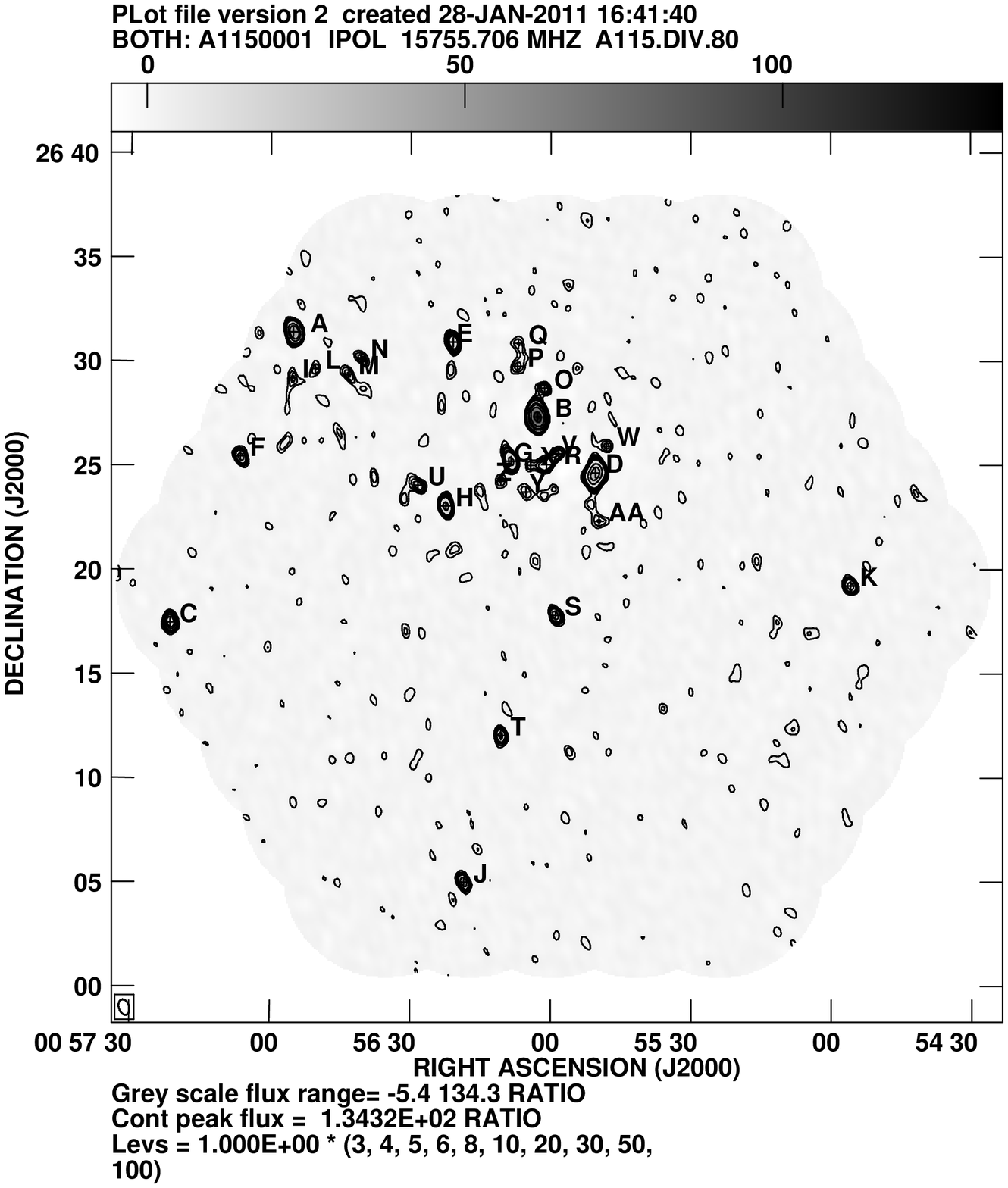} \qquad
            \includegraphics[width=7cm,clip=]{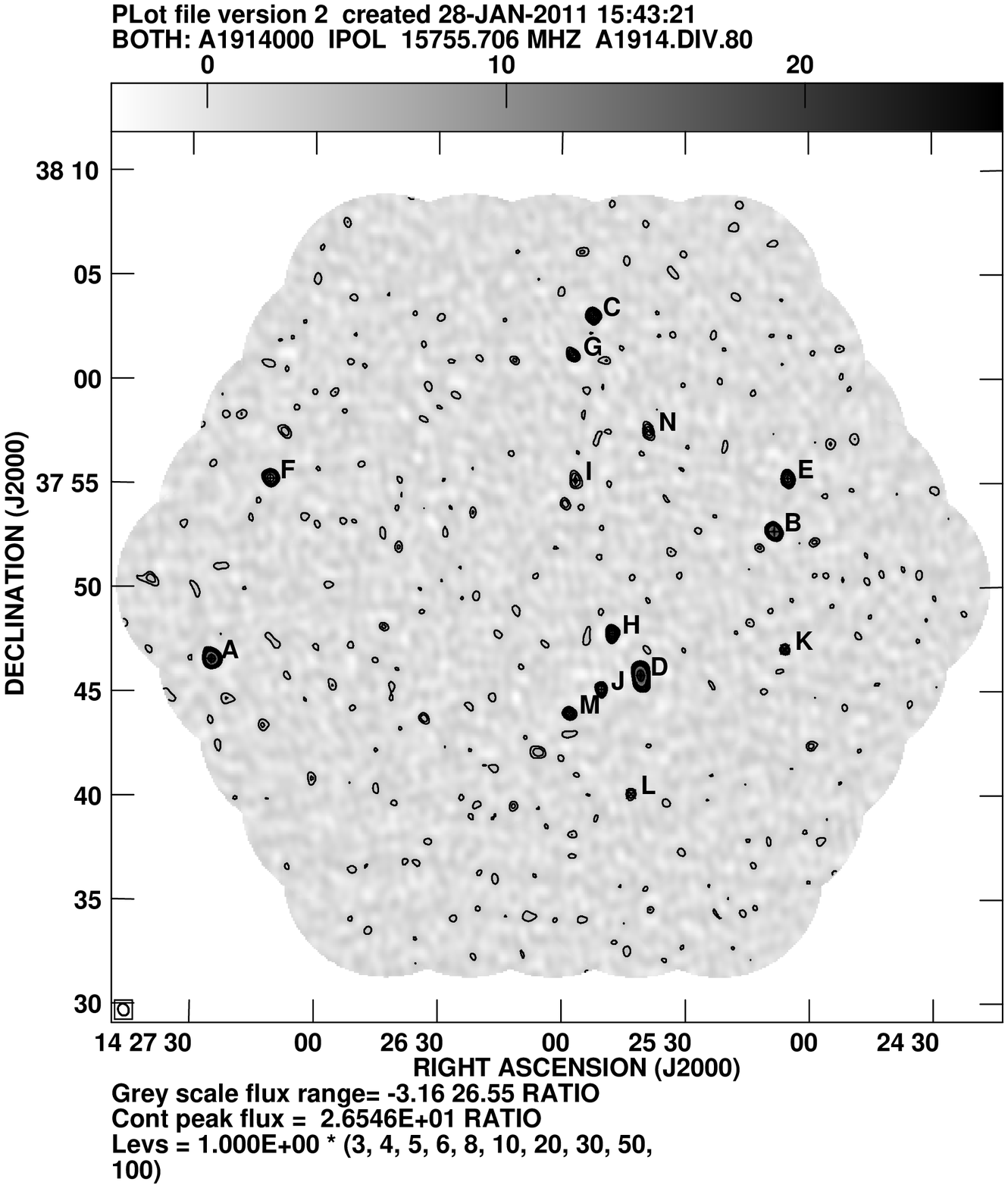}}
\centerline{\subcaption{(a) A115 (Table~\ref{tab:A115-sources}).} \qquad 
            \subcaption{(b) A1914 (Table~\ref{tab:A1914-sources}).}}
\smallskip
\centerline{\includegraphics[width=7cm,clip=]{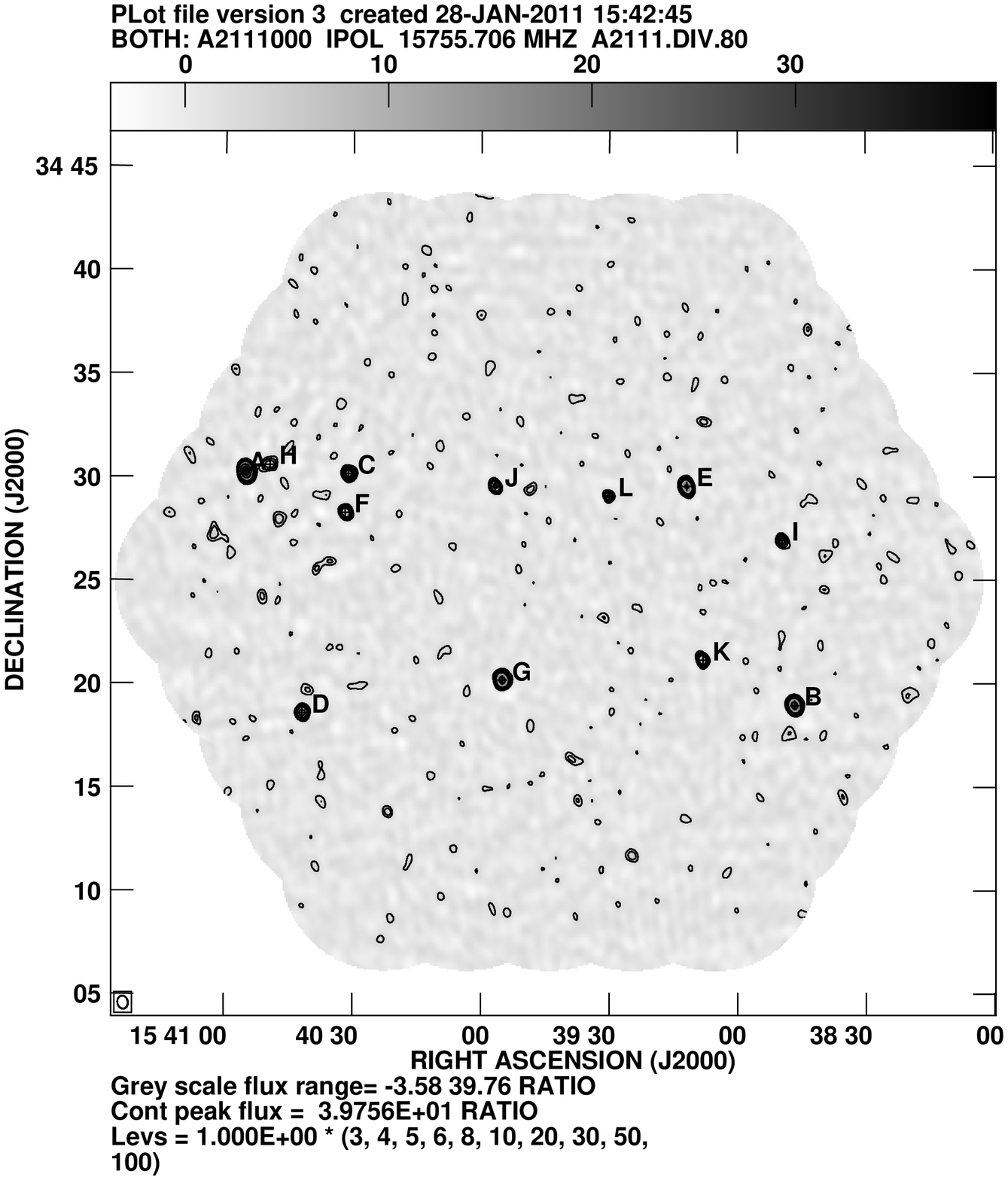} \qquad
            \includegraphics[width=7cm,clip=]{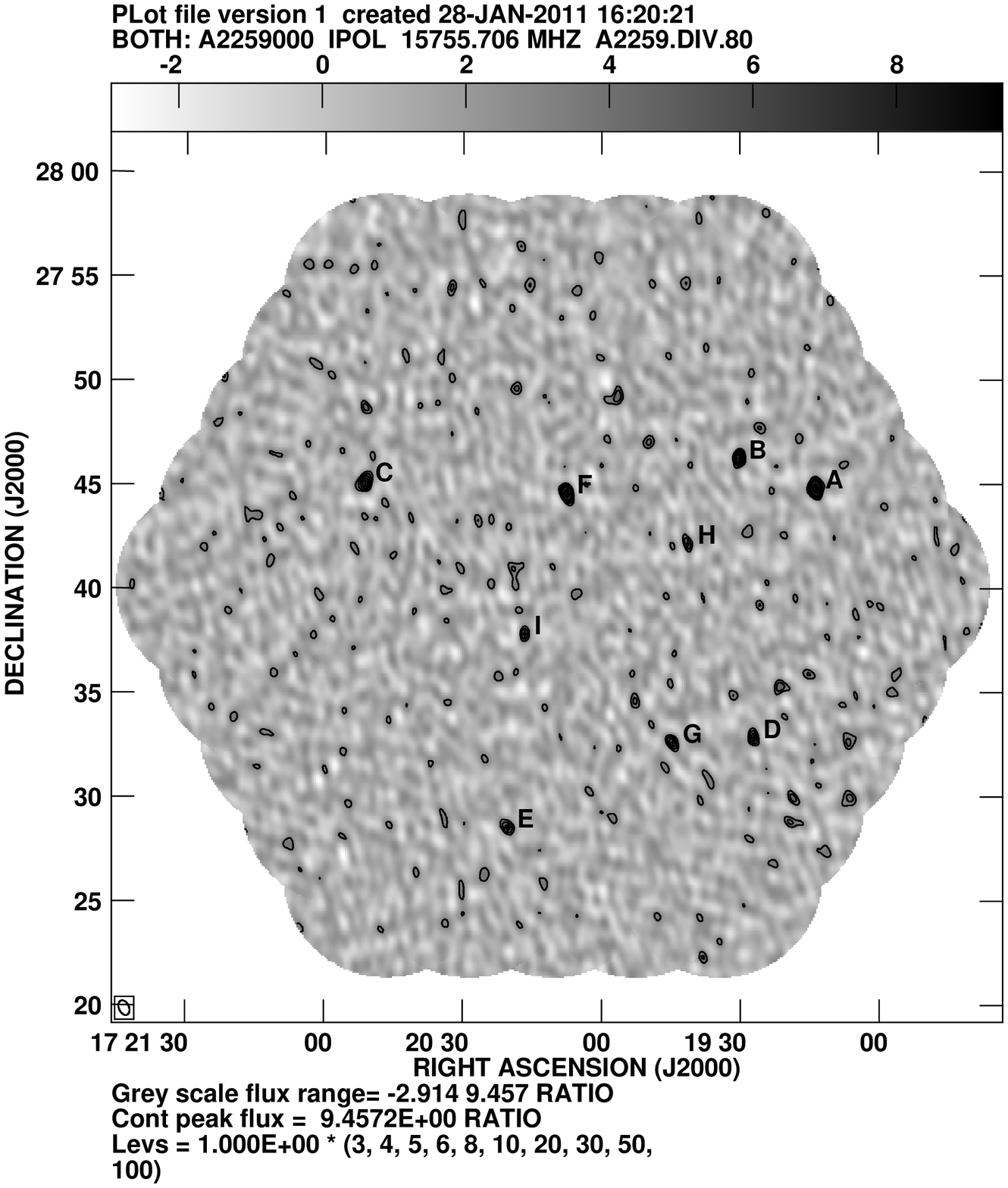}}
\centerline{\subcaption{(c) A2111 (Table~\ref{tab:A2111-sources}).}\qquad
            \subcaption{(d) A2259 (Table~\ref{tab:A2259-sources}).}}
\smallskip
\centerline{\includegraphics[width=7cm,clip=]{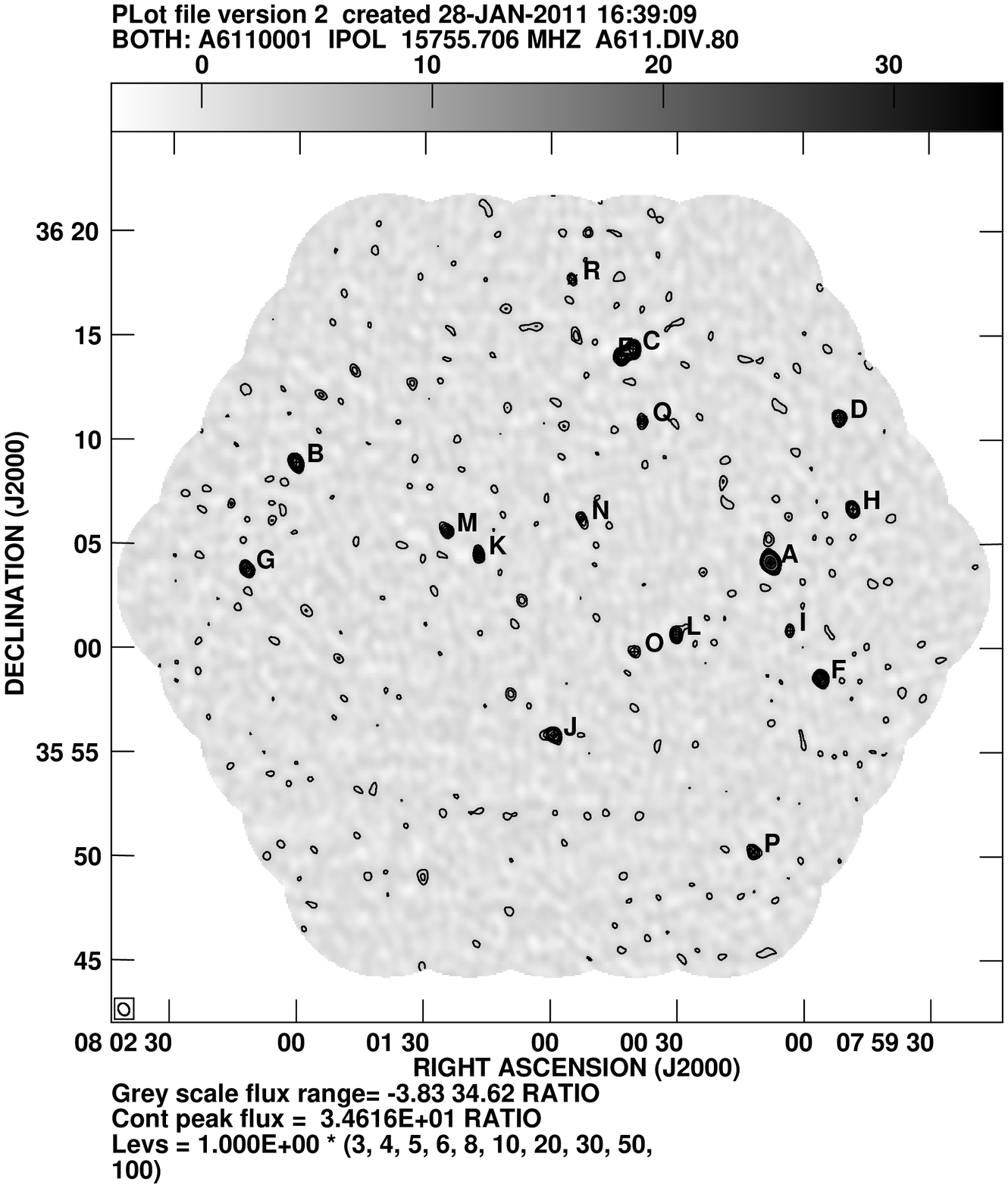} \qquad
            \includegraphics[width=7cm,clip=]{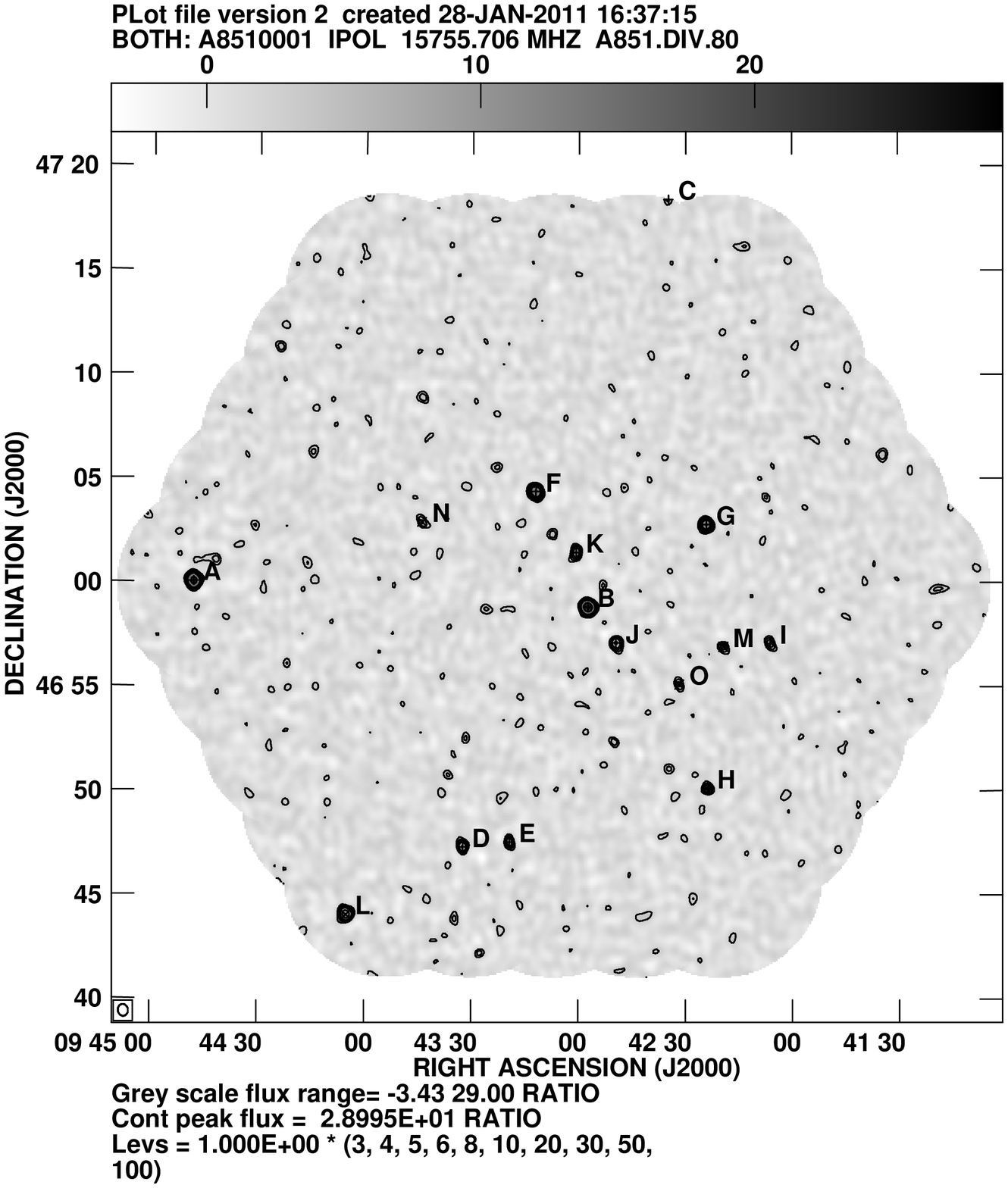}}
\centerline{\subcaption{(e) A611 (Table~\ref{tab:A611-sources}).} \qquad
            \subcaption{(f) A851 (Table~\ref{tab:A851-sources}).}}
\caption{LA raster signal-to-noise maps of A115, A1914, A2111, A2259, A611 and
A851, primary-beam-corrected, with sources detected at $4\sigma$ and
above labelled with identifying letters which refer to the sources as shown
in Tables~\ref{tab:A115-sources}--\ref{tab:A851-sources}.
Greyscale and contours are both signal-to-noise, with the latter in levels of
$+3$, $+4$, $+5$, $+6$, $+8$, $+10$, $+20$, $+30$, $+50$ and $+100$.
\label{fig:LA-cluster-maps}}
\end{figure*}

%Since the AMI antennas are sensitive to I+Q and are equatorially mounted,
%this polarization is fixed on the sky during the observation.
%Q may be positive or negative, but is expected to be small
%when integrated over the whole source.
%
To transform the SA data into a format suitable for our analysis, the unsmoothed
$uv$-data for all good observations were concatenated together to make
a single $uv$-\textsc{FITS} file for each channel. These were then transformed into
lists of visibilities for the purpose of generating a covariance matrix for the
data.

The \textsc{FITS} data were also binned into cells
of width $40\lambda$. This was determined by \citet{marshall03}
as a suitable binning scale which reduces the size of the data to
a manageable level without adversely affecting the resulting inference of cluster
properties.

Maps were made using \texttt{imagr} in \textsc{aips} for each channel of the SA and LA;
however we present only the combined-channel maps of the
SA and LA observations. \texttt{imean} was used on the LA individual maps to attach
the map noise to the map header, and \texttt{flatn} was then used to stitch the maps together, with
a primary beam correction applied using parameters shown in Table~\ref{tab:Fluxes-of-3C286}.

Source-finding was carried out on the LA data using software modified from that used in the
9C~survey \citep{waldram03}.
Spectral indices were fitted 
using LA maps for all six channels, assuming
source fluxes, $S$, follow a power-law relation of $S\propto\nu^{-\alpha}$ for the AMI frequencies, $\nu$.
The properties of point sources detected at $4\sigma$ and above by the LA
are used as priors when modelling the SA data in the analysis (see Section~\ref{sec:modelling}).
Fainter sources more than 5\,arcmin from the pointing centre are
directly subtracted in order to reduce the computational time needed; these are
marked with a `$\times$' instead of a `$+$' in the AMI maps.

The position prior is a delta function since the resolution of the LA
is around three times that of the SA. The flux densities $S$ are given
Gaussian priors: $\sigma$ is given by a conservative calibration
error of five~per~cent added in quadrature to the local map noise.
Spectral index ($\alpha$) priors are also Gaussians with $\sigma$ equal
to the error on the spectral index fit. These errors tend to be small ($\simeq0.4$)
for bright sources and large ($\simeq2$) for faint sources.

Fig.~\ref{fig:LA-cluster-maps} shows LA maps for the six
clusters with the detected radio point sources labelled with identifying letters. The positions
and properties of the sources are given in Tables~\ref{tab:A115-sources}--\ref{tab:A851-sources},
sorted by measured LA flux density.
\section{Weak gravitational lensing data}\label{sec:GL}
The clusters observed by AMI are of a few arcminutes angular extent.
High-quality optical data for the six clusters were retrieved from the
public Canadian Astronomy Data Centre archive (see \texttt{http://cadcwww.dao.nrc.ca/cadc/}).  All
clusters were observed for weak-lensing in the $r$-band ($\lambda_{\mathrm{eff}}=$\,623\,nm) and
for supplementary colour information in the $g$-band ($\lambda_{\mathrm{eff}}=$\,477\,nm), using MegaCam \citep{O/Bou++03}
on the Canada-France-Hawaii-Telescope (CFHT). MegaCam offers a field-of-view
of one square degree, which is more than sufficient for examining
the lensing signal from our six clusters.
After retrieval of the archival data they were astrometrically and photometrically calibrated and
finally co-added as described in \cite{2009A+A...493.1197E}, using textsc{swarp}
(see \texttt{http://www.astromatic.net/software/swarp}).
This resulted in approximately degree-square frames with pixel sizes of 0.186\,arcsec.

We give important characteristics of the finally co-added data in Table
\ref{tab:dataquality}. The limiting magnitude in this table is defined
as the 5-$\sigma$ detection limit in a 2\,arcsec aperture via
$m_{\textrm{lim}}=ZP-2.5\log(5\sqrt{N_{\textrm{pix}}}\sigma_{\textrm{sky}})$, where
$ZP$ is the magnitude zeropoint, $N_{\textrm{pix}}$ is the number of pixels
in a circle with radius 2\,arcsec and $\sigma_{\textrm{sky}}$ the sky
background noise variation. 
\begin{table*}
\begin{tabular}{ccccccc}
\hline
Cluster & obs. dates & P.I. & filter & exp. time [s] &
seeing [arcsec] & ZP mag. [AB mag] \\
\hline
A115 & 08/2004 -- 10/2004 & H. Hoekstra & $r$ & 6602.07 & 0.69 & 24.87 \\
A115 & 08/2004 -- 10/2004 & H. Hoekstra & $g$ & 1600.65 & 0.79 & 23.53 \\
A1914 & 05/2006 & H. Hoekstra & $r$ & 6001.9 & 0.71 & 24.56 \\
A1914 & 05/2006 & H. Hoekstra & $g$ & 3601.63 & 0.90 & 24.97 \\
A2111 & 05/2006 -- 06/2006 & H. Hoekstra & $r$ & 6482.13 & 0.66 & 24.94 \\
A2111 & 05/2006 -- 06/2006 & H. Hoekstra & $g$ & 1800.75 & 0.55 & 24.34 \\
A2259 & 08/2004 & H. Hoekstra & $r$ & 4001.29 & 0.82 & 24.18 \\
A2259 & 08/2004 & H. Hoekstra & $g$ & 1600.65 & 0.82 & 24.00 \\
A611 & 12/2004 -- 01/2005 & H. Hoekstra & $r$ & 4801.59 & 0.77 & 24.29 \\
A611 & 12/2004 -- 01/2005 & H. Hoekstra & $g$ & 2520.93 & 0.82 & 24.52 \\
A851 & 11/2004 & H. Hoekstra & $r$ & 6602.67 & 0.95 & 24.40 \\
A851 & 11/2004 & H. Hoekstra & $g$ & 2751.05 & 0.77 & 23.89 \\
\hline
\end{tabular}
\caption{Quality information on the co-added optical MegaCam data.\label{tab:dataquality}}
\end{table*}
\textsc{SExtractor} \citep{SEx} was run using default parameters on each of the images to extract all
sources.
Generally for each image, around $10^{5}$ objects were detected in the $r$-band, and around $7\times10^{4}$
in the $g$-band. An object was defined as consisting of 10 or more pixels above $3\times$ the local noise.
\subsection{Point spread function (PSF)\label{sub:Point-Spread-Function}}
The point spread function (PSF) describes the convolution of the
image due to the blurring effect of the atmosphere, telescope wind
shake, telescope optical distortions and CCD diffusion.
It is essential to correct for the effects of
the PSF, which can be at least as strong as the lensing signal. In
these CFHT images, the PSF ellipticity distorition is of the order of a few per~cent,
and is not always restricted to the edges of the field. Usefully,
foreground stars are unresolved and trace out the PSF, so by measuring
stellar ellipticities and sizes we can measure and then correct for the PSF.

Optically-detected objects can be described by five parameters,
their positions $x$, $y$, the ellipticity vectors $\epsilon_{1}$, $\epsilon_{2}$
and $ab$, the product of the semi-major ($a$) and semi-minor ($b$) axes at the FWHM
and thus describing the size
of the objects. The average FWHM of the stars for an image is taken as the
`seeing', and also has an effect on the lensing signal: poor seeing
reduces the ellipticity of objects and makes them appear more round: this tends to circularise
background galaxies and thus weaken the shear signal. $\epsilon_{1}$
and $\epsilon_{2}$ are the spin-2 tensors: $\epsilon_{1}=e\cos(2\theta)$;
$\epsilon_{2}=e\sin(2\theta)$,
where $e$ is the ellipticity $(a-b)/(a+b)$ and $\theta$ is the orientation, measured 
clockwise from south.

Within the source catalogues, stars form
distinct populations of low-FWHM objects and were easily extracted.
Stars with \textsc{SExtractor} flags of less
than 4 (i.e. their shapes are well-known and there are no nearby contaminating
objects) were used as the template stars for fitting the PSF. Typically 
for these images, around $3\times10^{3}$ stars were visible, with
the exception of A2259, whose location near the Galactic plane resulted
in the detection of $10^{4}$ stars. The $r$-band images of these stars
were further analysed using the program \textsc{im2shape}, which fits a Gaussian
to each object using a Bayesian method to find the most likely parameters
\citep{GL/Bri++01}.

The $r$-band images of A115, A2259 and A611 had discontinuous behaviour
along the joins between the individual frames. These joins, and areas directly around
bright stars and their diffraction patterns, were excluded
from the final lensing analysis, reducing the available image areas
by $\simeq20$\,per~cent. These regions are never directly over cluster
centres, are less than 0.5\,arcmin wide and are in a gridlike pattern so the
decrease in available field galaxies does not strongly affect any particular radial bin.
Objects with extreme ($e>0.95$) ellipticities were also discarded.

The functions of $\epsilon_{1}$, $\epsilon_{2}$ and $ab$ for the stars contained noise components
dependent on the individual frames and the \textsc{swarp} process, so the functional forms were not known.
Given the gaps discussed above and the inclusion of this
noise component, a simple interpolation scheme was not optimal. In order to fit
a smooth and continuous form for each function an artificial neural network was used to
`learn' the noise component and ultimately remove it, and interpolate over regions of
sparse or no data. Details of the network software used can be found from a previous
application to cosmology in \citet{2008MNRAS.387.1575A}.

\begin{figure*}
\begin{center}
\subfloat[A115.]{\includegraphics[width=5.51475cm,clip=]{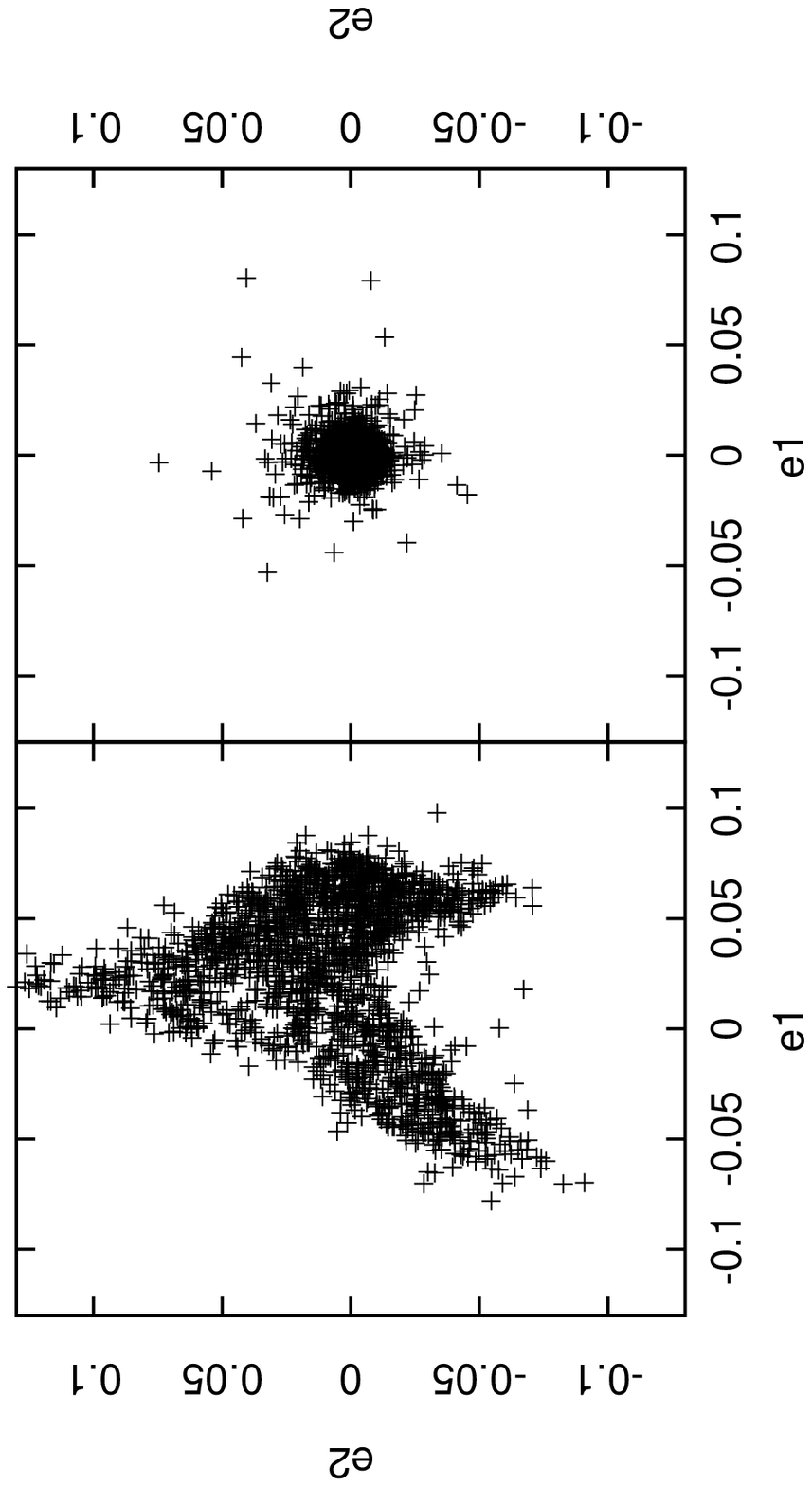}}\quad\subfloat[A1914.]{\includegraphics[width=4.959cm,clip=]{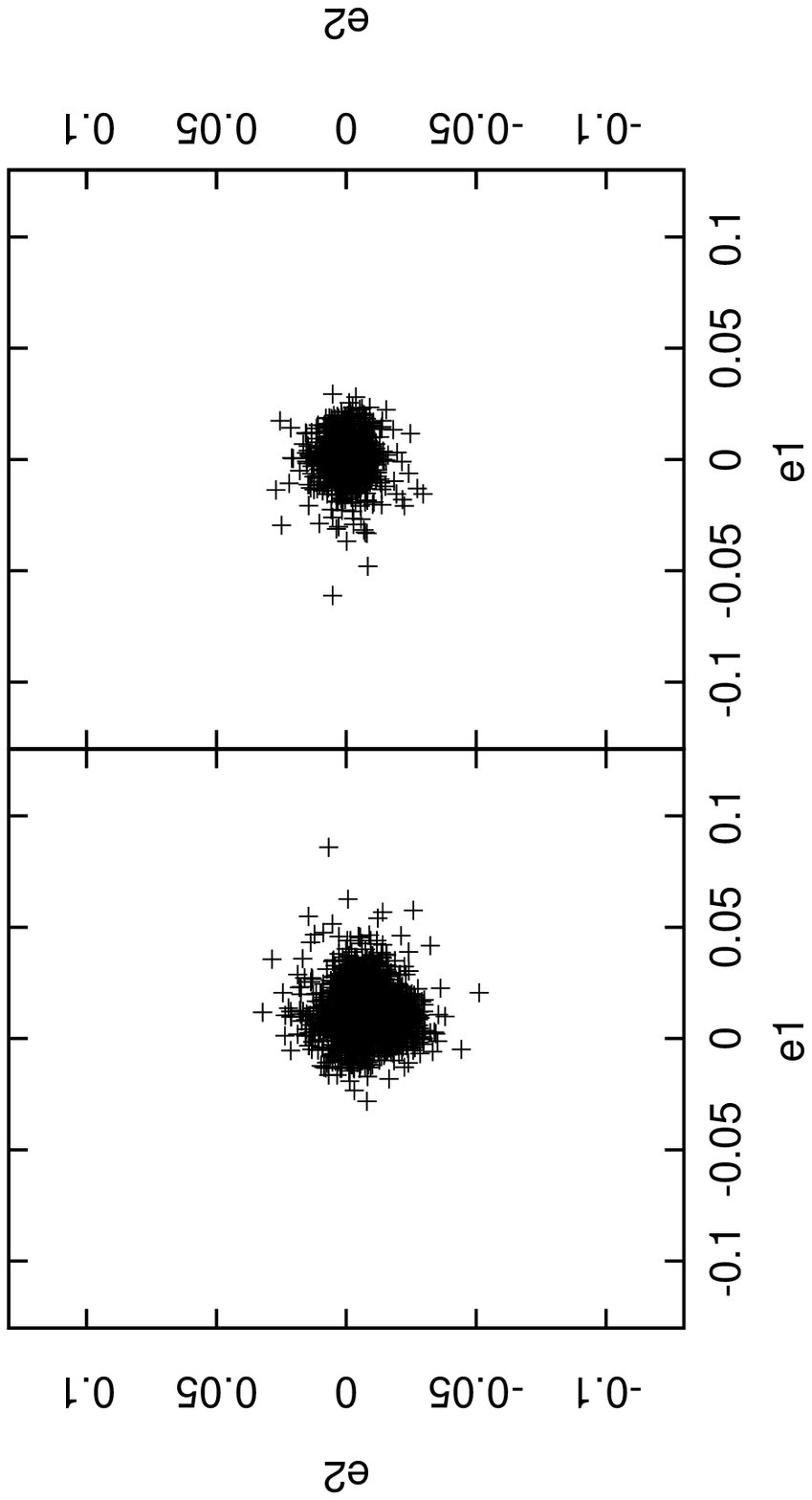}}\quad\subfloat[A2111.]{\includegraphics[width=5.68575cm,clip=]{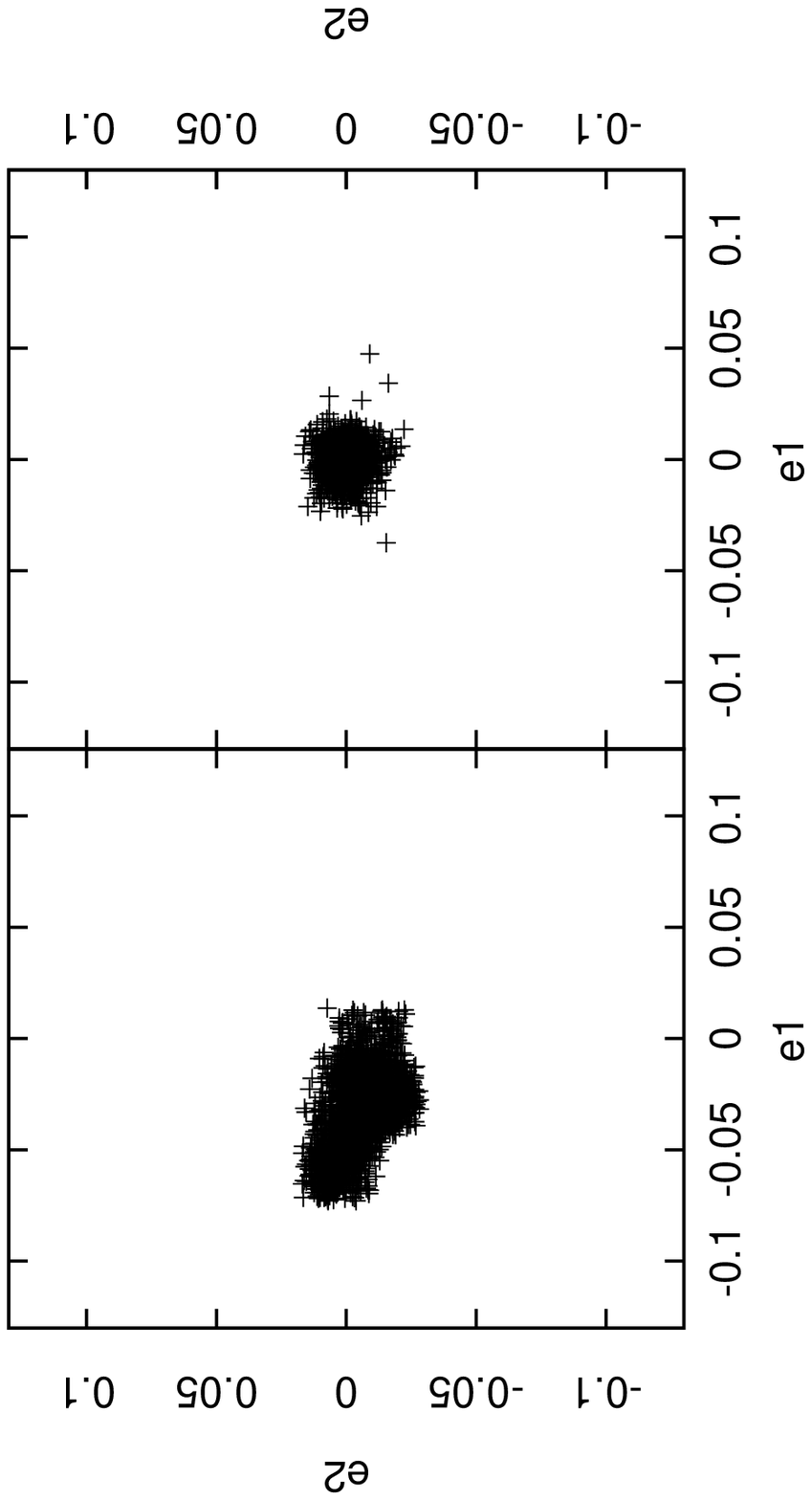}}\qquad
\subfloat[A2259.]{\includegraphics[width=5.51475cm,clip=]{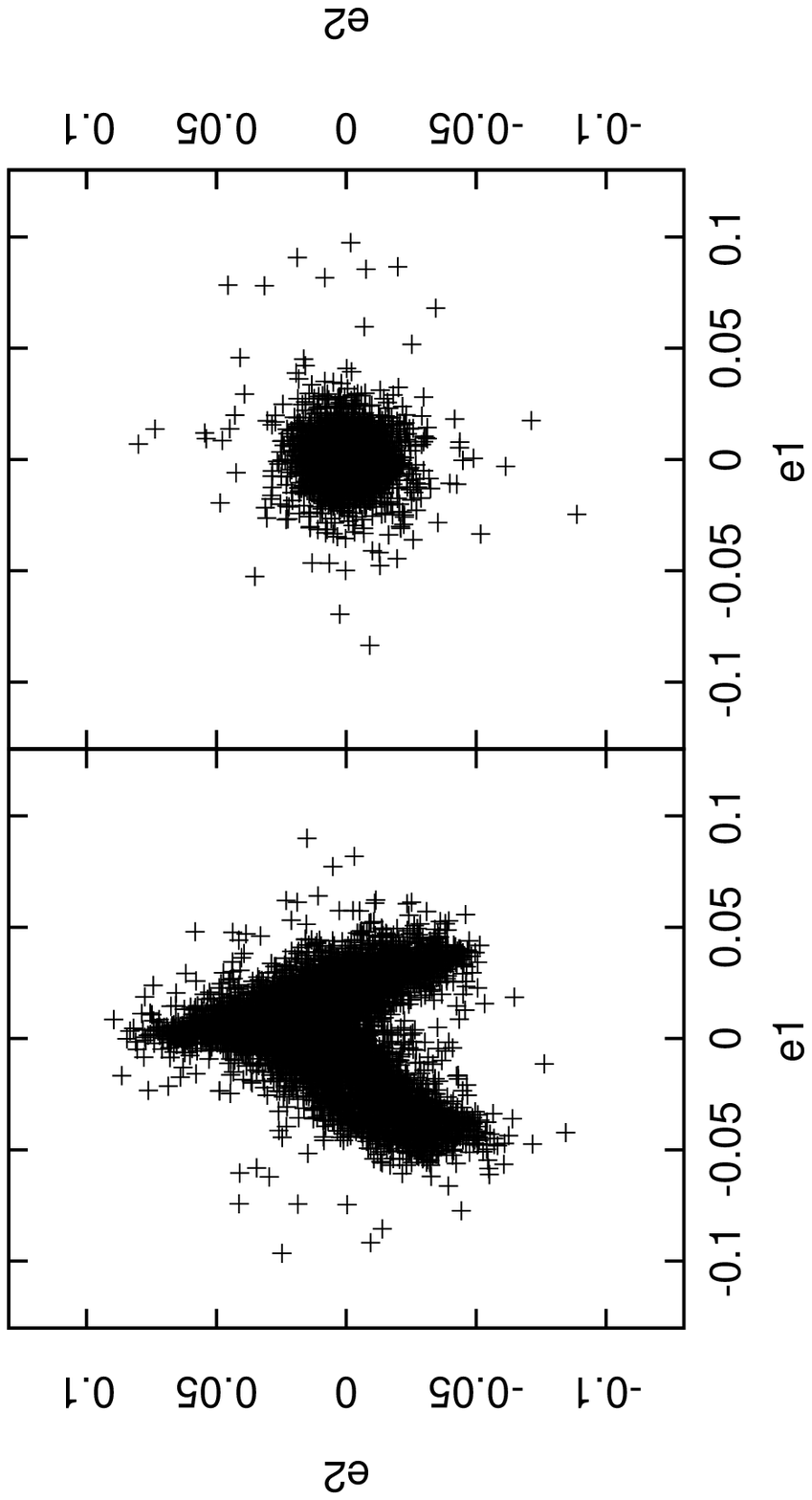}}\quad\subfloat[A611.]{\includegraphics[width=4.959cm,clip=]{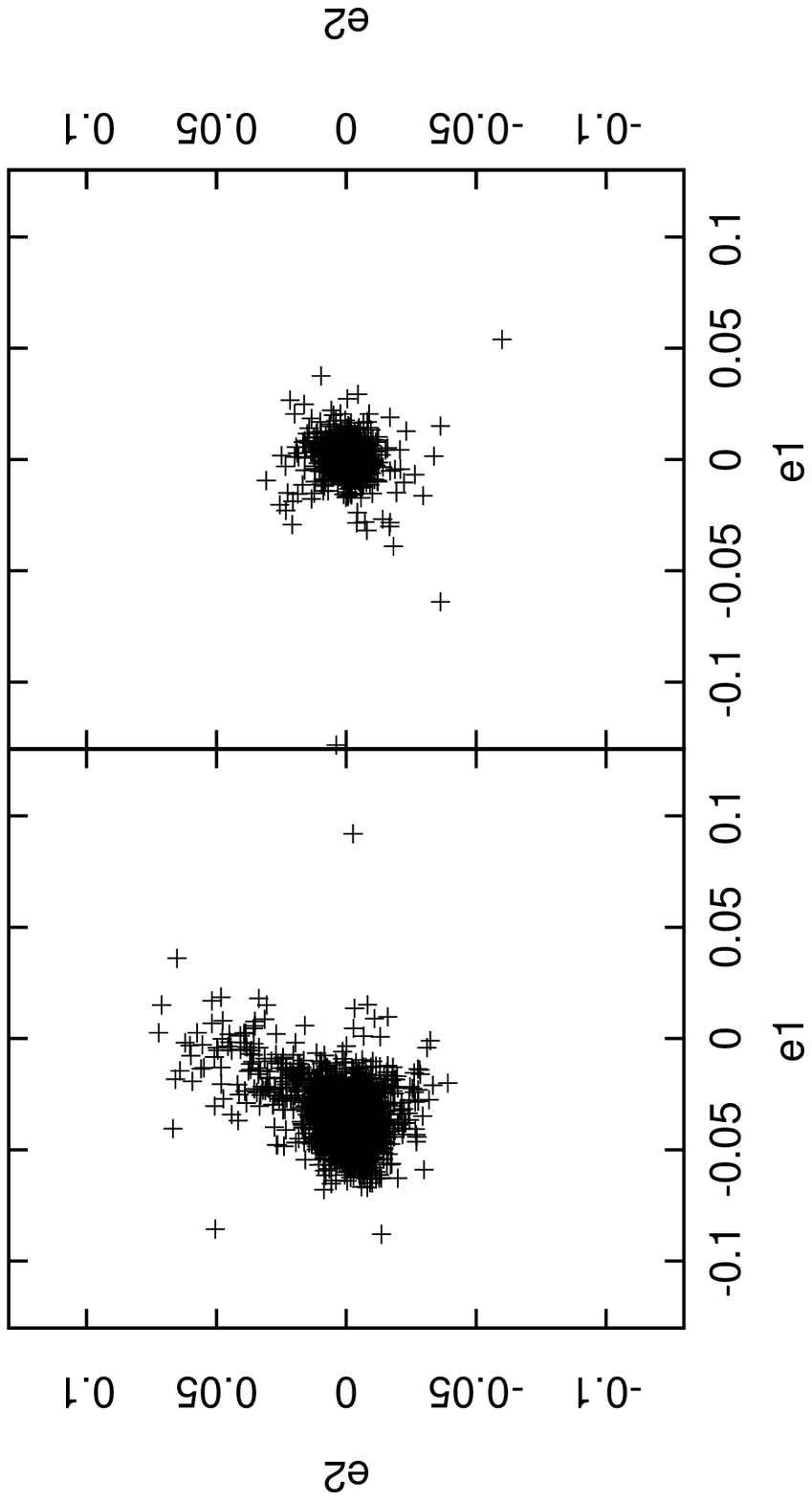}}\quad\subfloat[A851.]{\includegraphics[width=5.714cm,clip=]{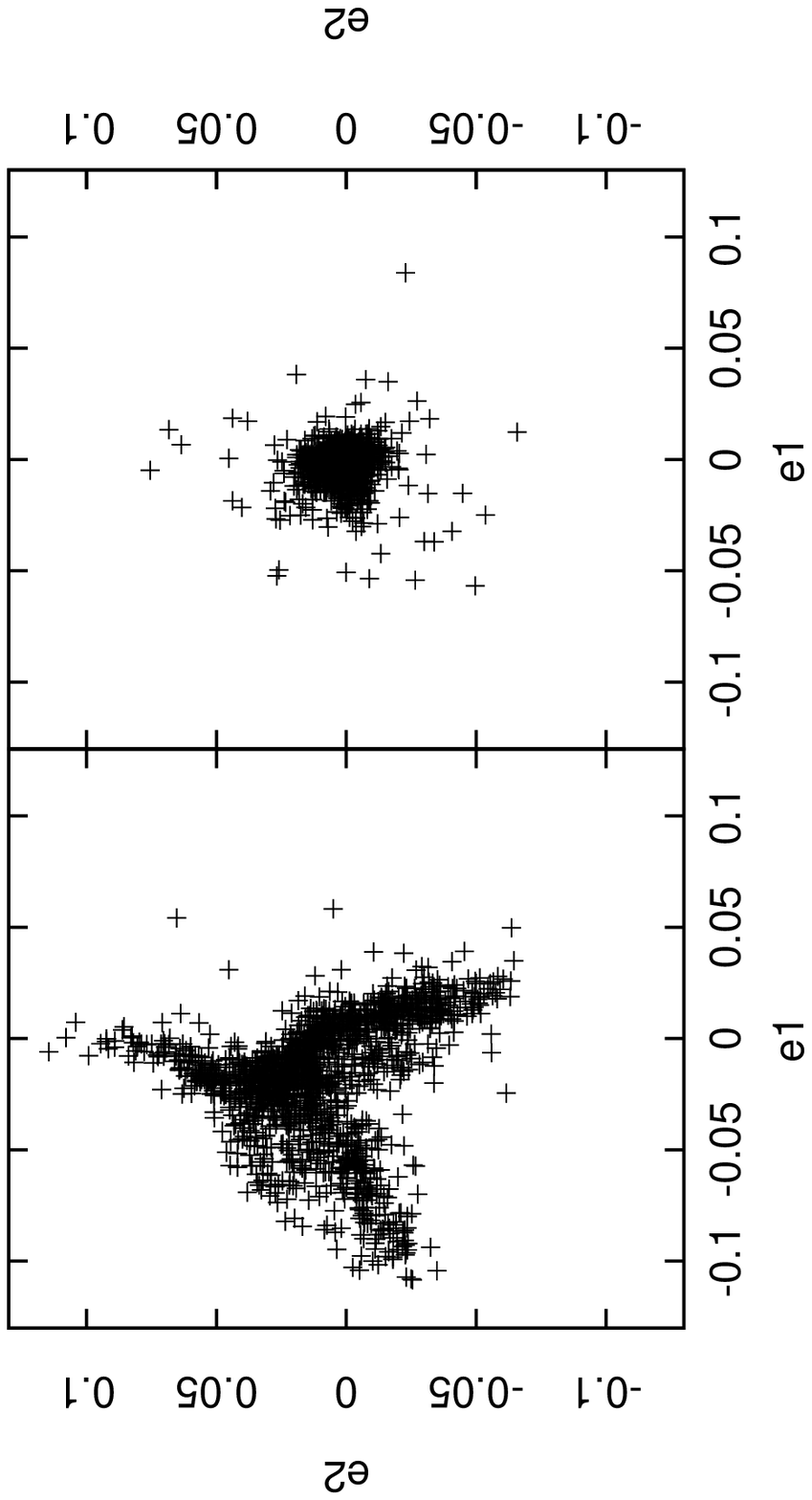}}
\caption{$\epsilon_{1}$ against $\epsilon_{2}$ for each stellar catalogue, before (left panels) and
after (right panels) PSF deconvolution. \label{fig:e1-against-e2}}
\end{center}
\end{figure*}

From these functions, a reference file describing the PSF was generated for the
background galaxy positions, to be deconvolved when fitting the galaxy shapes.
Residuals of $\epsilon_{1}$ and $\epsilon_{2}$ for the stellar catalogues
after PSF fitting
(calculated simply as $\epsilon_j^{\mathrm{star}} - \epsilon_j^{\mathrm{PSF}}$)
are shown in Fig.~\ref{fig:e1-against-e2}.
\subsection{Contaminant and cluster galaxy removal}
\begin{figure*}
\begin{center}
\subfloat[A115 \citep{2007A+A...469..861B}.]{\includegraphics[angle=0,width=5.5cm,clip=]{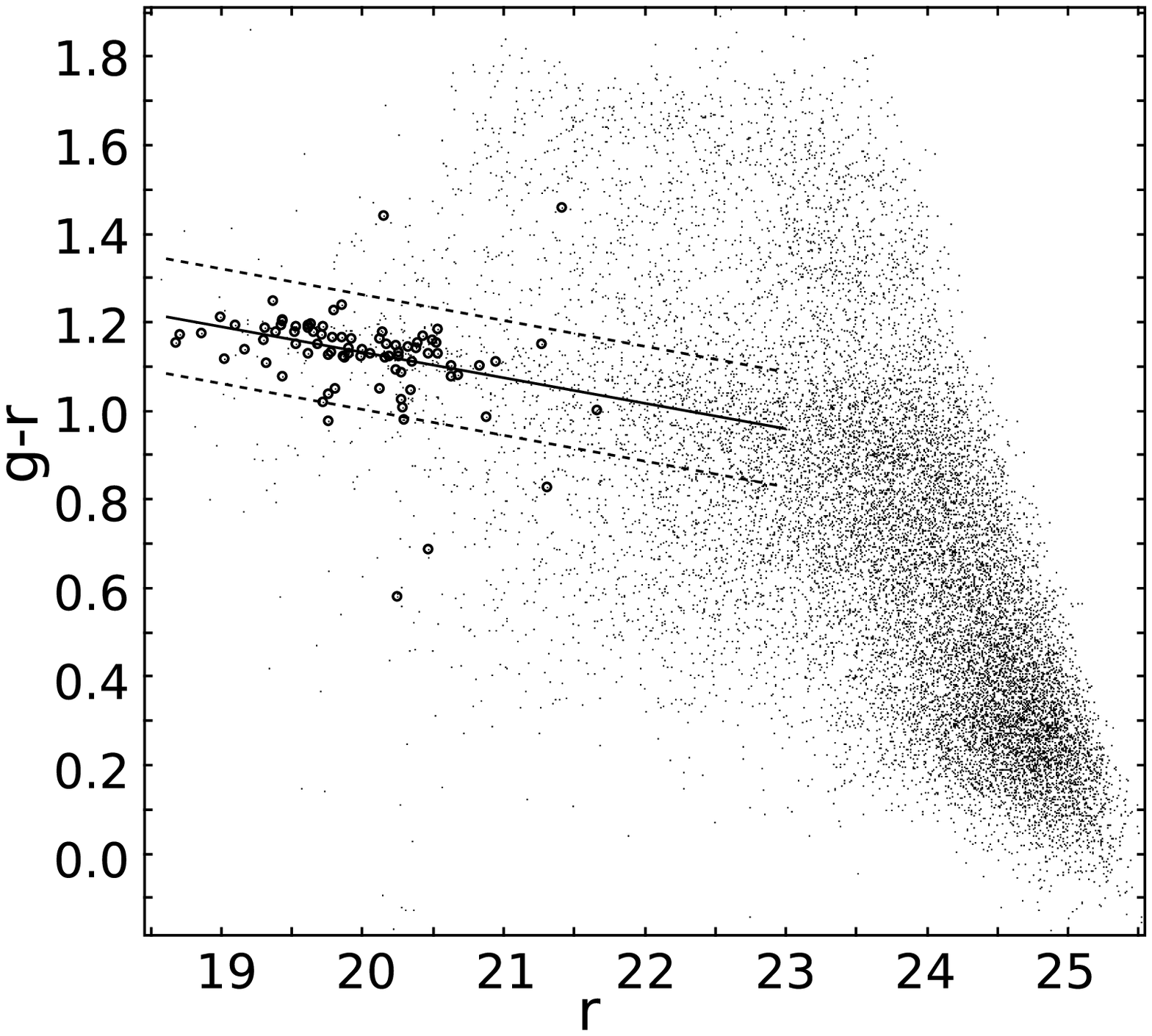}}\quad\subfloat[A1914 \citep{2008MNRAS.384.1502S}.]{\includegraphics[angle=0,width=5.5cm,clip=]{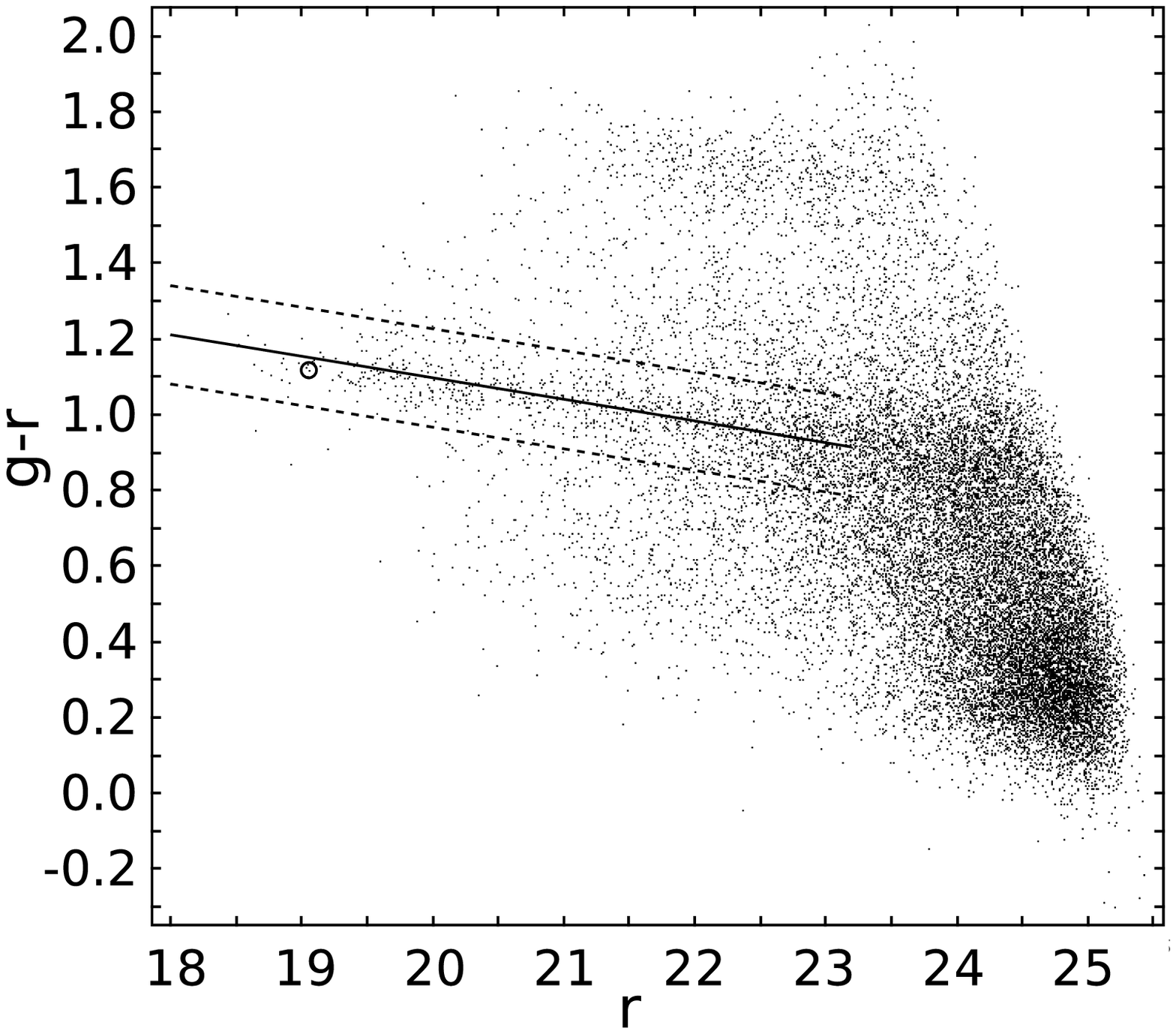}}\quad\subfloat[A2111 (\citealt{2006AJ....131.2426M}; \citealt{2008A+A...487..453A}).]{\includegraphics[angle=0,width=5.5cm,clip=]{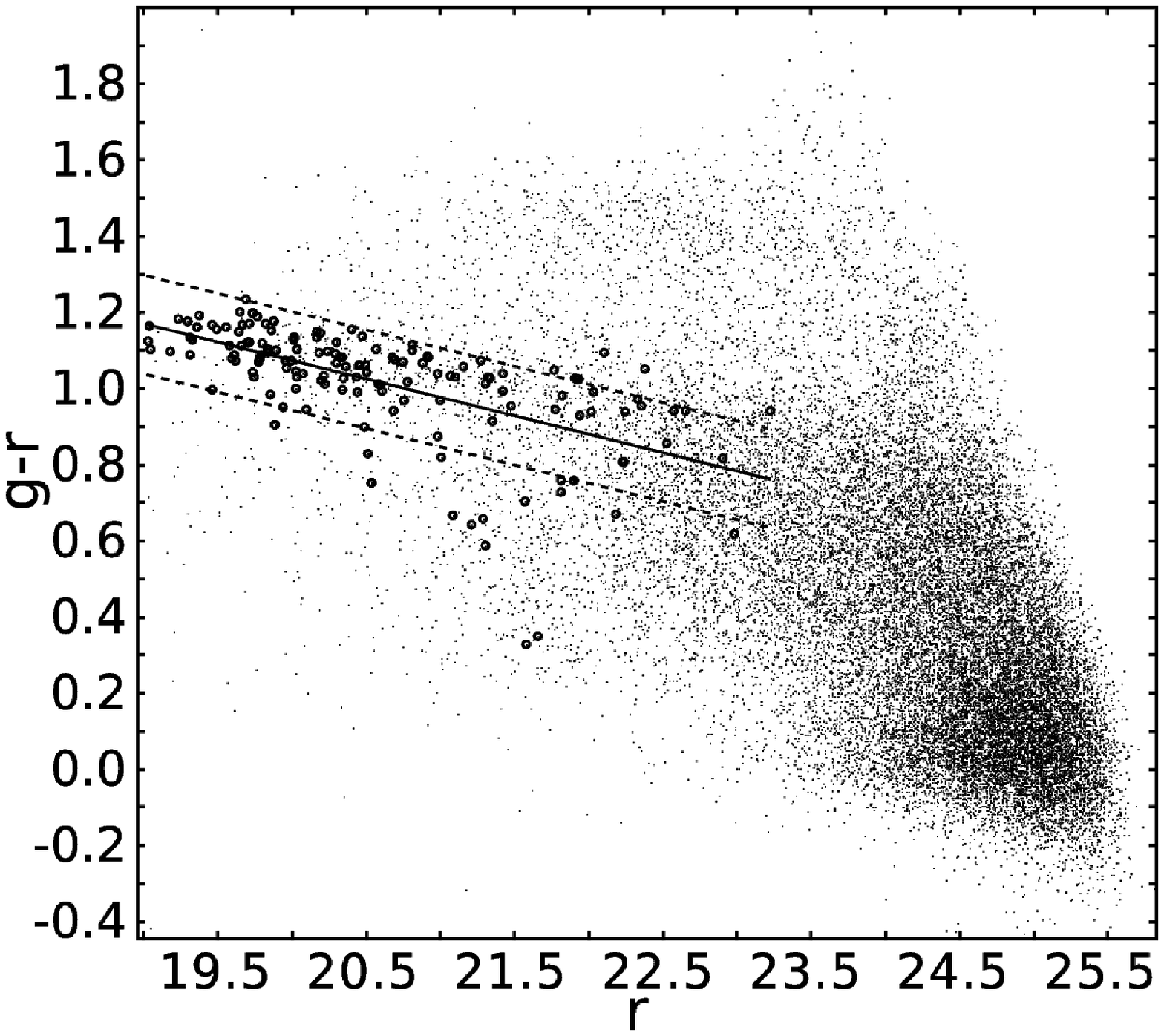}}\qquad
\subfloat[A2259 \citep{2008MNRAS.384.1502S}.]{\includegraphics[angle=0,width=5.5cm,clip=]{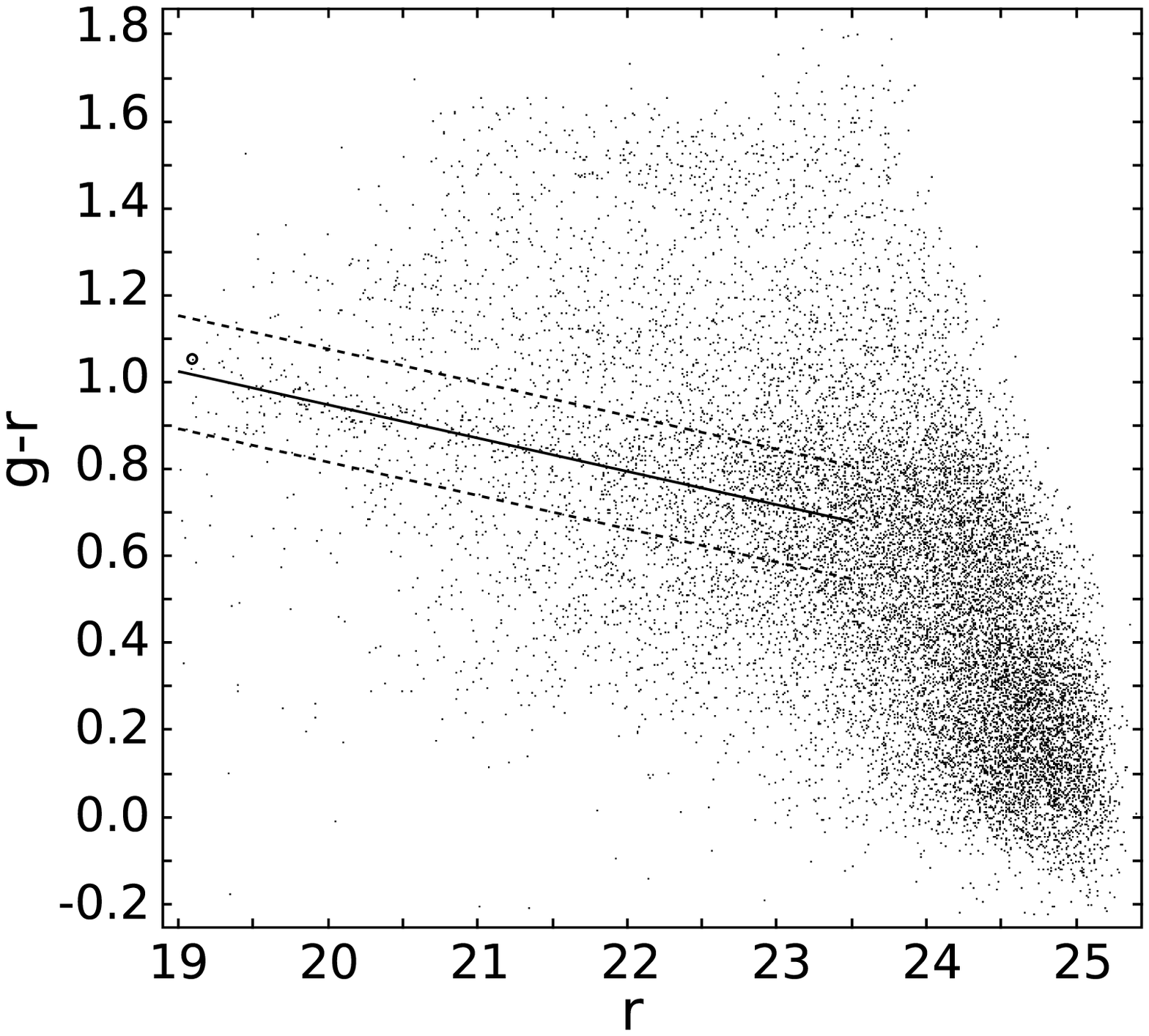}}\quad\subfloat[A611 \citep{2008MNRAS.384.1502S}.]{\includegraphics[angle=0,width=5.5cm,clip=]{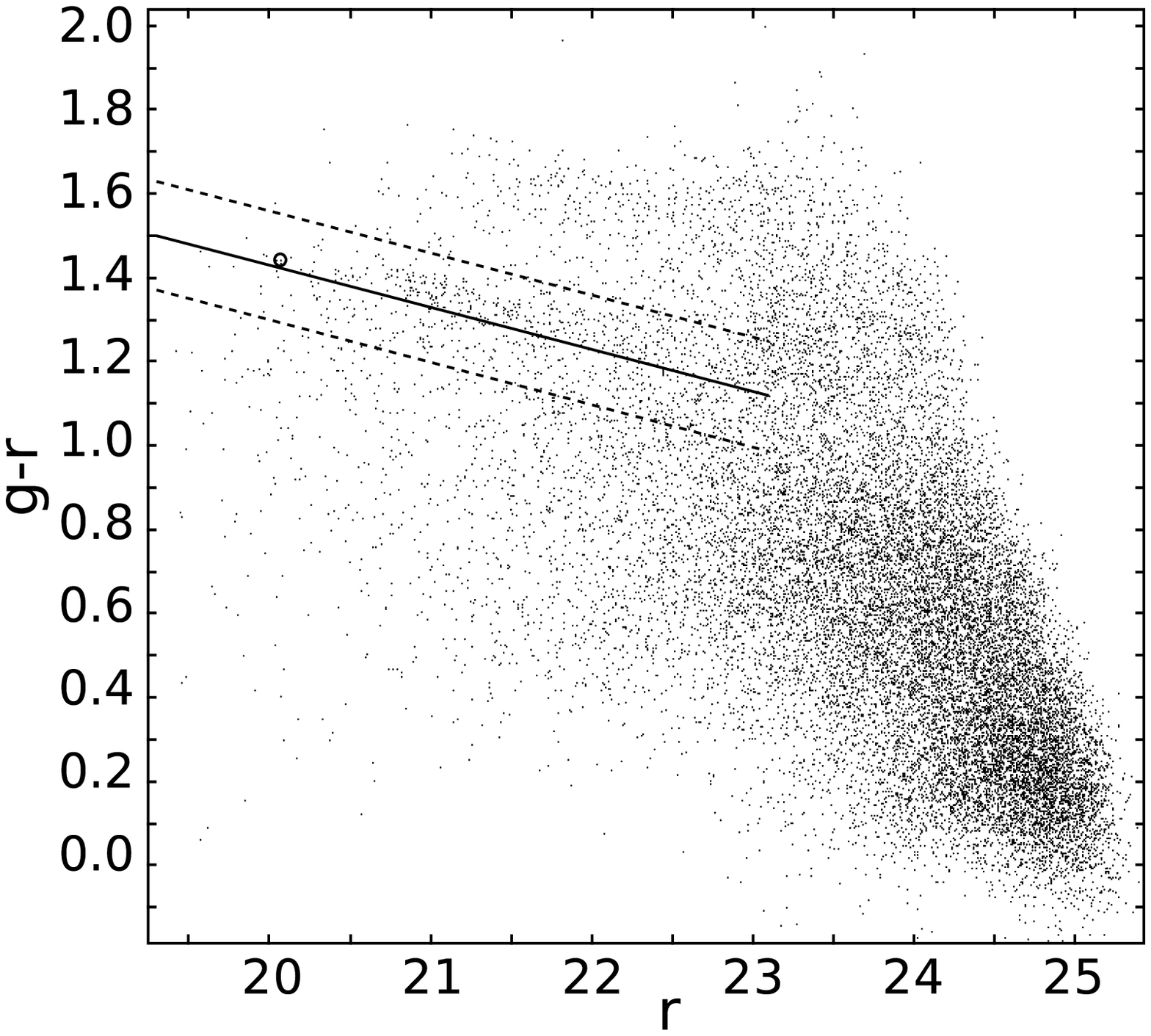}}\quad\subfloat[A851 \citep{1996A+AS..118...65B}.]{\includegraphics[angle=0,width=5.5cm,clip=]{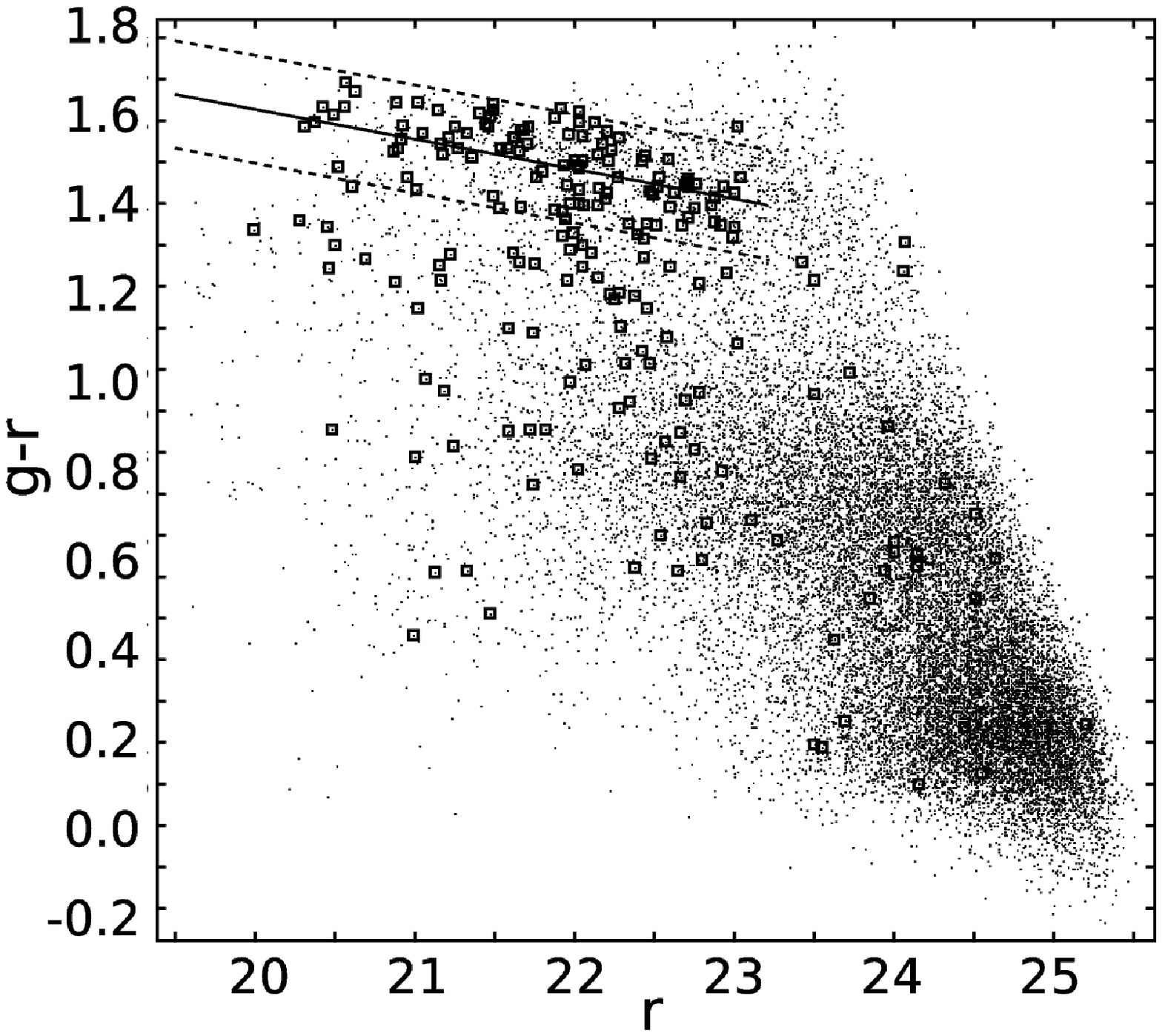}}
\caption{Red sequences identified by overdensity in the $g-r$:$r$ colour-magnitude plot. Reliably spectroscopically-identified
cluster galaxies are circled and references are given in the subfigure captions. In the case of A851, the identification of
cluster members is only accurate to 80~per~cent \citep{1996A+AS..118...65B}
and these are encased by $\Box$s. In each subfigure, the solid line shows the
red sequence fit described in Table~\ref{tab:red-seq-colours}, while the dashed
lines show the upper and lower limits on the space excluded
from the field galaxy selection.\label{fig:red-sequences}} 
\end{center}
\end{figure*}

Once PSFs had been determined for each $r$-band image, \textsc{im2shape} was
run again, this time on the background galaxy catalogue,
extracted from the initial full set of objects by:
\begin{itemize}
\item removing the stars and saturated objects, obvious from their FWHMs;
\item removing very blue objects, generally foreground galaxies with
large angular extent;
\item removing the cluster galaxies;
\item making a brightness cut in the $r$-band: $r>21$.
\end{itemize}
Cluster red sequence galaxies were found from their overdensity in the ($g-r$):$r$ colour-magnitude
diagram and examining those objects in the $x$--$y$ plane. We observed dense 
spatial clustering at the cluster positions indicating that the cut included the
cluster galaxies. Their distributions on the
sky also give rough estimates of the angular extents and thus sizes
and shapes of the galaxy cluster mass distributions, although we do not attempt
to use this information quantitatively in this paper. Table~\ref{tab:red-seq-colours}
shows the colour properties of the galaxies lying in the red sequence region and including the cluster
members, and Fig.~\ref{fig:red-sequences} shows the selections in ($g-r$):$r$ space.
A check on the selection method was made by examining the spectroscopically-confirmed
cluster members for A115 \citep{2007A+A...469..861B} and A2111
(\citealt{2006AJ....131.2426M} and \citealt{2008A+A...487..453A}) and matching their positions
with those in our catalogue; they were found to lie along the same red sequences.
Slightly older data from \cite{1999ApJS..122...51D} and \cite{1996A+AS..118...65B}
were used to find likely cluster members
of A851, with an estimated reliability of 80~per~cent.
The brightest cluster galaxies of the three less extensively-observed clusters A1914, A611 and A2259
are also marked in Fig.~\ref{fig:red-sequences}, originally identified by \cite{2008MNRAS.384.1502S}.
%Removing all galaxies in these ($g-r$):$r$-selected regions will also remove background galaxies,
%but as the majority of objects are faint ($r>23.5$), we estimate that we remove less than five~per~cent of the available lensing field galaxies.
%
\begin{table}
\begin{center}
\begin{tabular}{cccc}
\hline
Cluster & intercept at $g-r$ & gradient & max $r$-band magnitude\tabularnewline
\hline
A115 & 2.50 & $-0.0677$ & 23.0\tabularnewline
A1914 & 2.24 & $-0.0573$ & 23.2\tabularnewline
A2111 & 2.95 & $-0.0918$ & 23.2\tabularnewline
A2259 & 2.48 & $-0.0768$ & 23.5\tabularnewline
A611 & 3.44 & $-0.1007$ & 23.1\tabularnewline
A851 & 3.06 & $-0.0718$ & 23.2\tabularnewline
\hline
\end{tabular}

\caption{Colour properties of the red sequences for each galaxy cluster, plotted in Fig.~\ref{fig:red-sequences}.\label{tab:red-seq-colours}}
\end{center}
\end{table}

For each cluster, around $2\times10^{3}$ objects were excluded for being on
the red sequence, and a few hundred contaminating large,
blue, or very bright objects were removed. After a
cut at $r$-magnitude$=21$ to exclude the brightest objects,
catalogues of 5--$8\times10^{4}$ `background' galaxies were therefore obtained for each cluster,
corresponding to number densities of 14--22 galaxies per square arcminute.

\subsection{Background galaxies}\label{sec:bggals}

\textsc{im2shape} was run on the background galaxies, and its output
used to generate a shear catalogue of $x$, $y$, $\epsilon_{1}$, $\epsilon_{2}$,
$\sigma_{\epsilon_{1}}$and $\sigma_{\epsilon_{2}}$. The errors $\sigma_{\epsilon_{1}}$and
$\sigma_{\epsilon_{2}}$ are the fitting errors
of the respective ellipticity vectors added in quadrature with the
root-mean-squared (r.m.s.) of the intrinsic ellipticity distribution of galaxies, 0.25 \citep{GL/HFK00}.

\begin{figure*}
\begin{center}
%
%\subfloat[A115.]{\includegraphics[angle=-90,width=6cm,clip=]{A115_SIS}}\subfloat[A1914.]{\includegraphics[angle=-90,width=6cm,clip=]{A1914_SIS}}\subfloat[A2111.]{\includegraphics[angle=-90,width=6cm,clip=]{A2111_SIS}}\qquad
%
%\subfloat[A611.]{\includegraphics[angle=-90,width=6cm,clip=]{A611_SIS}}\subfloat[A851 -- left.]{\includegraphics[angle=-90,width=6cm,clip=]{A851_left_SIS}}\subfloat[A851 -- right.]{\includegraphics[angle=-90,width=6cm,clip=]{A851_right_SIS}}
%
{\includegraphics[width=16cm,clip=]{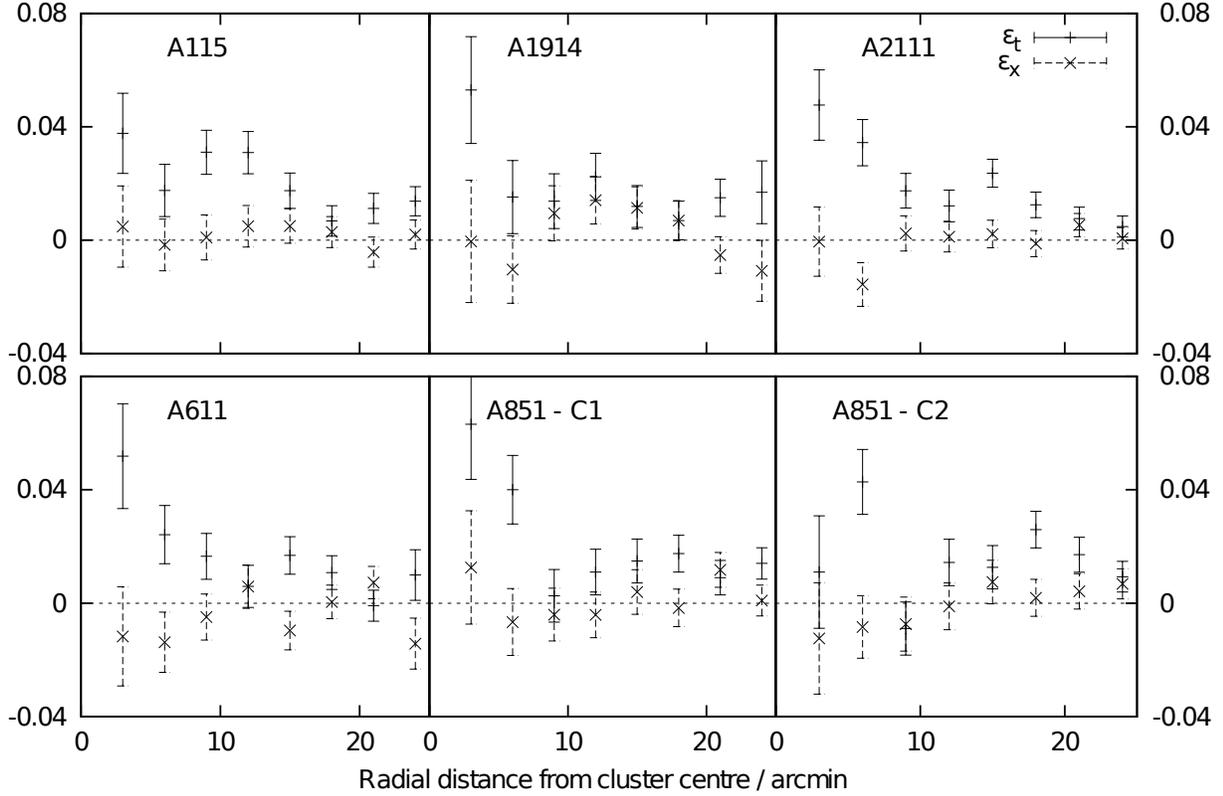}}
\caption{$\epsilon_{t}$ and $\epsilon_{x}$ of the field galaxies binned in 3\,arcmin bins from the cluster
centres (see Section~\ref{sec:bggals}).\label{fig:binned-signals}}
\end{center}
\end{figure*}

A simple test of whether the PSF deconvolution has been performed correctly,
and the lensing signal is the dominant component of the map, is to bin and compare
the E- and B-modes of the ellipticities: $\epsilon_{t}$ and $\epsilon_{x}$,
respectively. The former should be positive and
decreasing with angular distance from the cluster centre, and the latter
should be zero within the errors (see e.g. \citealt{gl-basics}).
This test was performed by binning the ellipticities
in radial bins of width 3\,arcmin, from the fitted cluster
centroid positions (Section~\ref{sec:results}) for A2111, A611 and A1914. For A115,
the centre used was the position of the highest-mass clump; for A851, two separate mass
overdensities were detected, so the signal is shown for both of the components.
The plot for A2259 is not shown as the lensing signal was too weak to recover good cluster parameters.
Fig.~\ref{fig:binned-signals} shows the binned ellipticity vectors and, as expected,
the B-mode signal is negligible given the errors, calculated as
the r.m.s. of each bin divided by the square root of the number of galaxies in that bin.
%
%To model the lensing effect of the cluster, we needed some estimate
%of the redshift of the lensed objects. This was calculated by averaging
%the photometric redshifts of the COSMOS catalogue \citep{COSMOScat}, for $r$-magnitude$<26$,
%the magnitude zero-point in the CFHT images used. The result was
%$z=0.979$ with a standard deviation of 0.714. Spectroscopic redshifts for the
%lensed sources would, of course, be ideal; however with catalogues
%of at least $5\times10^{4}$ objects per cluster, this is unfeasible.
%There are more thorough ways of estimating the photometric redshifts
%of the lensed galaxies, but with only two colour bands available,
%the accuracy is necessarily poor. We therefore assumed an
%average redshift of $z=1.0\pm0.7$.

To model the mass of the cluster via lensing, we needed some estimate
of the redshifts of these lensed field galaxies. Fortunately, the CFHT Legacy Survey Deep Field catalogues
(see \texttt{http://www.cfht.hawaii.edu/Science/CFHTLS/})
were produced using the same instrument and filters as our data, but to
a greater depth, and with much greater redshift accuracy. We were therefore able to make cuts within this catalogue
at the same levels of those of our own data, and then average the photometric
redshifts of the galaxies within this selection. This resulted in
a redshift distribution with a mean of 1.2, a median of 1.1 and an r.m.s. of 0.7. With the
Bayesian analysis method we employ (see Section~\ref{sec:modelling}) we
are able to incorporate this large uncertainty as a sampling parameter,
propagating the uncertainty through to the estimation of our cluster
parameters.
%
%COSMOS catalogue \citep{COSMOScat}, for $r$-magnitude$<26$,
%the magnitude zero-point in the CFHT images used. The result was
%$z=0.979$ with a standard deviation of 0.714. Spectroscopic redshifts for the
%lensed sources would, of course, be ideal; however with catalogues
%of at least $5\times10^{4}$ objects per cluster, this is unfeasible.
%There are more thorough ways of estimating the photometric redshifts
%of the lensed galaxies, but with only two colour bands available,
%the accuracy is necessarily poor. We therefore assumed an
%average redshift of $z=1.0\pm0.7$.
%
\section{Modelling and Priors}\label{sec:modelling}
To analyse the AMI cluster observations we use a Bayesian analysis methodology
(\citealt{marshall03} and \citealt{McAdam})
%implemented as
%Markov Chain Monte Carlo Astronomical Detection And Measurement (\textsc{McAdam}).
%This software
which implements \textsc{MultiNest}
(\citealt{feroz07} and \citealt{MultiNest}), an application of nested sampling
\citep{skilling04}, to efficiently explore multi-dimensional parameter space and
to calculate Bayesian evidence. This analysis has been applied to pointed observations
of known clusters (e.g. \citealt{bullet-like}; \citealt{the-eight-clusters}), 
and also to detect previously unknown clusters \citep{AMIblind}. As is standard in Bayesian
methods, priors are given for sampling parameters; the LA data is used to produce
priors for the radio sources present in the SA data.
\subsection{SZ}
The SZ effect from a cluster can be measured by its comptonisation parameter, $y$,
which is the integral of the gas pressure along the line of sight $l$ through the cluster:
\begin{equation}
y(s)=\frac{\sigma_{T}}{m_{\textrm{{e}}}c^{2}}\intop_{-\infty}^{\infty}n_{\textrm{{e}}}k_{\textrm{{B}}}T\textrm{d}l, %\propto\int_{-r_{\textrm{lim}}}^{+r_{\textrm{lim}}}\rho_{\textrm{{g}}}T\textrm{d}l,
\label{eq:compton} 
\end{equation}
where $\sigma_{\textrm{T}}$ is the Thomson scattering cross-section,
$T$ is the temperature of the ionised cluster gas, $n_{\textrm{e}}$ is the
electron number density, (measured using Equation~\ref{eq:ne}), $m_{\textrm{e}}$ is the electron mass, $c$ is the speed
of light and $k_{\textrm{B}}$ is the Boltzmann constant. $s=\theta D_{\theta}$
is the deprojected radius such that $r^{2}=s^{2}+l^{2}$ and
$D_{\theta}$ is the angular diameter distance to the cluster.
%which can be
%calculated for clusters at redshifts, $z$, using 
%\begin{equation}
%D_{\theta} =\frac{c\int^z_0 H^{-1}\left(z'\right)\textrm{d}z'}{\left(1+z\right)}. 
%\end{equation}
%
A cluster with comptonisation parameter $y$ appears as a surface brightness
fluctuation of magnitude
\begin{equation}
\delta I_{\nu}=f(\nu)yB_{\nu}\left(T_{\mathrm{CMB}}\right),
\end{equation}
where $B_{\nu}(T_{\mathrm{CMB}})$ is the CMB blackbody spectrum and $f(\nu)$
is the SZ spectrum.
Interferometers such as AMI effectively measure spatial
fluctuations in surface brightness $\delta I_{\nu}$ in the visibility
plane, so without any transformation to the map plane (and without
the associated problems wherein), the model prediction can be directly
compared to the data.

The likelihood function for a set of cluster parameters $\mathbf{x}$ can be written:
\begin{equation}
L_{{\rm SZ}} = \mathrm{\mathbf{Pr}} \left( \mathrm{\mathbf{d}} \mid \mathbf{x} \right) =
\frac{1}{Z_{L}}\exp\left(\frac{-\chi^{2}}{2}\right),\label{eq:sz-like}
\end{equation}
where $\chi^{2}$ is a statistic quantifying the misfit between the
observed data $\mathrm{\mathbf{d}}$ and predicted data $\mathrm{\mathbf{d}}^{\mathrm{p}}$ (the latter of which
is a function of the model SZ surface brightness $\delta I_{\nu}$):
\begin{equation}
\chi^{2}=\left(\mathrm{\mathbf{d}}-\mathrm{\mathbf{d}}^{\mathrm{p}}\right)^{T}\mathrm{\mathbf{C}}^{-1}\left(\mathrm{\mathbf{d}}-\mathrm{\mathbf{d}}^{\mathrm{p}}\right),
\end{equation}
and the normalisation factor is:
\begin{equation}
Z_{L}=\left(2\pi\right)^{N_{\mathrm{vis}}}\mid \mathrm{\mathbf{C}}\mid.
\end{equation}
$N_{\mathrm{vis}}$ is the number of visibilities and $\mathrm{\mathbf{C}}$ is the covariance
matrix, which describes the
terms that contribute to the data but are not part of the model:
the CMB, thermal noise from the telescope
and confusion noise from unresolved point sources. The last were modelled using
the Tenth Cambridge Radio Survey (10C: \citealt{10C1}; \citealt{10C2}), integrating
the confusion power from zero up to the source detection limit on the LA rasters.
%
%In order to calculate the contribution of the cluster SZ signal to interferometric
%visibility data, the Comptonization parameter of the cluster, $y(s)$,
%across the sky must be determined (see \cite{McAdam} for further details).

The cluster geometry, as well as two linearly independent functions of
its temperature and density profiles, must be specified in order to compute the Comptonization parameter.
For the cluster geometry we test both spherical and elliptical profiles; the elliptical model is not
properly triaxial, merely stretching out the model via an ellipticity parameter $\eta$ at an angle $\phi$
(measured anti-clockwise from west).
The temperature profile
is assumed to be constant throughout the cluster and a
$\beta$-model is assumed for the cluster gas density, $\rho_g$
\citep{cdz78}:
\begin{equation}
\rho_{\mathrm{g}}(r)=\frac{\rho_{\mathrm{g}}(0)}{\left[1+\left(\frac{r}{r_{\mathrm{c}}}\right)^{2}\right]^{\frac{3\beta}{2}}},\label{eq:beta}
\end{equation}
where 
\begin{equation}
\rho_{\mathrm{g}}(r)=\mu_{\mathrm{e}}n_{\mathrm{e}}(r), 
\label{eq:ne}
\end{equation} $\mu_{\mathrm{e}}=1.14m_{\mathrm{p}}$ is the gas mass per electron and $m_{\mathrm{p}}$ is the proton mass. The core radius $r_{\textrm{c}}$ gives the
density profile a flat top at low $\frac{r}{r_{\rm{c}}}$ and
$\rho_{\mathrm{{g}}}$ has a logarithmic slope of
$3\beta$ at large $\frac{r}{r_{\mathrm{c}}}$.
%We set $r_{\mathrm{lim}}$ in Equation~(\ref{eq:compton}) to $20h^{-1}\textrm{Mpc}$ -- this result has been tested and
%shown to be large enough even for small values of $\beta$
%\citep{marshall-thesis}. 

X-ray temperature measurements can be used to break the degeneracy in the SZ $y$ parameter between gas mass and temperature,
but these are often overestimates of the temperature of the gas detected via SZ, if, as is usually the case,
they are made at much
smaller radius. Instead we use a theoretical mass-temperature relation to constrain the degeneracy:
this was introduced in \cite{bullet-like}
and assumes that the cluster is virialised, and that all kinetic energy is in the internal energy
of the cluster gas; it does not assume hydrostatic equilibrium.
\cite{malak-param2010} show that this is a useful model for SZ observations, testing it on simulated data
and extracting physical cluster parameters, and finding these to be
in agreement with true input values in simulations. Thus we have:
\begin{align}
\mathrm{k}_{\mathrm{B}}\it{T} &= \frac{\mathrm{G}\mu_{\mathrm{e}} \it{M}}{2r_{200}} \\
&= \frac{\mathrm{G}\mu}{2\left(\frac{3}{4\pi\left(200\rho_{\mathrm{crit}}\right)}\right)^{1/3}}M^{2/3} \\
&= 8.2 \textrm{keV}\left(\frac{M}{10^{15}h^{-1}\mathrm{M}_{\odot}}\right)^{2/3}\left(\frac{H(z)}{H_{0}}\right)^{2/3}.
\label{eq:virtemp}
\end{align}
The center of the cluster profile is also fitted for via two position parameters, $x$ and $y$; the former
equal to negative Right Ascension, in arcseconds, and the latter equal to Declination in arcseconds.
Thus the final gas model has seven parameters (nine in the case of elliptical geometry):
$x$, $y$, ($\eta$, $\phi$), $\beta$, $r_{\mathrm{c}}$, $f_{\mathrm{gas}}(r_{200})$ and $M_{\mathrm{T}}(r_{200})$.
The last parameter is the total mass inside a radius, $r_{200}$, at which the total density is $200\times$
$\rho_{\mathrm{crit}}(z)$, the critical density for closure of the Universe at redshift $z$.
For simplicity, this mass will henceforth be denoted $M$, and it was given a log uniform prior varying between
$10^{13}-2\times10^{15}h^{-1}\textrm{M}_\odot$, which is a physically reasonable range of cluster masses.
Gaussian priors were used for the position parameters, centred on the cluster catalogue positions in Table~\ref{tab:obslist}
with $\sigma$ set to 1\,arcmin. $\eta$, $\phi$, $\beta$ and $r_{\mathrm{c}}$ were given uniform priors 
between 0.5--1, 0--180$^{\circ}$, 0.3--2.5, and 10--1000\,$h^{-1}$kpc, respectively. 

The prior on $f_{\mathrm{gas}}(r_{200})$ was set to a narrow Gaussian centred at
the 90~per~cent of the Wilkinson Microwave Anisotropy Probe 7-year \citep{2010WMAP7}
best-fit value of the baryonic mass fraction,
$f_{\rm{g}}(r_{200})=0.123h^{-1}$, with $\sigma=0.02h^{-1}$.
%which translates to $f_{\rm{g}}=0.08h^{-1}$, $\sigma=0.02h^{-1}$, for our cosmology.
%This was derived from the results of \cite{2010WMAP7} taking into account our value for $h$ and that the gas-mass
%fraction is $\simeq0.9$ of the baryonic mass fraction.

%Using \textsc{McAdam}
Sample visibilities are generated for the model within the prior ranges,
compared to the data and the process is iterated until the sampling converges and no further improvement
can be made.
\subsection{Weak gravitational lensing}
The model used for dark matter distributions throughout this analysis is
the Navarro-Frenk-White profile (NFW: \citealt{NFW95}):
\begin{equation}
\frac{\rho_{\mathrm{mass}}(r)}{\rho_{\mathrm{crit}}(z)}=\frac{\delta_{c}}{\left(r/r_{\mathrm{s}}\right)+\left(1+r/r_{\mathrm{s}}\right)^{2}},
\end{equation}
where $r_{\mathrm{s}}$ is a scale radius and $\delta_{c}$ is the
characteristic overdensity of the halo and is related to the concentration
parameter $c$. The equation can be usefully rewritten in terms of
a parameter $x$ related to the virial radius $r_{200}$ by $x=r/r_{200}$: 
\begin{equation}
\frac{\rho_{\mathrm{mass}}(x)}{\rho_{\mathrm{crit}}(z)}=\frac{\delta_{c}}{cx+(1+cx)^{2}}.
\end{equation}
The lens potential $\psi$ for the NFW profile is derived in \citet{GL/MBM03}:
\begin{equation}
\psi=\frac{4\rho_{\mathrm{s}}r_{\mathrm{s}}}{\Sigma_{\mathrm{crit}}}\left[\frac{1}{2}\left(\log\frac{d}{2}\right)^{2}-2\left(\mathrm{arctanh}\sqrt{\frac{1-d}{1+d}}\right)\right]
\end{equation}
where $d$ is the projected radius scaled by the scale radius:
$d=\theta/(r_{\mathrm{s}}/D_{\theta})$, $\Sigma_{\mathrm{crit}}$ is the critical
surface mass density for strong lensing and $\rho_{\mathrm{s}}$ is a scale density.
For an elliptical NFW profile, $d$ is scaled by the ellipticity $e$ as\footnote{We note that
the published version of \citet{GL/MBM03} contains a version of this equation that
is in error, and show the corrected version here.}:
\begin{equation}
d \rightarrow d' = \sqrt{{d_1^2(1-e)}+\frac{d_2^2}{(1-e)}}.
\end{equation}
In a similar way to the use of the Comptonisation
parameter $y$ in the SZ analysis, the convergence $\kappa$ and shear
$\gamma$ are generated at each galaxy position $\theta_{1}$, $\theta_{2}$
using the lens potential:
\begin{equation}
\kappa=\frac{1}{2}\left(\frac{\partial^{2}\psi}{\partial\theta_{1}^{2}}+
\frac{\partial^{2}\psi}{\partial\theta_{2}^{2}}\right),
\end{equation}
\begin{equation}
\gamma_{1}=\frac{1}{2}\left(\frac{\partial^{2}\psi}{\partial\theta_{1}^{2}}-
\frac{\partial^{2}\psi}{\partial\theta_{2}^{2}}\right),
\end{equation}
\begin{equation}
\gamma_{2}=\frac{\partial^{2}\psi}{\partial\theta_{1}\partial\theta_{2}},
\end{equation}
and compared to that generated by a model set of cluster parameters $\mathbf{x}$ in a similar way to the SZ likelihood (Equation~\ref{eq:sz-like})
\begin{equation}
\mathrm{\mathbf{Pr}}\left(\mathrm{\mathbf{d}}\mid \mathbf{x} \right)=\frac{1}{Z_{L}}\exp\left(-\frac{\chi^{2}}{2}\right),
\end{equation}
where $\chi^{2}$ is the misfit statistic testing the measured lensed ellipticity components $\epsilon_{j}$,
taken as having been drawn independently from a Gaussian distribution 
of $N$ lensed galaxies with mean $g_{j}$ and variance $\sigma$:
\begin{equation}
\chi^{2}=\sum_{i=1}^{N}\sum_{j=1}^{2}\frac{\left(\epsilon_{j,i}-g_{j,i}\right)^{2}}{\sigma^{2}}.
\end{equation}
The normalisation factor $Z_{L}$ is
\begin{equation}
Z_{L}=\left(2\pi\sigma^{2}\right)^{N}.
\end{equation}
The mass contained within the scaled radius $d$ is
\begin{equation}
M(d)=4\pi\rho_{\mathrm{s}}r_{\mathrm{s}}^{3}\left[\log\left(1+d\right)-\frac{d}{1+d}\right],
\end{equation}
and when $r=r_{200}$, $d$ becomes $c$, the concentration parameter. More detailed derivations
are given in \cite{marshall03}.

Thus the mass model we use has five parameters (seven in the case of an elliptical profile):
$x$, $y$, ($\eta$, $\phi$), $M$, $c$, and $z_{\mathrm{field}}$, the last of which is incorporated
into $D_{\theta}$. These model statistics are immediately
useful when comparing to data from the literature and our SZ measurements.
The geometry and $M$ priors were the same as for the SZ modelling; $c$ was
given a uniform prior between 0.1--15, covering a physically reasonable range.
$z_{\mathrm{field}}$ was given a Gaussian prior
centred on 1.2 with $\sigma=$0.7, as discussed in Section~\ref{sec:GL}. Our data put little
constraint on this parameter so the posterior resembles the prior, and thus is not plotted
in the results (Section~\ref{sec:results}).

Fitting multiple components is performed for
disturbed clusters; the highest-evidence model is discussed in the text.
For A611, where the large-scale gas and dark-matter distributions were relaxed, it
was also possible to run a joint analysis in which the mass was constrained by the lensing
data, and the gas fraction was given a uniform prior.
\section{Results}\label{sec:results}
Full unmarginalised posteriors are available from the analysis; % from \textsc{McAdam};
however, these are multi-dimensional and too large to display. In order
to highlight particular features, small sets of parameters are shown
plotted against each other: for instance, axis ratio $\eta$ against
angle $\phi$. The contours contain posterior probabilities at 68\% and 99\%.
The mean value for each parameter
is given in tables in the section discussing each cluster, with errors at the 68\% level.

Source-subtracted SA maps are generated by using the \textsc{aips} task
\texttt{uvsub} on the $uv$-data.
The source parameters used are the mean source fluxes
and spectral indices generated by fitting source models to SA data.
%in \textsc{McAdam}.
LA data are only used to provide priors for these
fits. The source-subtracted
maps are solely for display purposes; all cluster parameter fitting
is performed in the $uv$-plane.% by \textsc{McAdam}.

A useful tool for displaying the likely mass distributions causing
shear in lensing data is \textsc{LensEnt} \citep{GL/Mar++02}.
While it can also be used as a tool to extract cluster parameters,
it is used here only to reconstruct the mass distributions for
display purposes. We now examine the SZ and lensing measurements for each cluster in detail.
%Different `resolutions' can be used; these are the sizes
%of the mass clumps that \textsc{LensEnt} generates to try to fit the observed
%shear data. Low-resolution models require less runtime and are best when
%the signal-to-noise is low; high-resolution models can be used when
%the signal-to-noise is high, but may not always generate a better description,
%especially if the cluster is relaxed and lacking in complex structure.
%
\subsection{A115}\label{sub:Results-A115}
A115 is the most disturbed cluster in the sample: it is fully bimodal
in X-rays, with a brighter, triangular clump in the north and a
dimmer, elliptical clump in the south. The two distinct gas clumps are distorted and in
motion both tangentially
and along the line-of-sight \citep{2005ApJ...619..161G}.
\citet{2001A+A...376..803G} also note
that this cluster has a bright relic radio source trailing from the
northern clump out to the north-east of the cluster. The source coincident
with the highest X-ray emission of the northern clump is identified
as source~D in our data (Table~\ref{tab:A115-sources}; Fig.~\ref{fig:LA-cluster-maps}).
At higher resolution, this source has a double radio structure and has
been studied in detail by e.g. \citet{1987A+AS...69..171G}. From NVSS data at
1.4\,GHz, we find an integrated flux density of $1.35\pm0.07$\,Jy, whereas
at 15.7\,GHz it has fallen to just $5.4\pm0.3$\,mJy, implying a spectral index
of $\alpha=2.28\pm0.03$, in agreement with
 $\alpha=2.5\pm0.4$ from the LA separate-channel data alone.

The relic radio structure is
resolved out by the LA, but one can see immediately from the LA point-source
map (Fig.~\ref{fig:LA-cluster-maps}) that the region contains
many radio point sources, including two very bright sources of 55
and 26\,mJy (sources~A and B). Attempts were made to model the cluster
and sources;% in \textsc{McAdam}.
unfortunately, the region also contains $\simeq25$ sources,
leading to a very
high-dimensionality problem. The high density of sources on top of
the X-ray gas also means that there is a large probability space to explore
when modelling the source and SZ flux, so the algorithm is very
slow to converge on a solution. One problem that was immediately noticeable
in early attempts was that source~B appeared to have a much higher
flux density in the SA data than in the LA data. This is unusual because
in the other cluster analyses, the agreement between LA fluxes and
SA fluxes is very good.

\begin{figure*}
\begin{center}
\subfloat[SA map without source-subtraction.\label{fig:A115-noss}]
{\includegraphics[width=8.5cm,clip=]{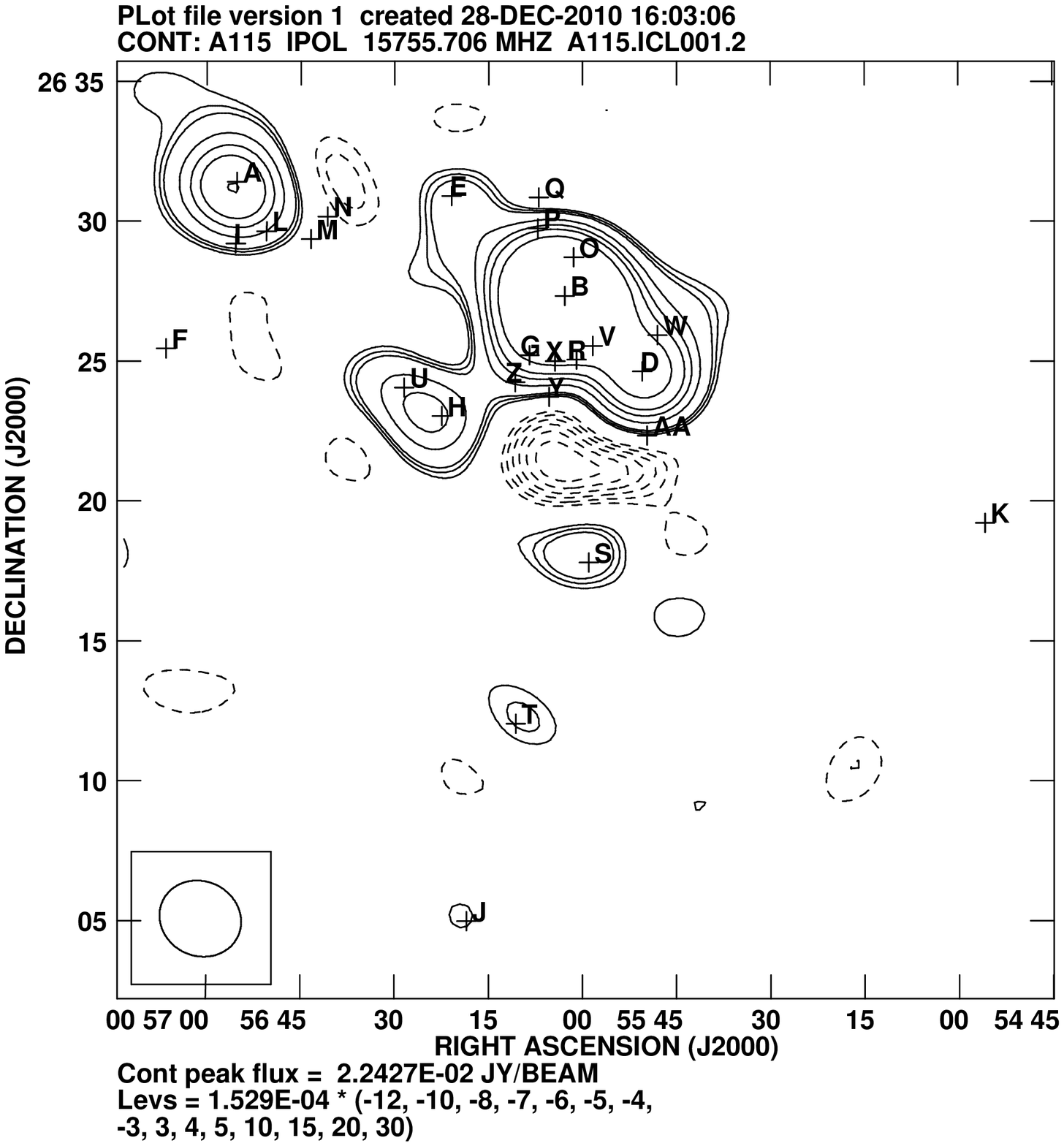}}\quad
\subfloat[\textsc{LensEnt} 2.5\,arcmin-resolution reconstruction of A115.\label{fig:A115-lensent-atoms}]
{\includegraphics[width=7.5cm,clip=]{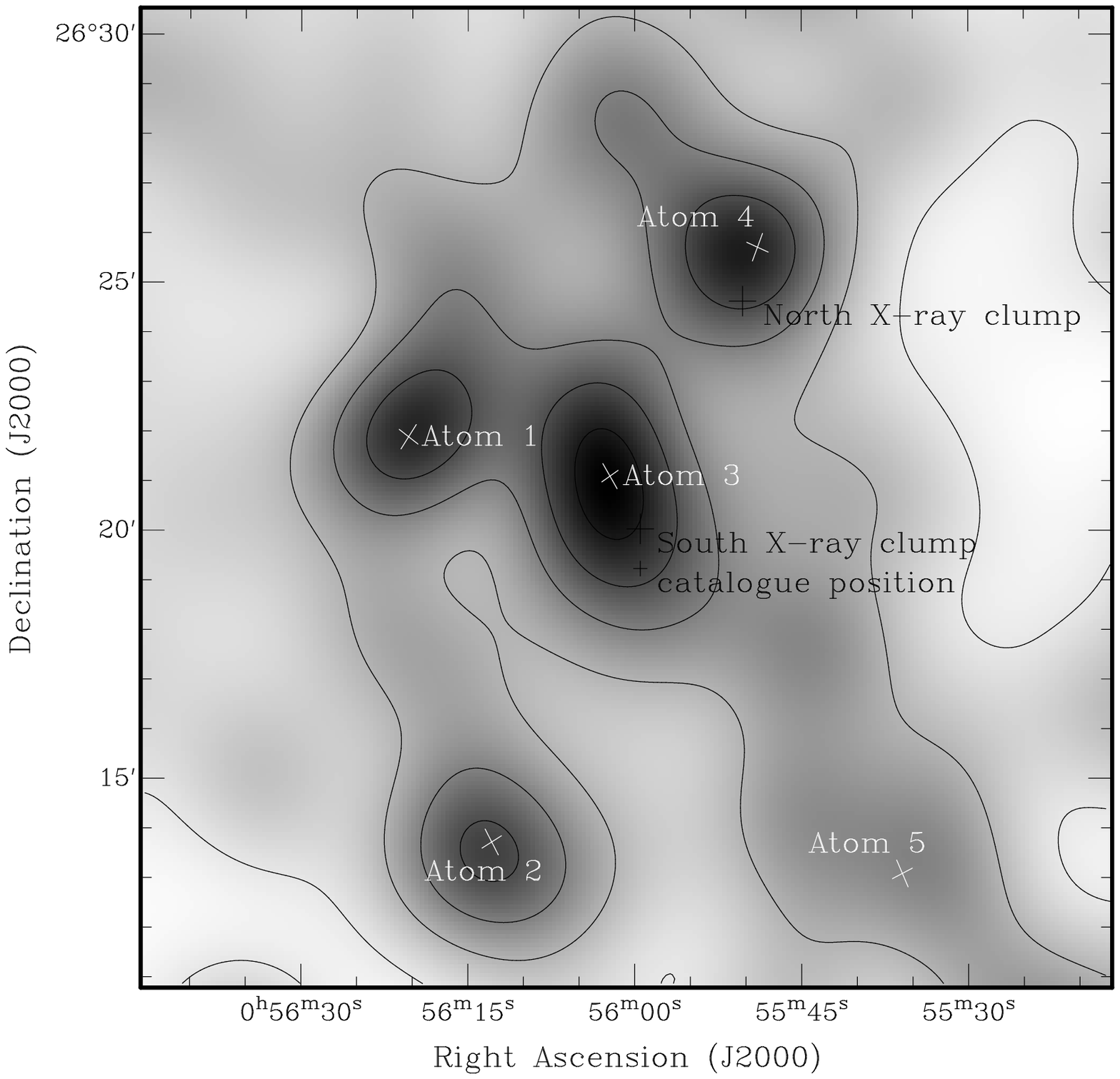}}\qquad
\subfloat[2D weak-lensing posteriors.\label{fig:A115-2DL}]
{\includegraphics[width=15cm,clip=]{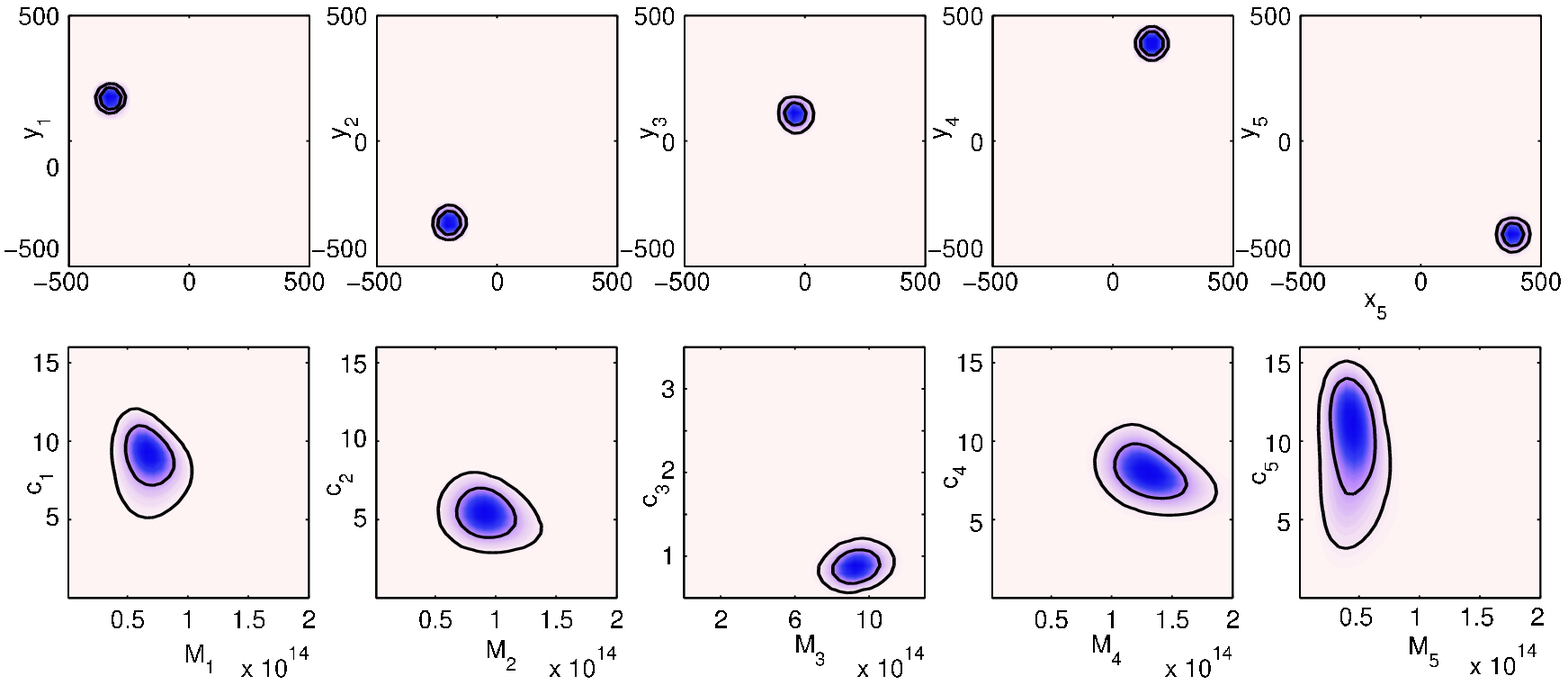}}
\caption{A115. The SA map is shown without source-subtraction; contours are every $3\sigma$,
and truncate at $+30\sigma$ for clarity. The parameters of the labelled
sources can be found in Table~\ref{tab:A115-sources}. In this and all
subsequent SA and LA maps in this section, the \textsc{clean} process
has been applied down to $3\sigma$, no correction for primary beam has been applied,
and the synthesised beam is represented by an ellipse in a box in the lower left-hand corner.
The second image shows
a \textsc{LensEnt} 2.5\,arcmin reconstruction of A115 overlaid with the positions and
labels of the components (Component~$x=$C$x$) whose fitted parameters are given in Table~\ref{tab:A115-results}.
The greyscale is linear and is overlaid with 20\% density contours. The
relative lengths of the cross arms indicate the axis ratios $\eta$, and the
inclinations indicate the angles $\phi$ of the components.
The final image shows the A115 weak-lensing 2D posteriors;
common ranges of $x,y=\pm500$\,arcsec, $c=1-15$,
and $M=10^{13}-10^{15}h^{-1}\textrm{M}_{\odot}$ are
shown for all components except component~3, and delineate the prior
space searched for the five components found. The posteriors for $\eta$ and $\phi$ are not
shown as they are only well-constrained for component~3. \label{fig:A115-all}}
\end{center}
\end{figure*}

Looking at the SA map of A115 (Fig.~\ref{fig:A115-noss}),
source~B is boosted to 36\,mJy after primary-beam correction, compared
to 27.6\,mJy measured by the LA. This could be for two reasons: a) dimmer
sources in the surrounding region are unresolved so their flux density
contributes to the flux density of source~B in the map plane; b) source~B
has extended structure that is resolved out by the LA, resulting
in a lower apparent flux density. If the first point were a problem,
%\textsc{McAdam} would increase the allocated flux density of the other surrounding
the flux density allocated to the other surrounding
sources would be increased in order to fit the data. However, in the $uv$-plane, the
power does not appear to be coming from these sources, as there is %\textsc{McAdam} finds
no evidence for these sources having flux densities larger than those
measured by the LA. Only source~B appears to have a boosted flux density.
This implies that the source is extended, which is not surprising
given the extended relic structure visible at lower frequencies. There
was no variation in the flux densities measured from observations at
different dates, so the difference is not due to variability.
The conclusion is that the source is an ellipsoid extended roughly north-south, with an
extent larger than the SA synthesised beam. 
Of course, extended positive sources are degenerate with the
SZ signal of the cluster gas, so it would be unlikely that an accurate
SZ signal could be extracted from the sky immediately behind and around
this source.

A negative signal is visible around the
location of the southern X-ray clump
($00^{\mathrm{h}} 55^{\mathrm{m}} 59\fs5$, $+26^{\circ} 20\arcmin 02\arcsec$)
in the SA source-unsubtracted data with a peak negative flux of $-1.4$\,mJy ($\simeq10\sigma$).
Some of the decrement may be
created by improperly \textsc{clean}ed sidelobes of nearby sources, particularly source~B. However
one does not see negative signal on all sides of the positive source conglomeration,
so it is likely that a large component of this negative signal is SZ.
Unfortunately, the data are insufficient to model the gas content of the cluster.

%The mass distribution was reconstructed with \textsc{LensEnt} at three
%different resolutions, in order to examine the structure on different
%scales. The 5\,arcmin reconstruction 
%showed an E-W extended mass overdensity overlying the south X-ray clump,
%and another overdensity atop the north X-ray clump.
%At 2.5\,arcmin resolution, shown in Fig.~\ref{fig:A115-lensent-atoms} the mass clumps are more distinctly
%The 1\,arcmin-resolution
%map has much higher noise and lower evidence, but also shows five distinct mass clumps in the cluster. 
%In the \textsc{LensEnt} mass reconstruction in Fig.~\ref{fig:A115-lensent-atoms} the mass clumps are
%resolved , with little increase in noise.
The \textsc{LensEnt} mass reconstruction is shown in Fig.~\ref{fig:A115-lensent-atoms} and
is in visual agreement with that made by \cite{Okabe30GL}, with five mass clumps resolved;
their positions are overlaid.
%\textsc{McAdam} was used to
We also fitted elliptical NFW profiles to the data; the model with the highest evidence
contained five components and the resulting parameters are given in Table~\ref{tab:A115-results}.
%The atom positions are overlaid on the \textsc{LensEnt} reconstruction in
%Fig.~\ref{fig:A115-lensent-atoms}.
The position posteriors of the
fits to the five profiles were examined for evidence that they were composed
of yet further components, but it appears that the position posterior
distributions, visible in Fig.~\ref{fig:A115-2DL}, are unimodal.
\citeauthor{Okabe30GL} measure $M_{200}$ by fitting a NFW profile to the central
%mass peak and find $M_{200}=4.45^{+1.75}_{-1.35}h^{-1}\times 10^{14} \textrm{M}_{\odot}$, which agrees extremely
%well with our measurement of atom~3, in which the bulk of the mass is found, and
mass peak and found $M_{200}=4.45^{+1.75}_{-1.35}\times 10^{14} h^{-1}\textrm{M}_{\odot}$, the upper limit
of which is in agreement with our measurement of component~3, in which the bulk of the mass is found, and
which is coincident with the southern X-ray clump.
This mass is highly unconcentrated, with $c<1$; this
is in visual agreement with the \textsc{LensEnt} reconstruction,
but may show that the NFW profile is not well-suited to fitting this
irregular distribution.

\begin{table*}
\begin{center}
\begin{tabular}{c|ccccc}
\hline
Component & 1 & 2 & 3 & 4 & 5\tabularnewline
\hline
$x$ / arcsec & $-327\pm6$ & $-200\pm10$ & $-49\pm9$ & $161\pm8$ & $355\pm44$\tabularnewline
$y$ / arcsec & $169\pm5$ & $-328\pm8$ & $133\pm24$ & $389\pm9$ & $-394\pm51$\tabularnewline
$\phi/^{\circ}$  & $85\pm60$ & $100\pm50$ & $123\pm3$ & $75\pm53$ & $113\pm45$\tabularnewline
$\eta$ & $0.88\pm0.08$ & $0.85\pm0.09$ & $0.71\pm0.07$ & $0.87\pm0.07$ & $0.83\pm0.09$\tabularnewline
$c$ & $8.8\pm1.7$ & $5.3\pm0.9$ & $0.4\pm0.1$ & $8.0\pm1.0$ & $9.8\pm2.5$\tabularnewline
$M/10^{14}h^{-1}\textrm{M}_{\odot}$ & $0.7\pm0.2$ & $0.9\pm0.2$ & $9.3\pm0.7$ & $1.3\pm0.2$ & $0.4\pm0.1$\tabularnewline
\hline
\end{tabular}

\caption{Mean posterior values for the five elliptical NFW profiles fitted
to the A115 weak-lensing data. Here and in Table~\ref{tab:cluster-parameters}, $x$ and $y$ positions
are given with respect to the cluster catalogue positions in Table~\ref{tab:obslist}, with positive $x$
corresponding to positive on the sky, i.e. negative in RA. \label{tab:A115-results}}
\end{center}
\end{table*}

Interestingly, component~1
is has a similar mass to component~4, despite only the latter clump coinciding with
X-ray emission. This implies that the bulk of the X-ray emission in
the northern clump is from the X-ray-emitting double radio source,
and there may not be a large dark matter
component to this clump. Component~1 might be a bullet-cluster-like dark-matter blob
that has passed through the cluster leaving behind its stripped gas. An elliptical
galaxy at $00^{\mathrm{h}} 56^{\mathrm{m}} 19\fs7$, $26^{\circ} 21\arcmin 53\arcsec$
is within the mass overdensity of component~1. It lies on the red sequence of A115,
which spatially overlaps the area of component~1.

Component~2 is clearly detected and it is uncertain what relation this object bears to the
bulk of the cluster. At a total mass of just $\simeq9\times10^{13}h^{-1}\textrm{M}_{\odot}$,
its gas mass should be of order $\simeq10^{13}h^{-1}\textrm{M}_{\odot}$, rendering it
undetectable by AMI in the SZ. Visual examination of the $r$-band image shows no unusual
galaxy overdensity at this position, and galaxies there do not seem to lie on the red
sequence of A115. Spectroscopic follow-up, or observations in another optical
band to compare the densest region of galaxies of A115 with this area might indicate
whether this mass concentration is sub-cluster-like, related to A115, or an unrelated
mass overdensity along the line-of-sight. Component~5's position is not well-constrained
compared to the other components and it has a low concentration and uncertain mass.
\subsection{A1914}\label{sub:A1914-results}
This cluster is a fairly complex merger with two distinct X-ray clumps
which appear to be moving in opposite directions east-west. At the
same time, the overall mass distribution is irregular and elongated NE--SW.
\citet{2004ApJ...605..695G} find filaments of hot gas connecting the mass
peaks.
\begin{figure*}
\begin{center}
\subfloat[Before source-subtraction.\label{fig:A1914-SA-noss}]{\includegraphics[width=8cm,clip=]{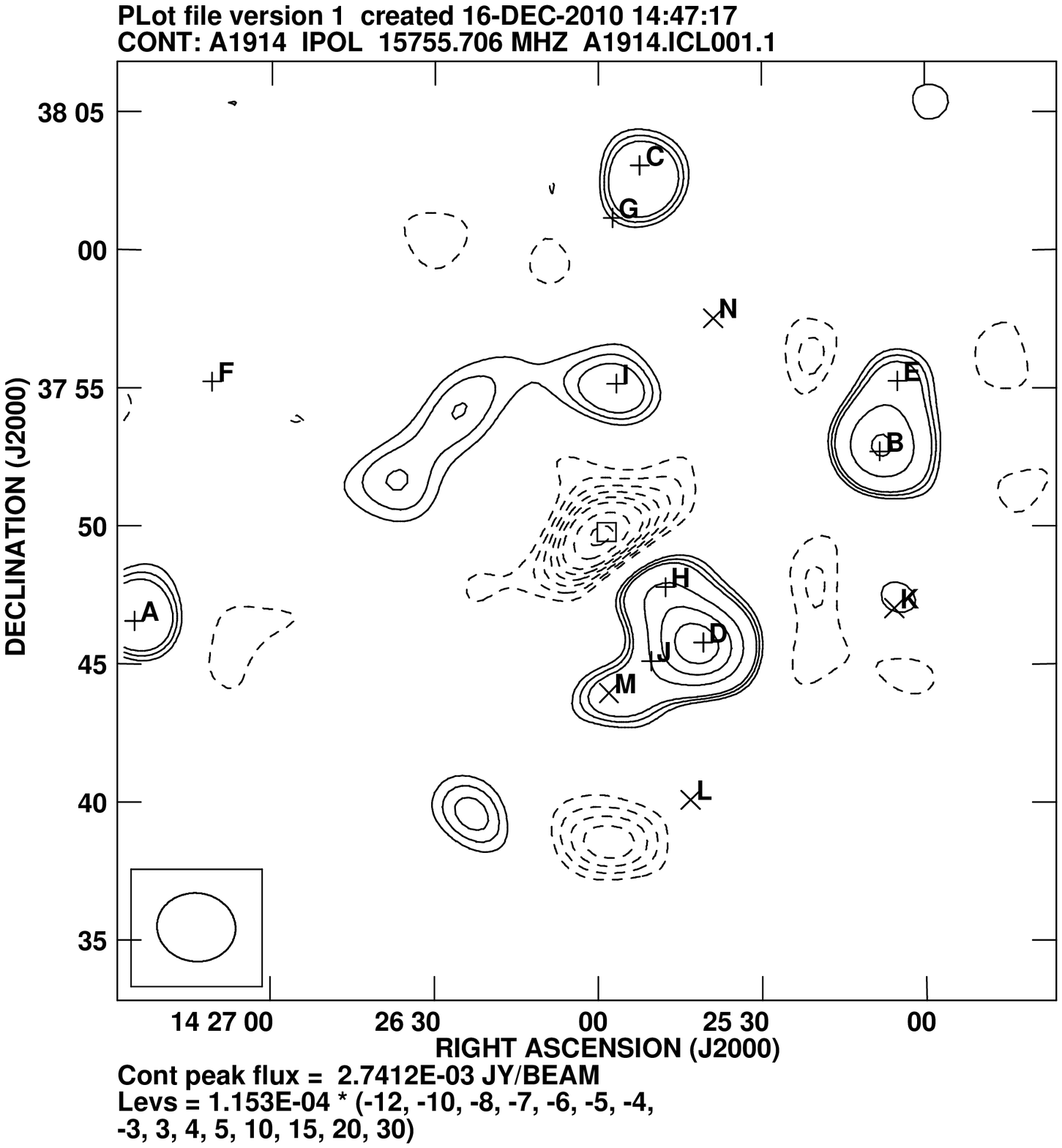}}\quad
\subfloat[After source-subtraction.\label{fig:A1914-SA-ss}]{\includegraphics[width=8cm,clip=]{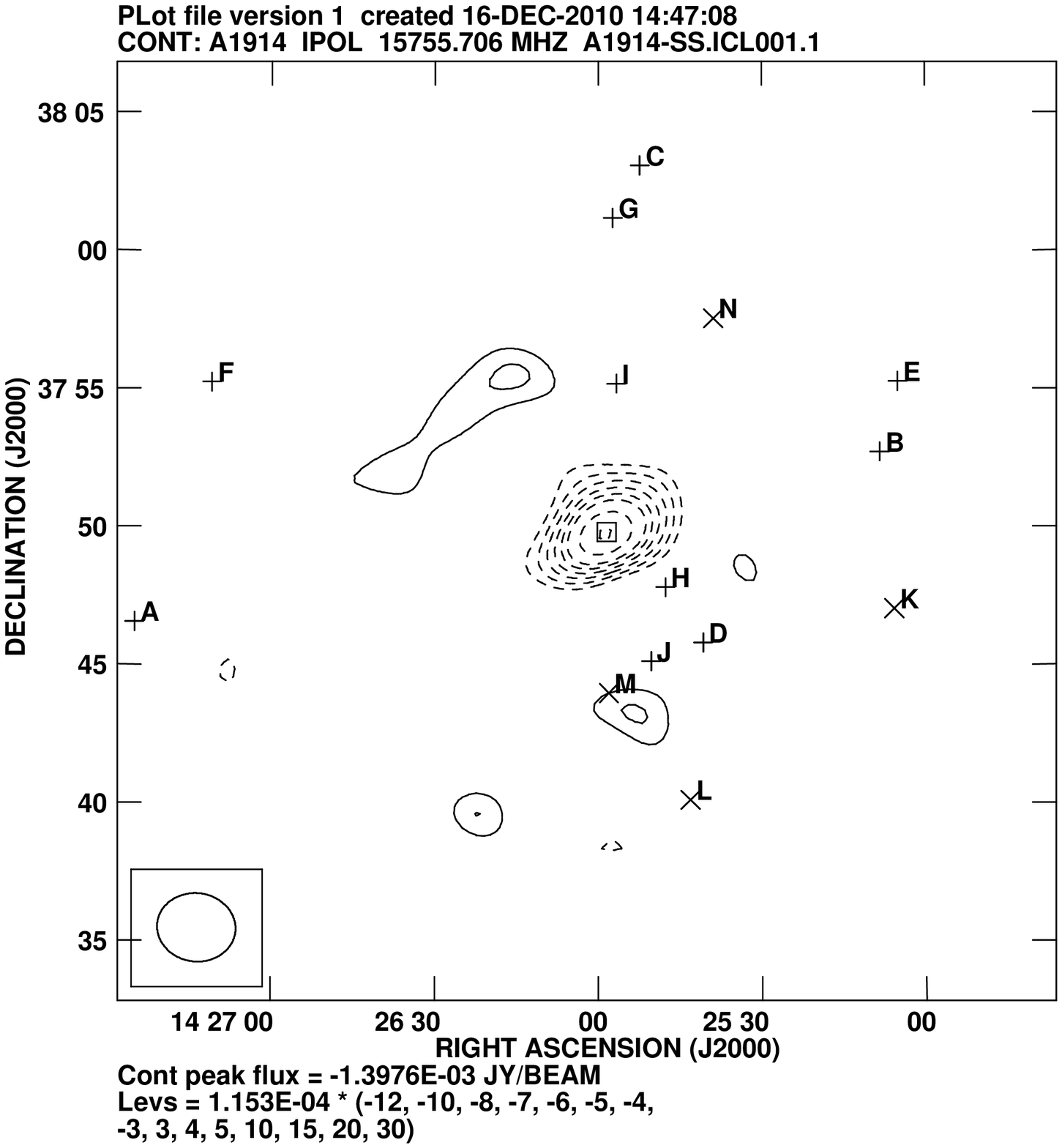}}\qquad
\subfloat[Weak lensing 2D posteriors.\label{fig:A1914-lensing}]{\includegraphics[width=11cm,clip=]{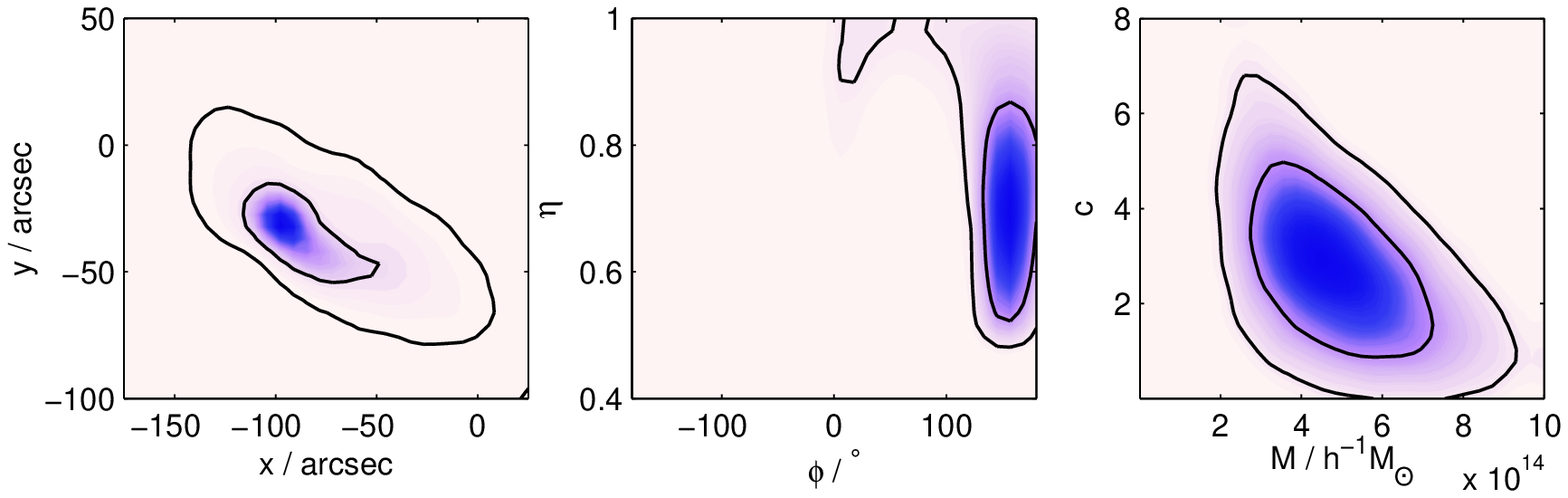}}\qquad\qquad\qquad
\subfloat[SZ 2D posteriors.\label{fig:A1914-SZ}]{\includegraphics[width=7.5cm,clip=]{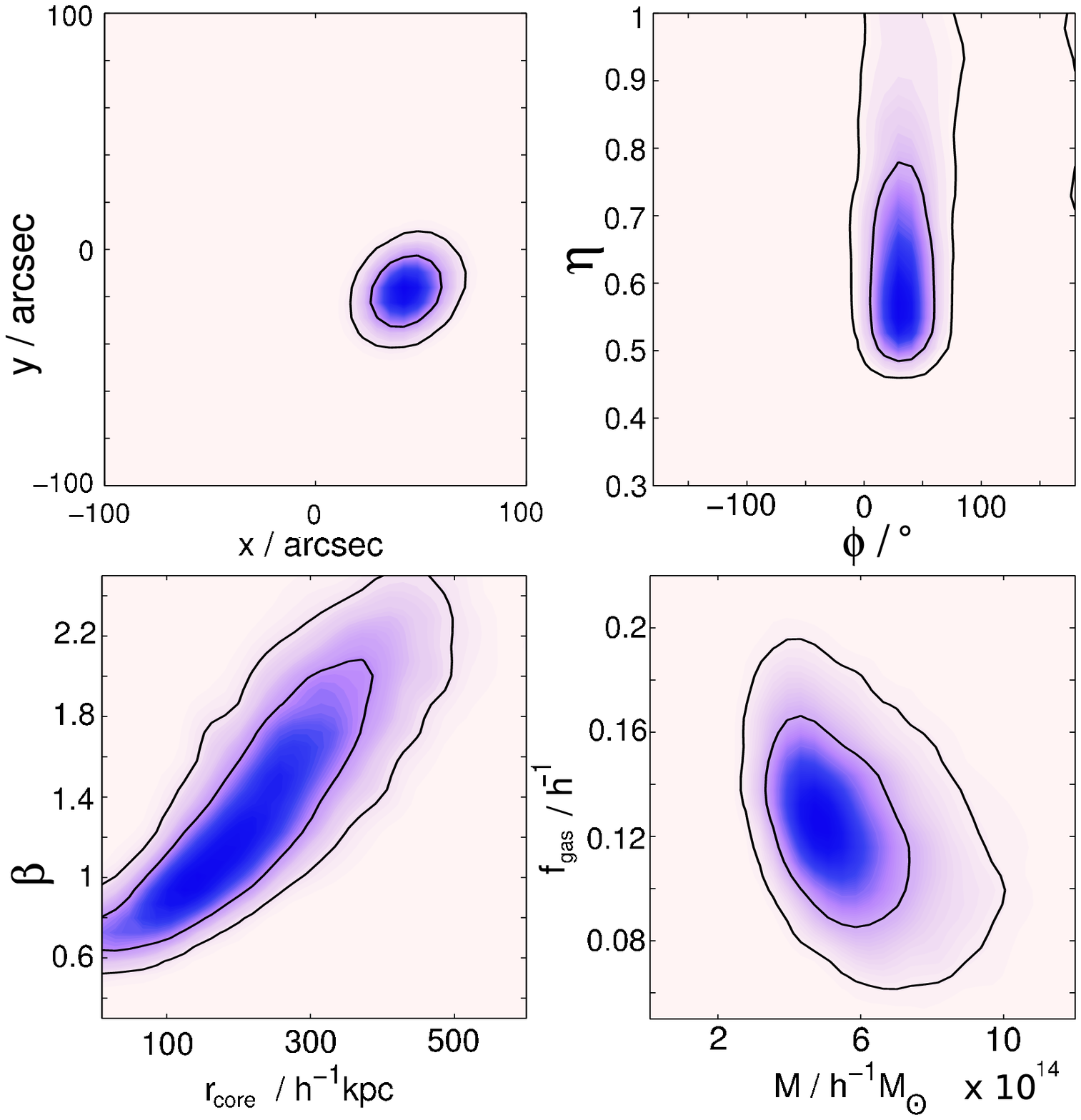}}\quad
\subfloat[X-ray, SZ and weak-lensing composite image.\label{fig:A1914}]{\includegraphics[width=9cm,clip=]{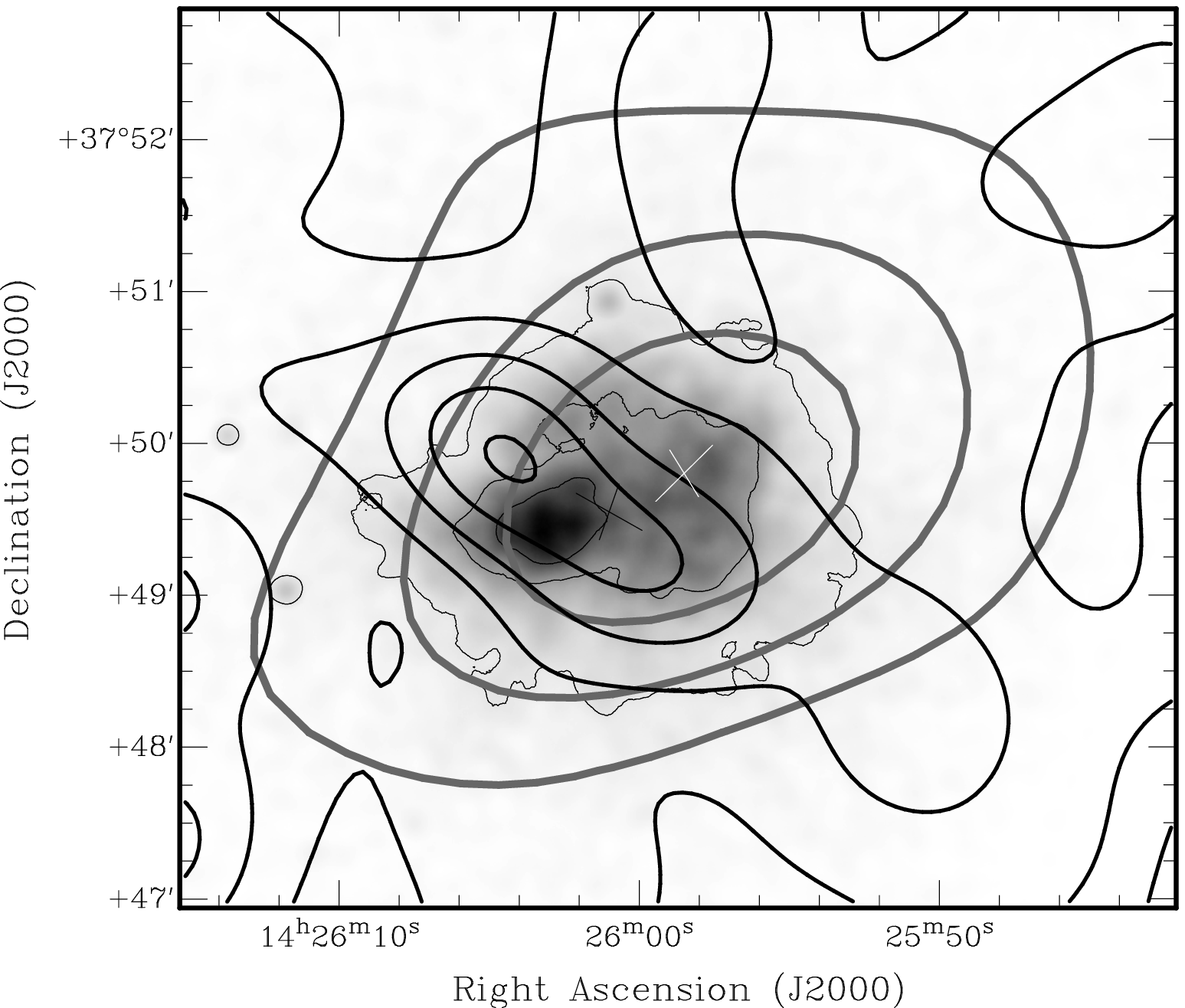}}
\caption{A1914. The maps at the top-left and top-right are from the SA before
and after source-subtraction;
in this and all subsequent SA figures, contours are every $1\sigma$, and follow levels
$-12$, $-10$, $-8$, $-7$, $-6$, $-5$, $-4$, $-3$, $3$, $4$, $5$, $10$, $15$, $20$, $30$.
Sources marked with a `$+$' were modelled,
%in \textsc{McAdam},
while those marked with
a `$\times$' were directly subtracted. The `$\Box$' marks the position of the centre of the fitted isothermal
$\beta$-model. The parameters of the labelled sources can be found in Table~\ref{tab:A1914-sources}.
Subfigure~(c) shows
A1914 weak-lensing 2D posteriors; in this and all subsequent weak-lensing 2D posterior plots,
$x$ and $y$ are in units of arcseconds, $c$ is dimensionless and $M$ is total mass (integrated to $r_{200}$).
in $h^{-1}\textrm{M}_{\odot}$.
Subfigure~(d) shows A1914 SZ 2D posteriors. In this and all subsequent SZ 2D posterior plots,
$x$ and $y$ are in units of arcseconds, $r_{\textrm{c}}$ in $h^{-1}$\,kpc,
$f_{\textrm{gas}}$ in $h^{-1}$, and total mass (integrated to $r_{200}$ as described
in Section~\ref{sec:modelling}) in $h^{-1}\textrm{M}_{\odot}$.
The bottom-right image shows a composite X-ray, mass and SZ image. The greyscale shows \textit{Chandra}
X-ray data \citep{2004ApJ...605..695G} smoothed by a Gaussian
of radius 15\,arcsec, with 30\% contours.
Thicker black contours show the \textsc{LensEnt} mass reconstruction
in 20\% levels of density. Thick grey contours show the SA SZ decrement
in $3\sigma$ intervals, where the map noise $\sigma=115\,\mu$Jy.
The white cross indicates the position, orientation and axis ratio of the elliptical $\beta$-model
fitted to the SZ data.  The black cross indicates the maximum-likelihood
position, orientation and axis ratio of the elliptical NFW profile fitted
to the lensing data.\label{fig:A1914-all}}
\end{center}
\end{figure*}
%
%\begin{table}
%\begin{center}
%\begin{tabular}{c|cc}
%\hline
%Parameter & Lensing & SZ\tabularnewline
%\hline
%$x$ / arcsec & $28\pm36$ & $43\pm8$\tabularnewline
%$y$ / arcsec & $-37\pm27$ & $-17\pm7$\tabularnewline
%$\phi$/ $^{\circ}$ & $150_{-60}^{+40}$ & $38^{+16}_{-8}$ \tabularnewline
%$\eta$ & $0.7\pm0.1$ & $0.7\pm0.1$ \tabularnewline
%$c$ & $2.8\pm1.4$ & ---\tabularnewline
%$M$/ $10^{14}h^{-1}M_{\odot}$ & $4.9\pm1.4$ & $5.5\pm1.5$\tabularnewline
%$r_{\mathrm{core}}$/ $h^{-1}$kpc & --- & $240\pm110$\tabularnewline
%$\beta$ & --- & $1.4\pm0.4$\tabularnewline
%\hline
%\end{tabular}
%
%\caption{Mean posterior values for the parameters of A1914. \label{tab:A1914-parameters}}
%
%\end{center}
%\end{table}
%
Even before source-subtraction is performed, it is clear that A1914
has an elliptical shape in the SZ (Fig.~\ref{fig:A1914-SA-noss}).
It is surrounded by a large number of point sources which are clearly
visible in the SA data. After source-subtraction (Fig.~\ref{fig:A1914-SA-ss}),
the majority of source contamination is removed
cleanly. There are some $2\sigma_{\mathrm{SA}}$ residuals but these are more likely
to be noise or faint extended sources than subtraction artifacts. The only significant
residuals are near two sources of flux$<4\sigma_{\mathrm{LA}}$.
The image here is consistent with that made during the early stages
of science observations with AMI \citep{ami-a1914}.

Fitting an elliptical $\beta$-model to the data
%via \textsc{McAdam}
gives the parameters shown in Table~\ref{tab:cluster-parameters};
a circular model was also fitted but gave results with lower evidence.
The centroid position is slightly offset south and east from the catalogue
position, which agrees well with the overall structure of the X-ray
data. It is also clear that SZ data traces out a much larger region of
the extension of the gas, perhaps produced from the merger of the two
sub-clumps visible in X-ray observations \citep{2004ApJ...605..695G}.
However, the individual clumps are not resolved.

Lensing analyses of A1914 are difficult due to its highly distorted
mass distribution. The contours of Fig.~\ref{fig:A1914}
show the \textsc{LensEnt} mass reconstruction from our optical data. The distribution
is clearly distorted and elongated, and looks similar to the distribution
reconstructed by \citet{2008PASJ...60..345O}, but with lower
resolution, so that the two different mass peaks remain unresolved.

%Using \textsc{McAdam}
An elliptical NFW profile was fitted to the data, resulting
in a mass estimate of $4.9\pm1.4\times10^{14}h^{-1}\textrm{M}_{\odot}$, which
is just in agreement with the lower end of \citeauthor{2008PASJ...60..345O}'s
stated estimate of $M_{\mathrm{virial}}=6.14\pm3.19\times10^{14}h^{-1}\textrm{M}_{\odot}$, but
we find that the mass is rather less concentrated than their estimate:
$c=2.0\pm1.3$, as opposed to $c=4.13\pm2.79$. These results are
still consistent within the errors. The model that agrees best with
the data has an elliptical geometry of $\eta=0.7\pm0.1$ inclined
at $\phi=150\pm30^{\circ}$. This fits with the mass reconstruction;
essentially we are fitting a single profile to the distribution that
\citeauthor{2008PASJ...60..345O} separate into two peaks.

An attempt was made to fit two components to this data; 
however the evidence for two components was lower than that for a single
ellipsoid. This could be due to a lower-resolution
catalogue than that available to \citeauthor{2008PASJ...60..345O}, who do
see two mass peaks. However it also appears that their detected peaks overlie
a mass ellipsoid, rather than being two distinct clumps, so
an ellipsoid might simply be a better description of the (clearly irregular) mass
distribution.

A reconstruction of the cluster mass and gas is shown in Fig.~\ref{fig:A1914}
along with the SZ data and the locations of the X-ray clumps.
It is interesting to note from the SZ data that on the largest scales,
aside from the high ellipticity, the gas is fairly uniformly distributed. Unlike
the complex temperature-dependent structures picked out by the X-ray
data, the SZ map provides a smooth and large-scale picture of the
gas distribution. Of note is the north-west extension of the
gas that is invisible in the X-ray data.

In conclusion, the distributions of both gas and mass are highly irregular
and detailed hydrodynamic simulations are necessary to improve
understanding of the merger in A1914.
\subsection{A2111}
A2111 is widely agreed to be a head-on merger between two smaller
clusters. The X-ray data presented by \citet{Wang97} show an elongated
morphology with two components. The subcluster was determined to have
entered from the north-west, its core heating as it entered the gas
in the centre of the main cluster. The outskirts
of the cluster gas are fairly relaxed and the most disturbed gas lies
only in the centre.

\begin{figure*}
\begin{center}
\subfloat[Before source-subtraction.\label{fig:A2111-SA-noss}]{\includegraphics[width=8cm,clip=]{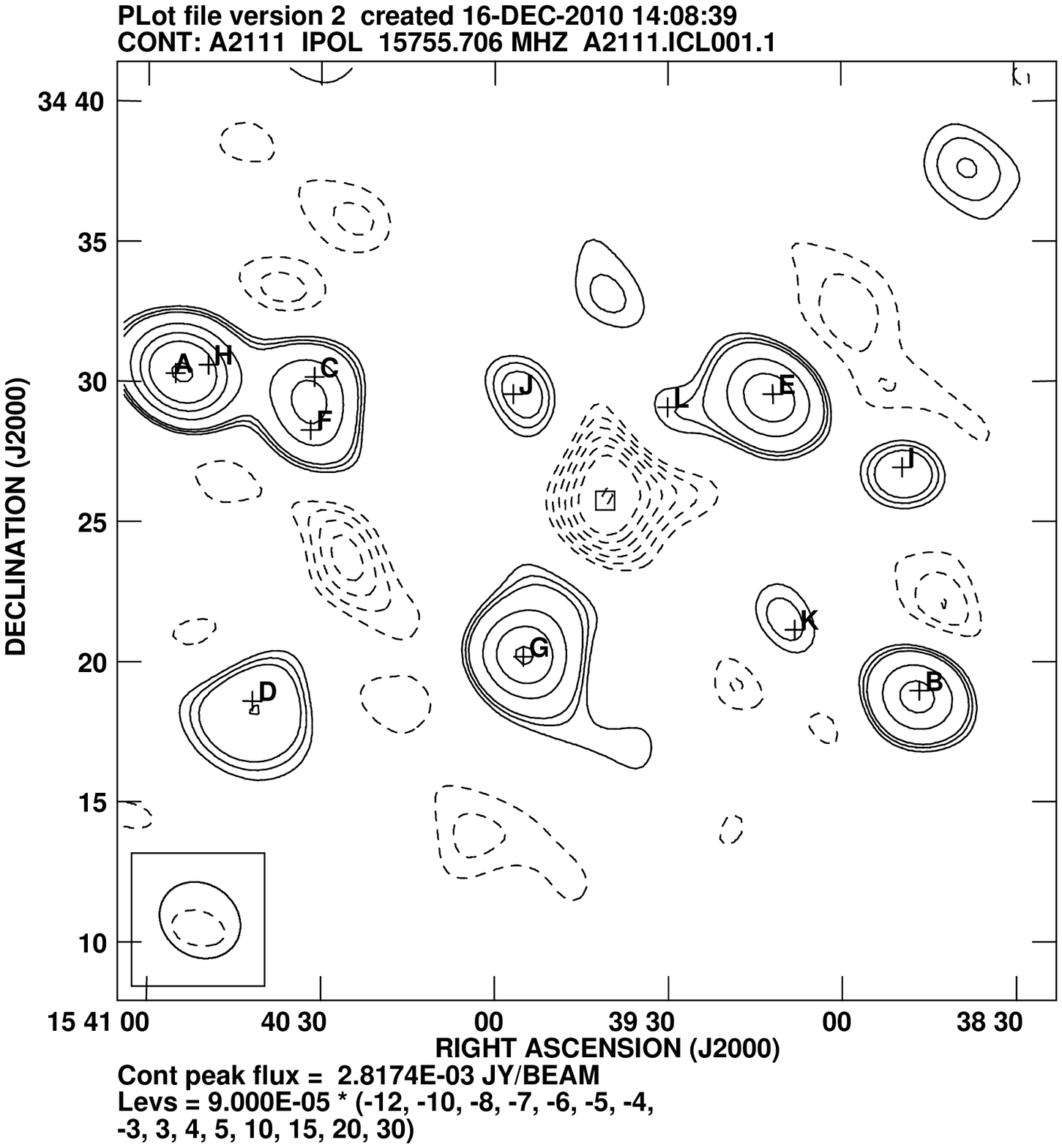}}\quad
\subfloat[After source-subtraction.\label{fig:A2111-SA-ss}]{\includegraphics[width=8cm,clip=]{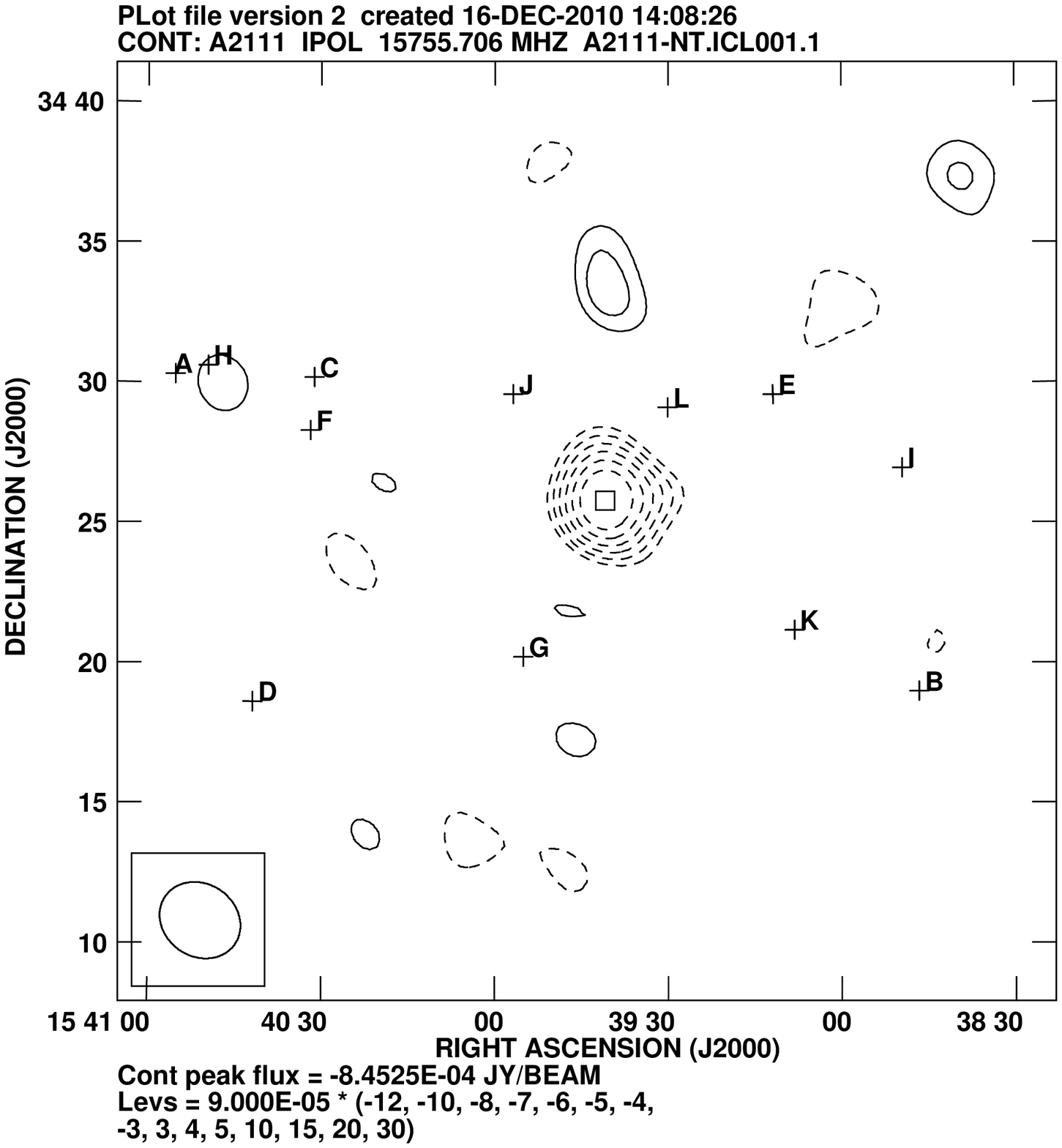}}\qquad
\quad\subfloat[SZ 2D posteriors.\label{fig:A2111-SZ}]{\includegraphics[width=12cm,clip=]{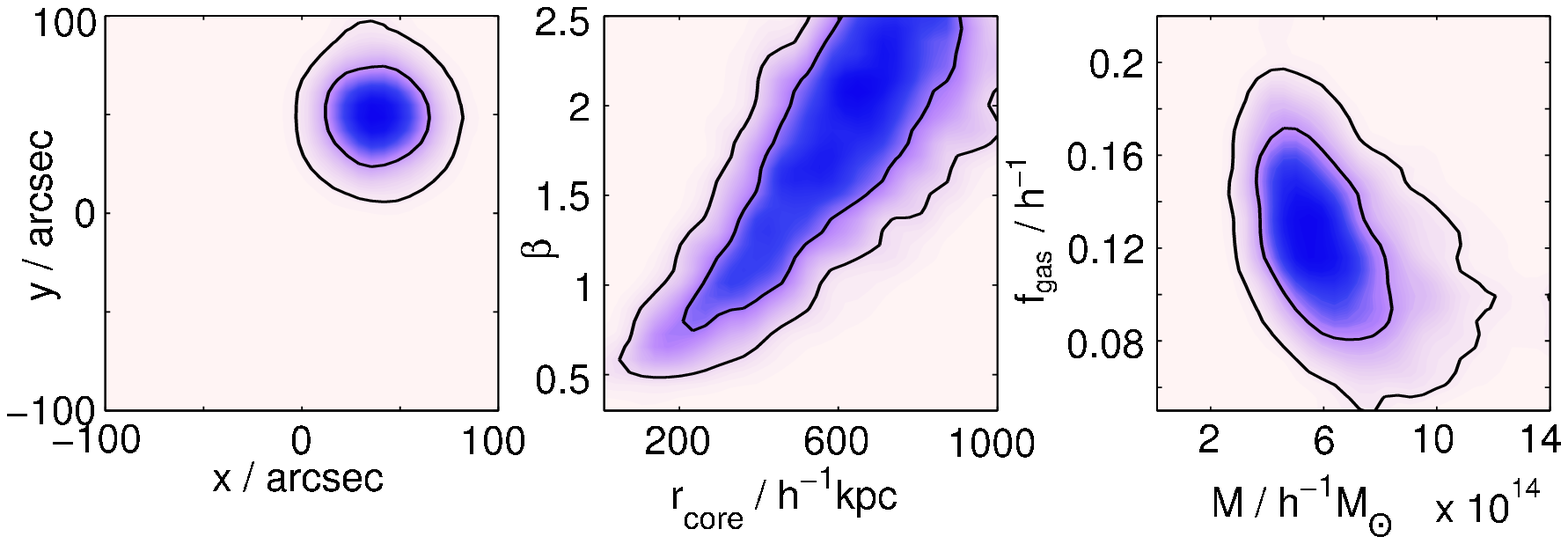}}\quad\quad\quad\quad\quad\quad\quad\qquad\qquad
\subfloat[Weak lensing 2D posteriors.\label{fig:A2111-lensing}]{\includegraphics[width=4cm,clip=]{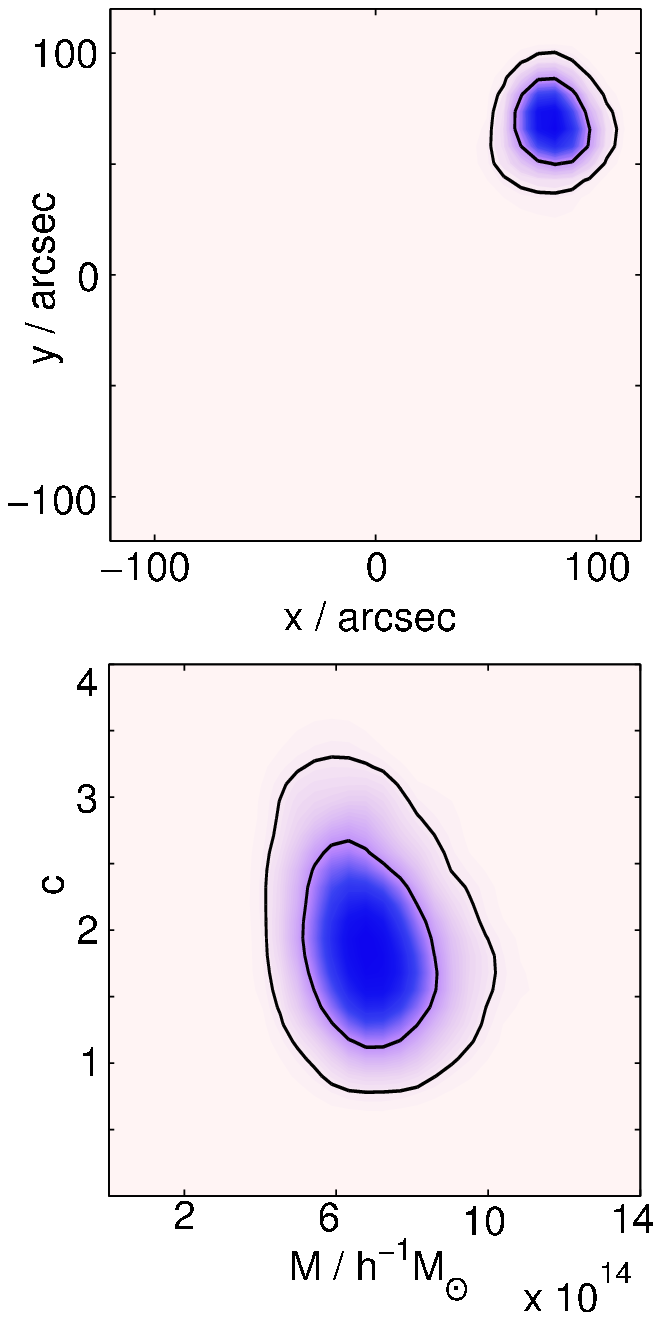}}\quad
\subfloat[X-ray, SZ and weak-lensing composite image.\label{fig:A2111}]{\includegraphics[width=10cm,clip=]{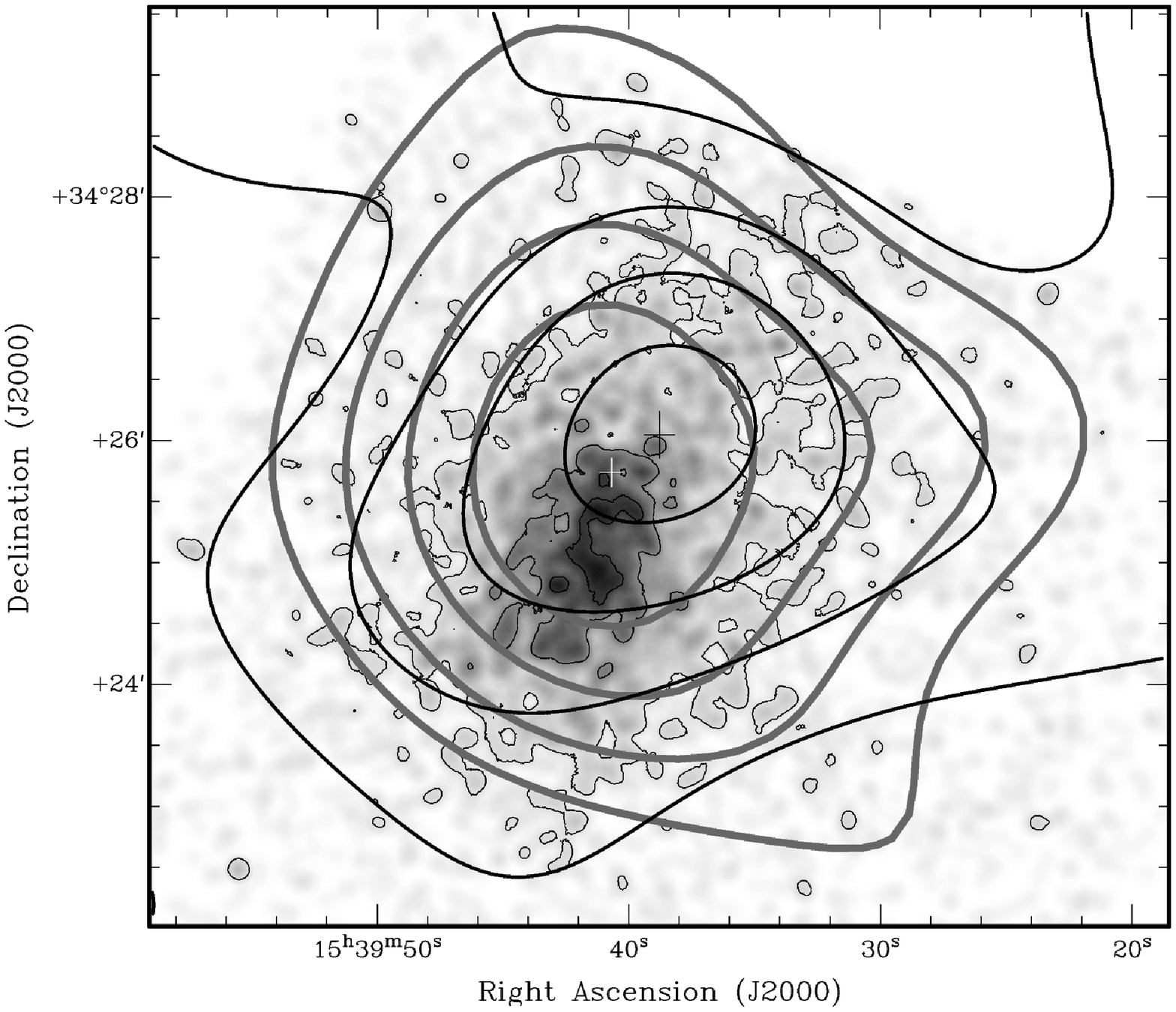}}

\caption{A2111.
For the maps and posterior distributions, annotations and axis ranges and labels are as in Fig.~\ref{fig:A1914-all}.
The parameters of the labelled sources in the SA maps can be found in Table~\ref{tab:A2111-sources}. 
For the compositie image, the greyscale shows \textit{Chandra}
X-ray data smoothed by a Gaussian
of radius 15\,arcsec, with 30\% contours. Thicker black contours show the \textsc{LensEnt}
mass reconstruction
in 20\% levels of density. Thick grey contours show the SZ decrement
in $2\sigma$ intervals, where the map noise $\sigma=90\,\mu$Jy. The
white cross indicates the fitted centre of the circular $\beta$-model with arm lengths
of $1\sigma$ error on the position. The black cross indicates the
fitted centre of the circular NFW profile with arm lengths of $1\sigma$
error on the position. \label{fig:A2111-all}}
\end{center}
\end{figure*}

Despite the large number of contaminating radio sources, the SZ decrement
of A2111 is clearly visible in the source-unsubtracted SA
map (Fig.~\ref{fig:A2111-SA-noss}).
A circularly-symmetric $\beta$-model was fitted to the gas distribution using
the SZ data, resulting in a centroid with parameters given in Table~\ref{tab:cluster-parameters}.

A NFW profile was fitted to the lensing data: a circularly-symmetric
model was slightly preferred
over an elliptical model by the evidence values: the parameters from
the resulting model are given in Table~\ref{tab:cluster-parameters}.

Fig.~\ref{fig:A2111} shows a composite
image of the X-ray, lensing and SZ data. The gas and mass
appear relaxed and circularly symmetric; the total mass is found to be
$6.9\pm1.1\times10^{14}h^{-1}\textrm{M}_{\odot}$ from the
lensing data; this is consistent within the errors with the value measured
via the SZ, $6.3\pm2.1\times10^{14}h^{-1}\textrm{M}_{\odot}$.
Unlike X-ray observations and the long-baseline SZ observations by
\citet{laroque06}, we are mapping the gas on the edges of the cluster, as well
as the denser core. We thus find higher values for $\beta$ and $r_{\mathrm{c}}$
than are typical in the literature.
%
%
%\begin{table}
%\begin{center}
%\begin{tabular}{c|cc}
%\hline
%Parameter & Lensing & SZ  \tabularnewline
%\hline
%$x$ / arcsec & $80\pm9$ & $38\pm16$\tabularnewline
%$y$ / arcsec & $68\pm11$ & $50\pm16$ \tabularnewline
%$c$ & $1.9\pm0.5$ & --- \tabularnewline
%$M$/ $10^{14}h^{-1}M_{\odot}$ & $6.9\pm1.1$ & $6.3\pm2.1$\tabularnewline
%$r_{\mathrm{core}}$/ $h^{-1}$kpc & --- & $565\pm195$ \tabularnewline
%$\beta$ & --- & $1.7\pm0.5$\tabularnewline
%\hline
%\end{tabular}
%%
%\caption{Mean posterior values for the parameters of A2111. \label{tab:A2111-parameters}}
%%
%\end{center}
%\end{table}
%
\begin{table*}
\begin{center}
\begin{tabular}{c|cc|cc|c|cc|cc|cc}
\hline
Cluster & \multicolumn{2}{c|}{A1914} & \multicolumn{2}{c|}{A2111} & A2259 & \multicolumn{2}{c|}{A611} & \multicolumn{4}{c}{A851} \tabularnewline
\hline
          & Lensing & SZ & Lensing & SZ           & SZ & Lensing & SZ       & \multicolumn{2}{c|}{Lensing}  & \multicolumn{2}{c}{SZ} \tabularnewline
Parameter           &    &                    &    &                  &               &  &  & C1 &  C2 & C1 & C2 \tabularnewline
\hline
$x$ / arcsec & $28\pm36$ & $43\pm8$ & $80\pm9$ & $38\pm16$ & $38\pm21$ & $47\pm9$ & $30\pm15$ & $33\pm5$  & $107\pm5$& $-88\pm5$ & $115\pm30$ \tabularnewline
$y$ / arcsec & $-37\pm27$ & $-17\pm7$ & $68\pm11$ & $50\pm16$ & $2\pm17$ & $13\pm5$ & $15\pm16$ & $-18\pm2$ & $-52\pm2$ & $-27\pm4$ & $120\pm25$ \tabularnewline
$\phi$ / $^{\circ}$ & $150_{-60}^{+40}$ & $38^{+16}_{-8}$ & -- &  -- & -- & -- & -- & -- & -- & -- & -- \tabularnewline
$\eta$              & $0.7\pm0.1$ & $0.7\pm0.1$ & -- & -- & -- & -- & -- & -- & -- & -- & -- \tabularnewline
$c$ & $2.8\pm1.4$ & -- & $1.9\pm0.5$ & --  & -- & $4.0\pm1.3$ & -- & $5.1\pm1.3$ & $5.1\pm0.9$ & -- & -- \tabularnewline
$M$/ $10^{14}h^{-1}\textrm{M}_{\odot}$ & $4.9\pm1.4$ & $5.5\pm1.5$ & $6.9\pm1.1$ & $6.3\pm2.1$ & $3.8^{+1.4}_{-1.2}$ & $4.7\pm1.2$ & $6.0\pm1.9$ & $3.2\pm1.0$ & $3.4\pm1.0$ & $3.9\pm0.5$ & $4.1\pm1.0$ \tabularnewline
$r_{\mathrm{c}}$/ $h^{-1}$kpc & -- & $240\pm110$ & -- & $565\pm195$  & $400\pm160$ & --- & $490\pm230$ & -- & -- & $155\pm55$ & $440^{+220}_{-200}$ \tabularnewline
$\beta$ & -- & $1.4\pm0.4$\ & -- & $1.7\pm0.5$ & $1.6\pm0.4$ & --- & $1.6\pm0.5$ & -- & -- & $2.0\pm0.3$ & $1.8\pm0.3$ \tabularnewline
\hline
\end{tabular}
\caption{Mean posterior values for the $\beta$-profile fits to the SZ data and NFW-profile fits to the lensing data for A1914, A2111, A2259, A611 and A851. For the last, values for the two different components fitted to each dataset are shown.\label{tab:cluster-parameters}}
\end{center}
\end{table*}
\subsection{A2259}
\begin{figure*}
\begin{center}
\subfloat[Before source-subtraction.\label{fig:A2259-SA-noss}]{\includegraphics[width=8cm,clip=]{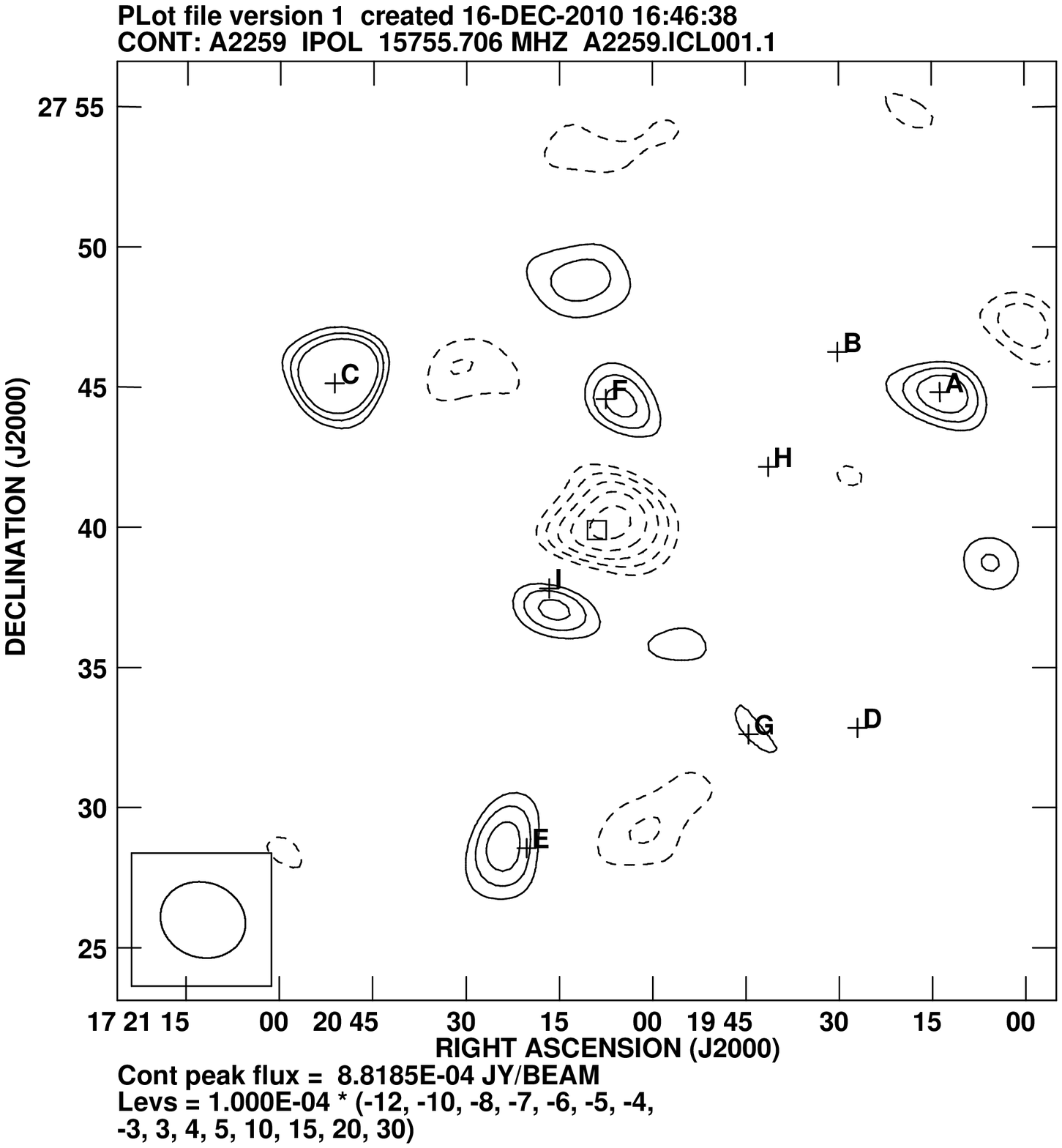}}\quad
\subfloat[After source-subtraction.\label{fig:A2259-SA-ss}]{\includegraphics[width=8cm,clip=]{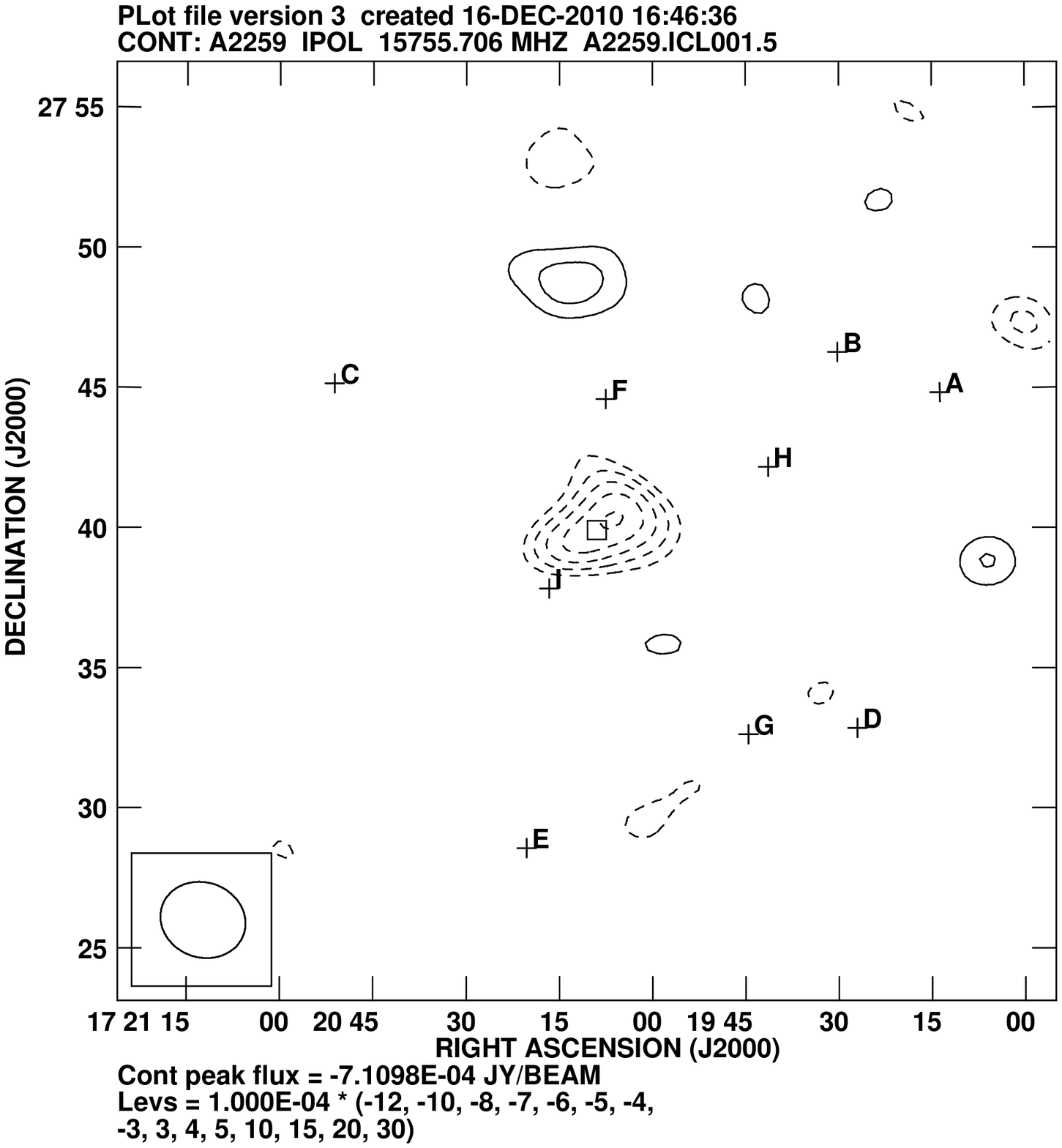}}\qquad
\subfloat[SZ 2D posteriors.\label{fig:A2259-SZ}]{\includegraphics[width=3.5cm]{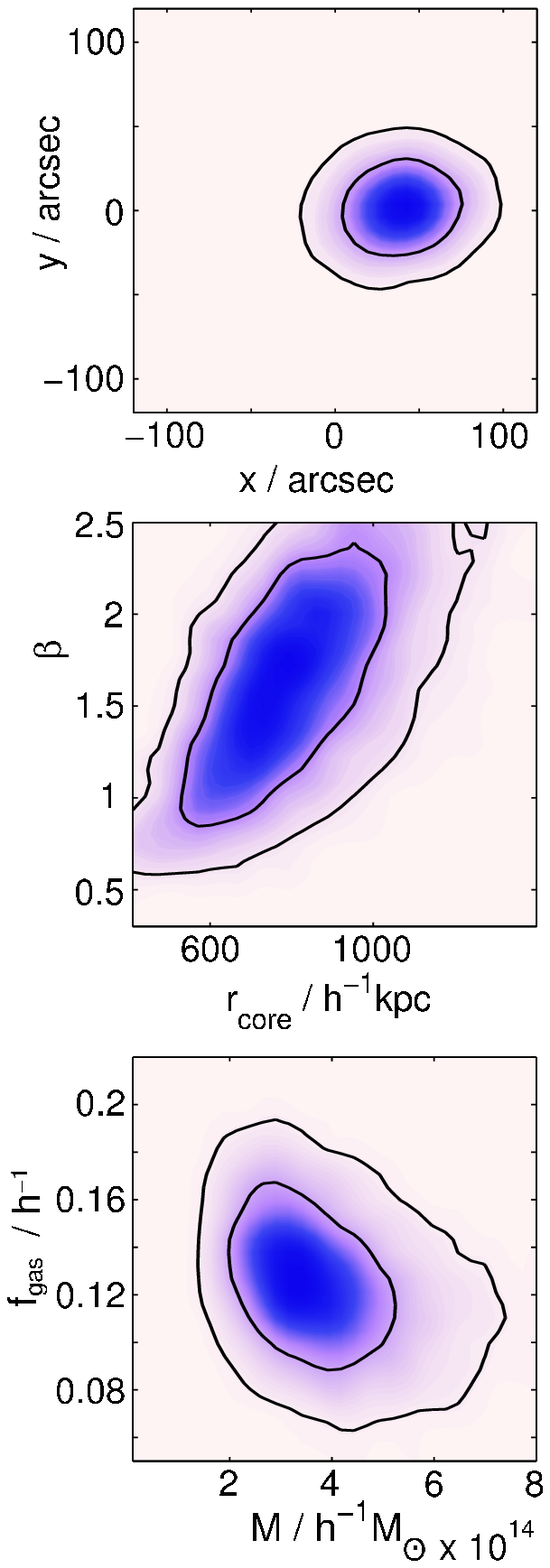}}\quad
\subfloat[X-ray and SZ composite image.\label{fig:A2259}]{\includegraphics[width=10cm,clip=]{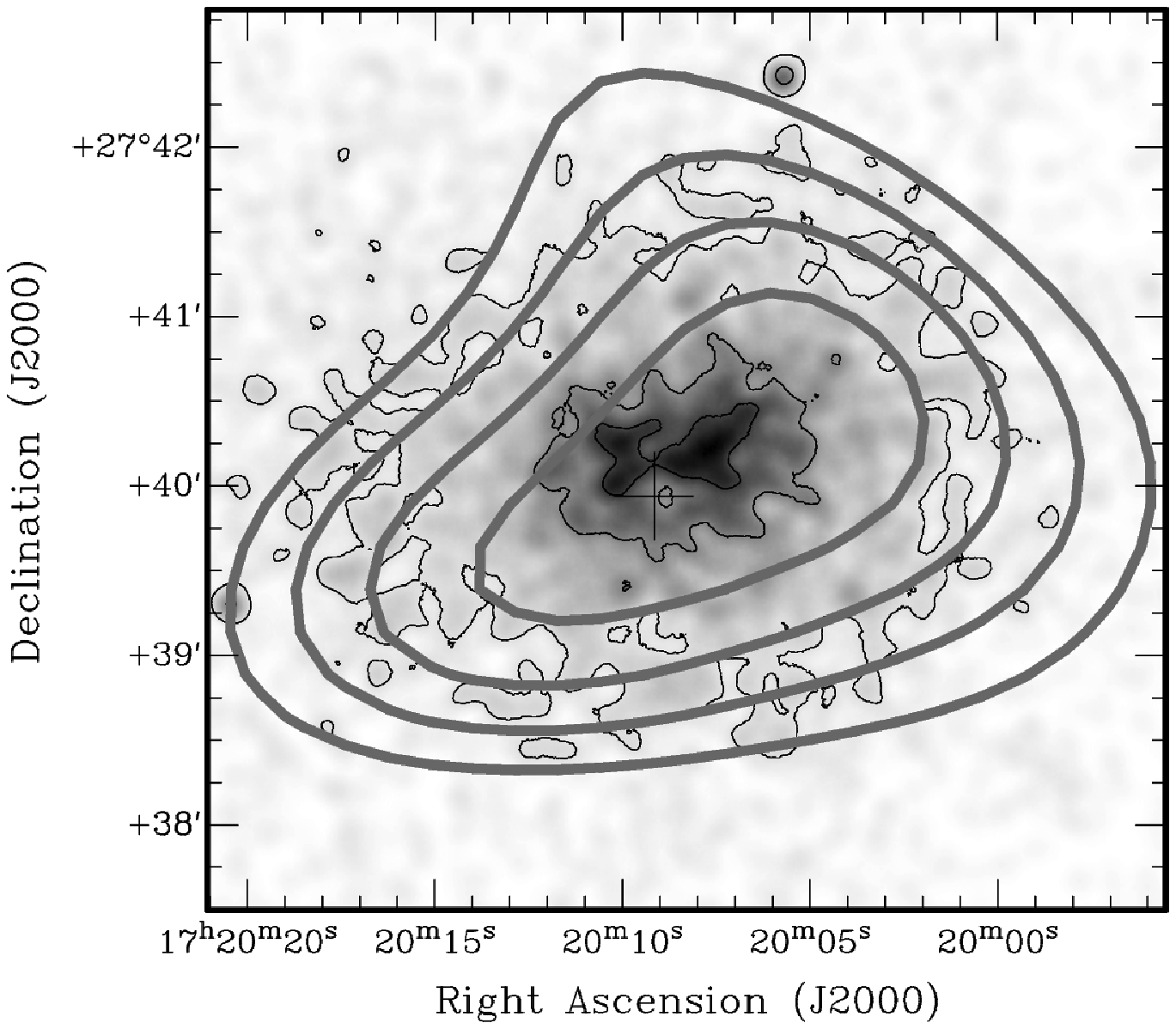}}
\caption{A2259.
For the maps and posterior distributions, annotations and axis ranges and labels are as in Fig.~\ref{fig:A1914-all}.
The parameters of the labelled sources in the SA maps can be found in Table~\ref{tab:A2259-sources}.
In the composite image, the greyscale shows \textit{Chandra}
X-ray data smoothed by a Gaussian
of radius 15\,arcsec, with 30\% contours. Thick grey contours show the SZ decrement in $1\sigma$
intervals, where the map noise $\sigma=100\,\mu$Jy. The black cross
indicates the fitted centre of the circular $\beta$-model with arm lengths of $1\sigma$
error on the position.\label{fig:A2259-all}}
\end{center}
\end{figure*}
%
%
%\begin{table}
%\begin{center}
%\begin{tabular}{c|c}
%\hline
%Parameter & SZ\tabularnewline
%\hline
%$x$ / arcsec & $38\pm21$\tabularnewline
%$y$ / arcsec & $2\pm17$\tabularnewline
%$M$/ $10^{14}h^{-1}M_{\odot}$ & $3.8^{+1.4}_{-1.2}$\tabularnewline
%$r_{\mathrm{core}}$/ $h^{-1}$kpc & $400\pm160$\tabularnewline
%$\beta$ & $1.6\pm0.4$\tabularnewline
%\hline
%\end{tabular}
%
%\caption{Mean posterior values for the parameters of A2259. \label{tab:A2259-parameters}}
%
%\end{center}
%\end{table}
%
%
The SZ decrement of A2259 is clearly visible; indeed it is more striking
than any of the nearby point sources (Fig.~\ref{fig:A2259-SA-noss}).
After these are fitted and removed, the decrement (Fig.~\ref{fig:A2259-SA-ss})
appears slightly less elliptical and more reminiscent of the X-ray
image: a composite is shown in Fig.~\ref{fig:A2259}. The results of the $\beta$-model
fitted to the SZ data are shown in Table~\ref{tab:cluster-parameters}.

A2259 lies in the Galactic plane, and this was immediately noticeable
when analysing the lensing data. Despite attempting various spatial,
magnitude and colour cuts to the background galaxies, the evidence
for a lensing effect around A2259 was very low, so that when fitting
a NFW profile, the results
from the posterior highly resemble the priors. It appears that contamination from Galactic
foregrounds prevented us from detecting the small lensing signal from this low-mass cluster.
\subsection{A611}
\begin{figure*}
\begin{center}
\subfloat[Before source-subtraction.\label{fig:A611-SA-noss}]{\includegraphics[width=8cm,clip=]{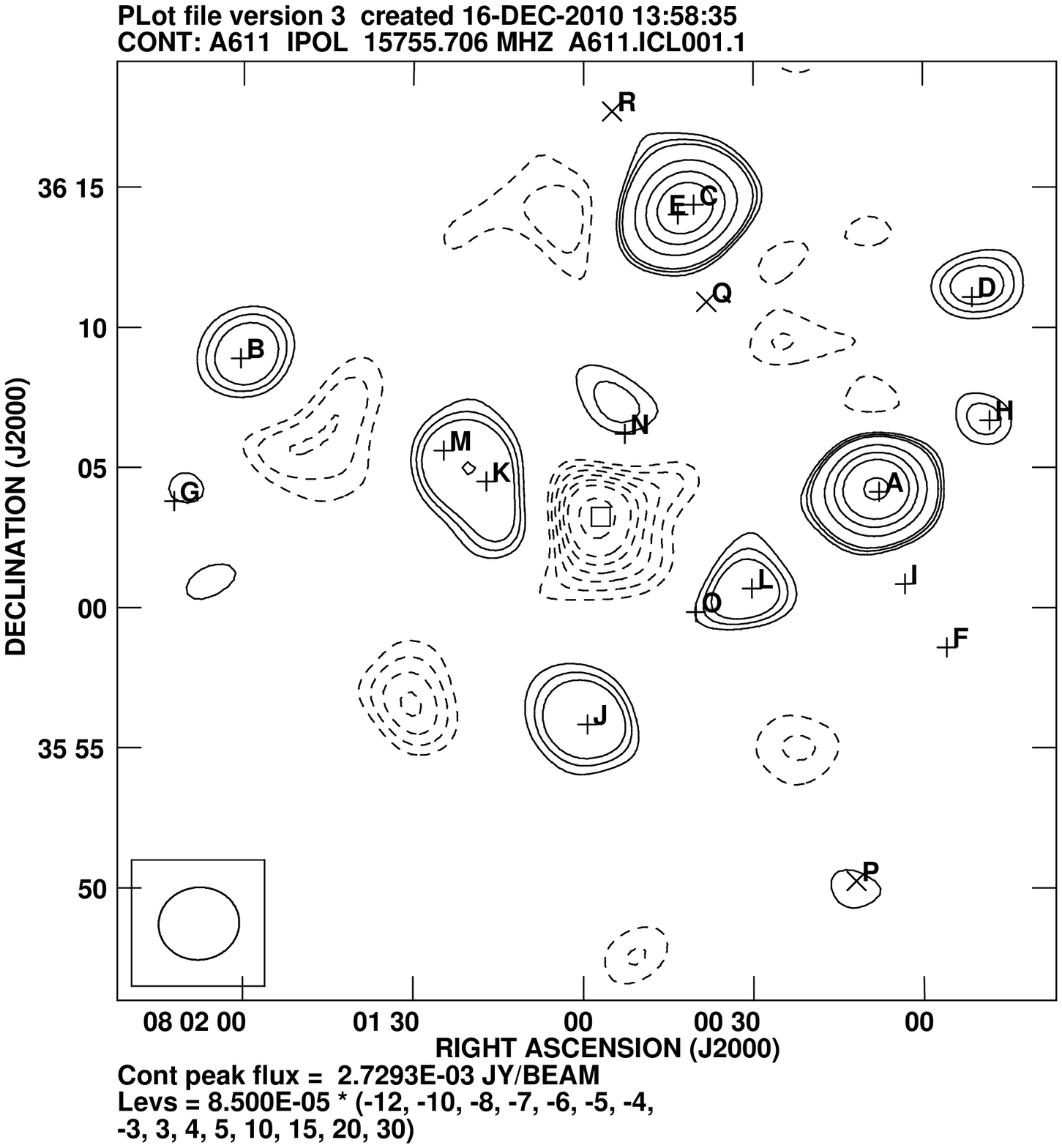}}\quad
\subfloat[After source-subtraction.\label{fig:A611-SA-ss}]{\includegraphics[width=8cm,clip=]{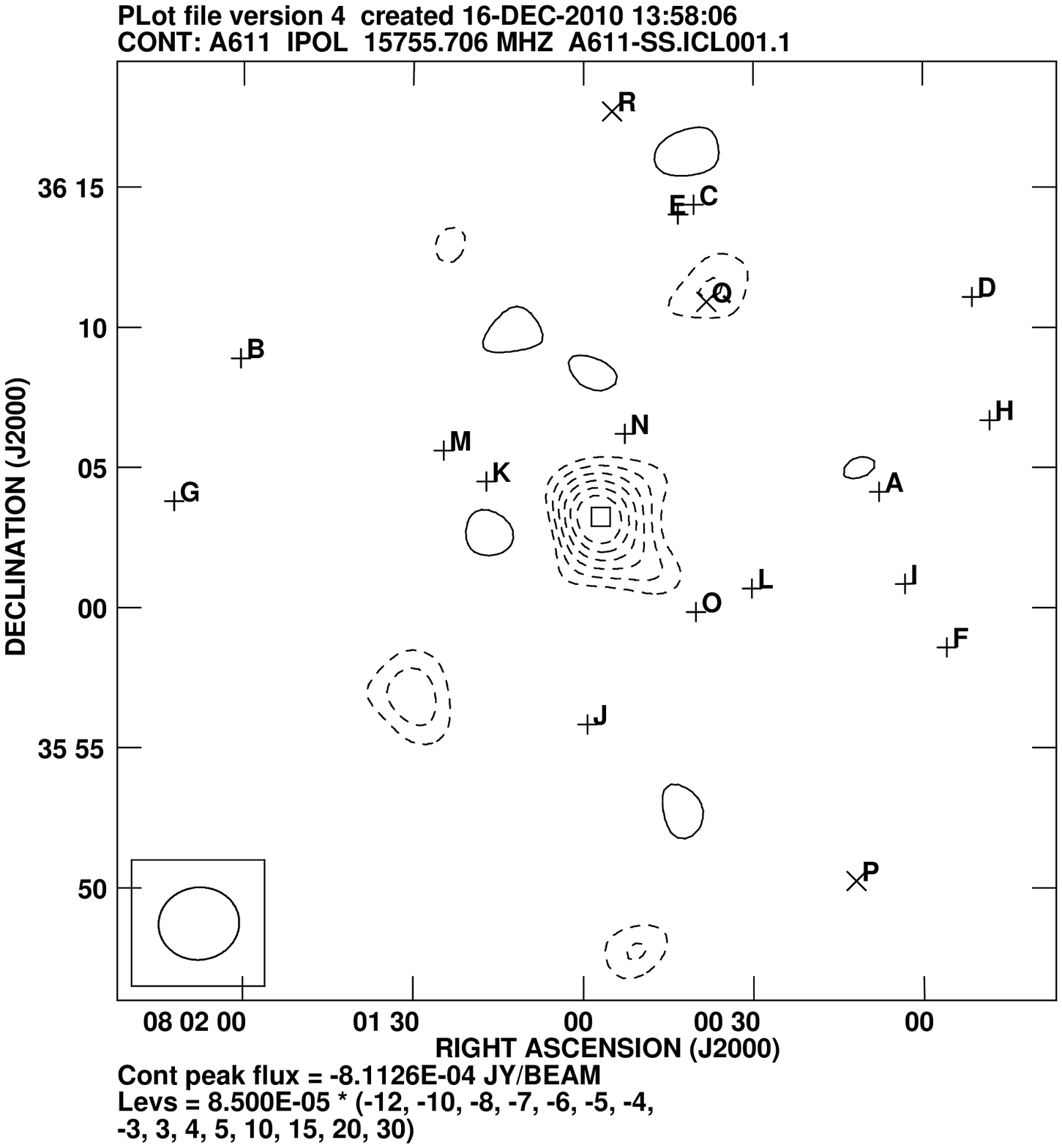}}\qquad
\subfloat[SZ 2D posteriors.\label{fig:A611-SZ}]{\includegraphics[width=13cm]{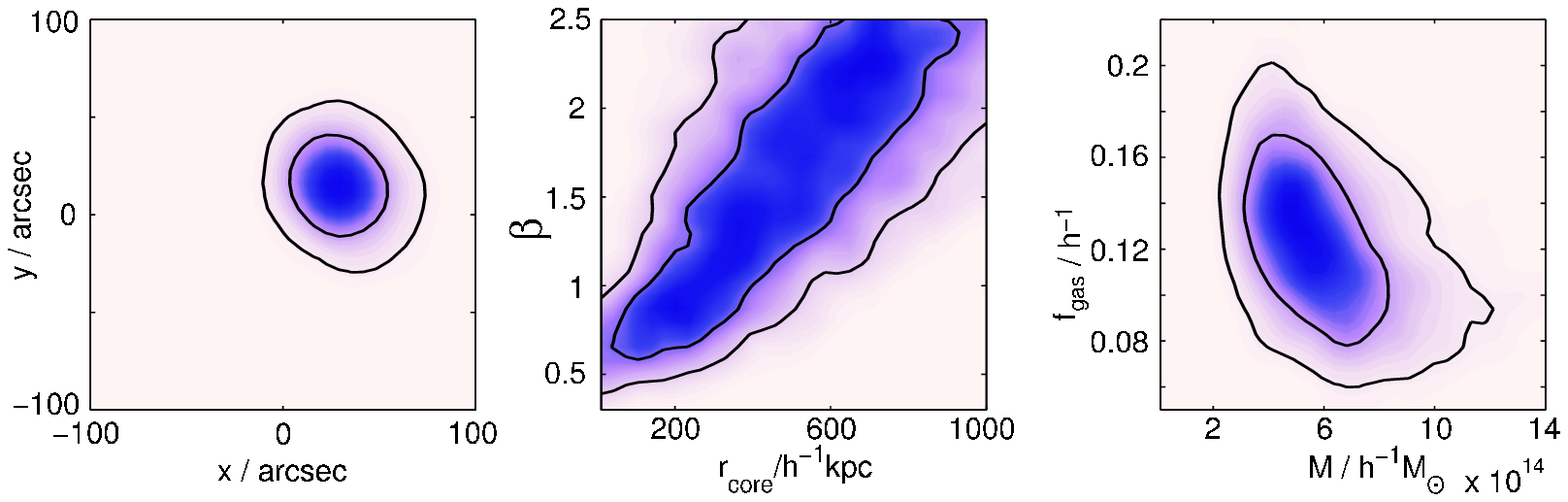}}\quad\quad\qquad\qquad\qquad\qquad\qquad
\subfloat[Weak lensing 2D posteriors.\label{fig:A611-lensing}]{\includegraphics[width=4.2cm]{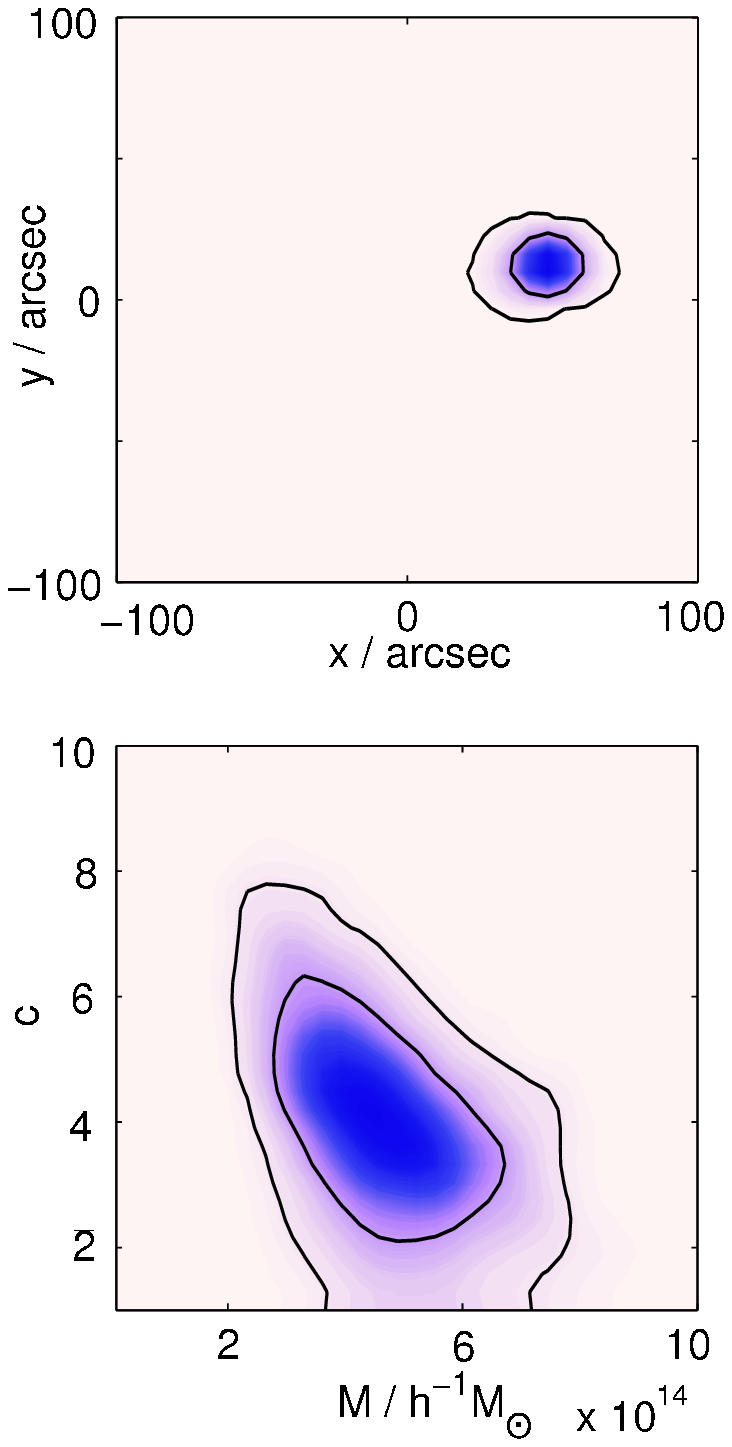}}\quad
\subfloat[X-ray, SZ and weak-lensing composite image.\label{fig:A611}]{\includegraphics[width=9.5cm,clip=]{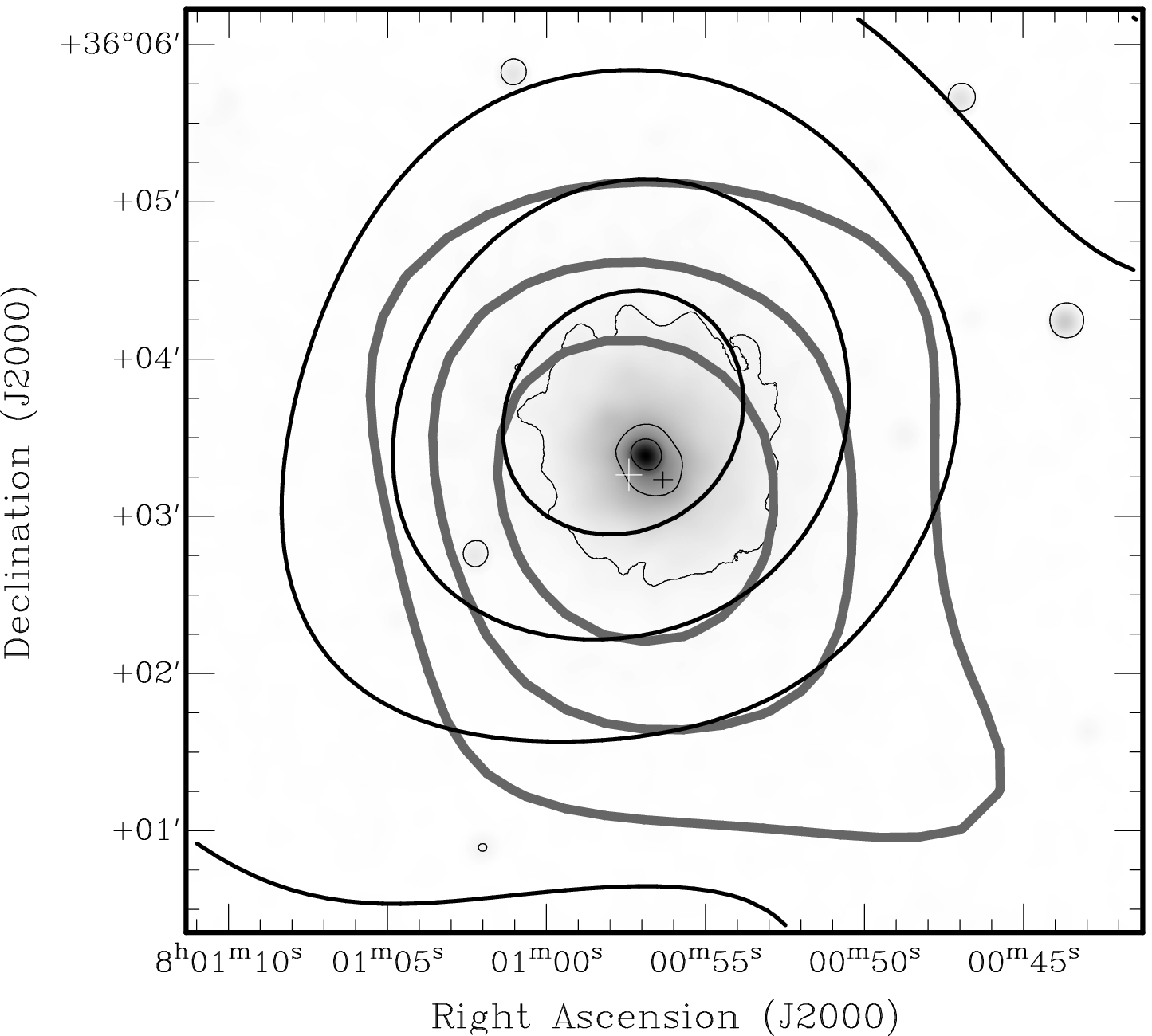}}
\caption{A611.
For the maps and posterior distributions, annotations and axis ranges and labels are as in Fig.~\ref{fig:A1914-all}.
The parameters of the labelled sources in the SA maps can be found in Table~\ref{tab:A611-sources}.
In the composite image, the greyscale shows \textit{Chandra}
X-ray data \citep{laroque06} smoothed by a Gaussian
of radius 15\,arcsec, with 30\% contours. Thicker black contours show the \textsc{LensEnt} mass
reconstruction in 20\% levels of density. Thick grey contours show the AMI
SZ decrement in $2\sigma$ intervals, where the map noise $\sigma=80\,\mu$Jy.
The black cross indicates the cluster catalogue position. The white
cross indicates the fitted centre of the circular $\beta$-model with
arm lengths of $1\sigma$ error on the position. The black cross indicates
the fitted centre of the circular NFW profile with arm lengths
of $1\sigma$ error on the position. \label{fig:A611-all}}
\end{center}
\end{figure*}

A611 is the most dynamically-relaxed cluster in the sample, having
a very uniform X-ray map with little substructure \citep{laroque06}.
The 15.7-GHz source environment
is clean; the decrement is immediately obvious even in the
source-unsubtracted SA map (Fig.~\ref{fig:A611-SA-noss}).
Fitting a $\beta$-model to the gas distribution results in
the parameters in Table~\ref{tab:cluster-parameters}.
%
%\begin{table}
%\begin{center}
%\begin{tabular}{c|ccc}
%\hline
%Parameter & Lensing & SZ \tabularnewline
%\hline
%$x$ / arcsec & $47\pm9$ & $30\pm15$ \tabularnewline
%$y$ / arcsec & $13\pm5$ & $15\pm16$ \tabularnewline
%$c$ & $4.0\pm1.3$ & --- \tabularnewline
%$M$/ $10^{14}h^{-1}M_{\odot}$ & $4.7\pm1.2$ & $6.0\pm1.9$\tabularnewline
%$r_{\mathrm{core}}$/ $h^{-1}$kpc & --- & $490\pm230$ \tabularnewline
%$\beta$ & --- & $1.6\pm0.5$\tabularnewline
%\hline
%\end{tabular}
%\caption{Mean posterior values for the parameters of A611. \label{tab:A611-parameters}}
%\end{center}
%\end{table}

Fitting a circularly-symmetric NFW profile to the lensing data results in
the parameters in Table~\ref{tab:cluster-parameters}.
An elliptical profile was also fitted, but the evidence was not higher
than that for the circular profile, so the extra model parameters were not
justified.
The SZ and lensing masses agree well with each other
at $M\simeq5\times10^{14}h^{-1}\textrm{M}_{\odot}$.
Fig.~\ref{fig:A611} is a composite image of the X-ray
data, \textsc{LensEnt} mass contours and SZ decrement.
\cite{Okabe30GL} find a more distorted mass-distribution for this cluster
but do see their mass distribution peak close to the position we find.

\cite{RomanoA611GL} also perform a weak-lensing analysis of A611
using data from the Large Binocular Telescope and we note that they also find
a relaxed shape to the mass distribution. Fitting a NFW profile, 
they estimate $M_{200}=5.6_{-2.7}^{+4.7}\times 10^{14}\textrm{M}_{\odot}(=3.9_{-1.9}^{+3.3}\times10^{14}h^{-1}\textrm{M}_{\odot})$,
with a concentration parameter $c=3.9^{+5.6}_{-2.1}$ and $r_{200}= 1545^{+345}_{-306}$\,kpc. 
Our measured mass and concentration parameter are higher but not significantly so given the errors (Table~\ref{tab:cluster-parameters}).
We derive, straightforwardly from $M_{200}$, $r_{200}=1560\pm160$\,kpc, which is in extremely good agreement with \citeauthor{RomanoA611GL} despite the strong degeneracy between
$c$ and $r_{200}$ in their model. Our measurement is also in agreement
with \cite{Okabe30GL}, who find $M_{200}=5.47^{+1.31}_{-1.11}\times 10^{14} h^{-1}\textrm{M}_{\odot}$.
%This is slightly higher than as measured by \citeauthor{RomanoA611GL}, but we note
%the low precision in their measurement of $c$ and the strong degeneracy between
These recent results and our own are lower than the estimate of
\citet{COS/All++03}, who use \textit{Chandra} data to calculate
$M_{200}=6.6^{+23.3}_{-5.5}\times10^{14}h^{-2}\mathrm{M}_{\odot}(=9.4^{+33.2}_{-7.9}\times10^{14}h^{-1}\textrm{M}_{\odot})$.
%$M_{total,200}=9.4^{+23.3}_{-5.5}\times10^{14}h^{-1}M_{\odot}$; our result is
%
%but can be
%difficult to compare with masses derived in X-ray and other SZ studies.
%For instance, \citet{laroque06} calculate the gas mass at $r_{2500}$, and
%warn that their $\beta$-model is not a good description of the outskirts
%of clusters. This has been seen in other studies, e.g. \cite{Vikh2006a}
%perform observations and then follow up with numerical simulations
%\citep{Vikh2006b} to explore the problem.
%
\subsection{A851}\label{sub:A851-results}
\begin{figure*}
\begin{center}
\subfloat[Before source-subtraction.\label{fig:A851-SA-noss}]{\includegraphics[width=7cm,clip=]{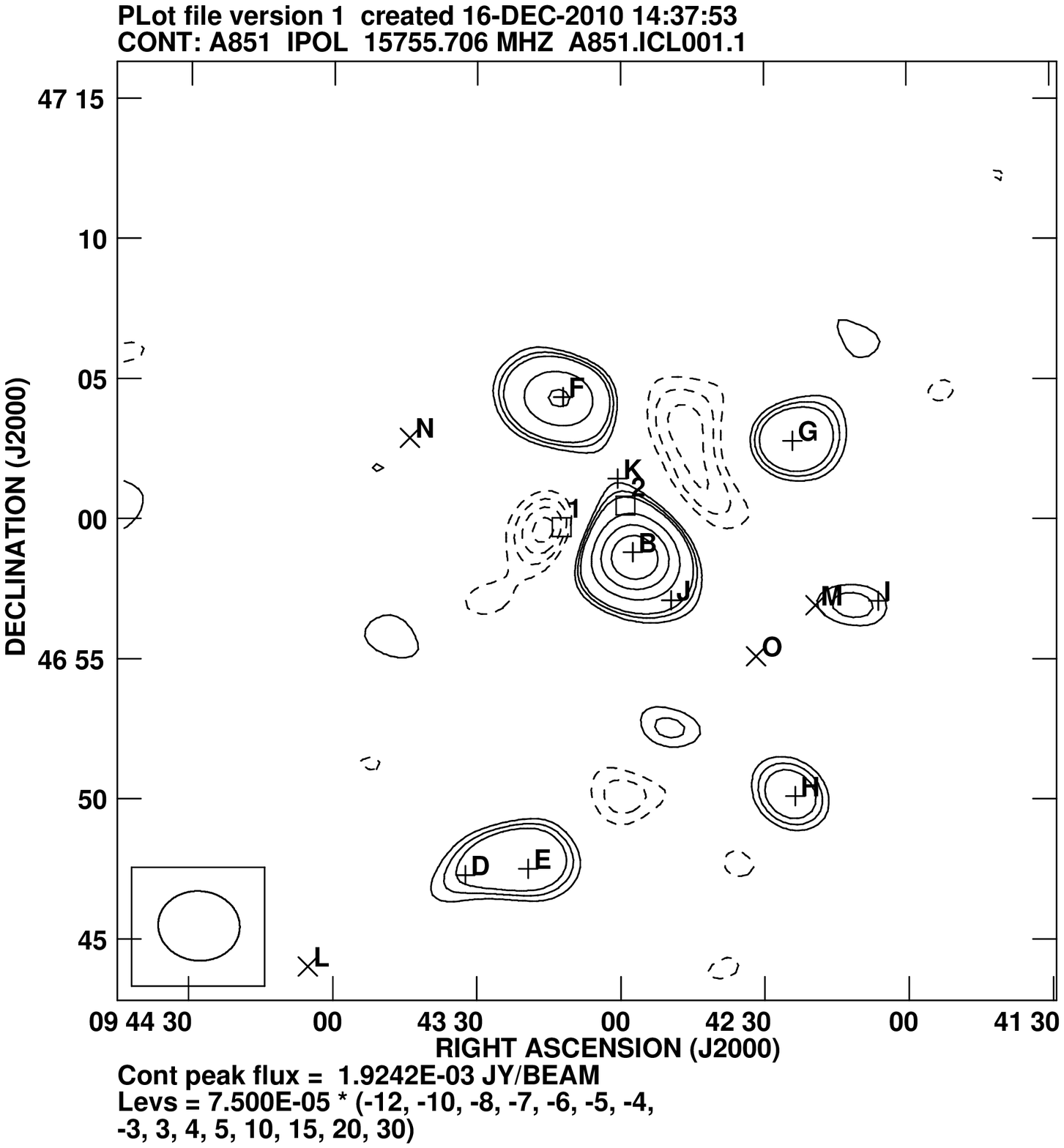}}\quad
\subfloat[After source-subtraction.\label{fig:A851-SA-ss}]{\includegraphics[width=7cm,clip=]{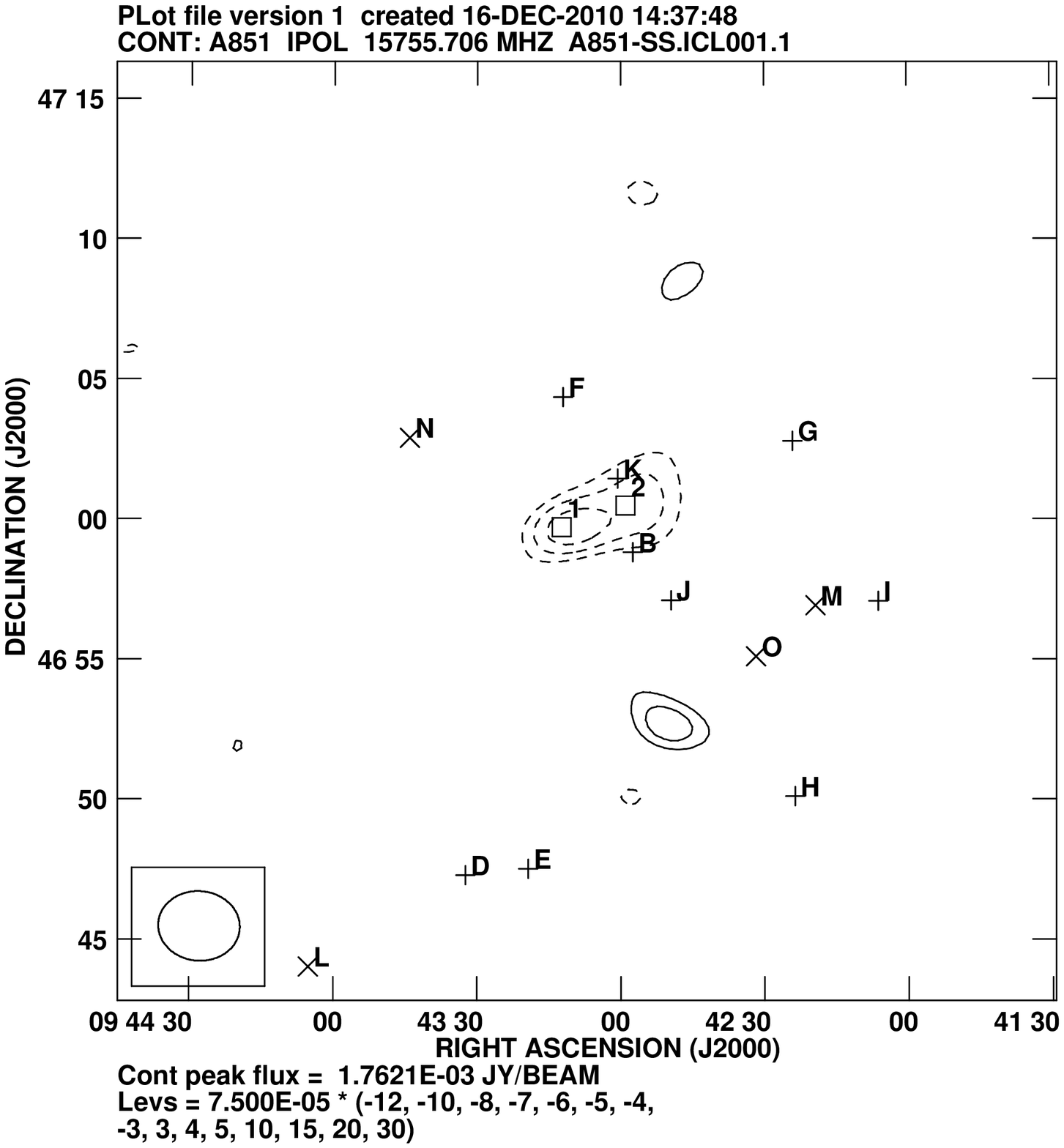}}\qquad
\subfloat[SZ 2D posteriors.\label{fig:A851-SZ}]{\includegraphics[width=10.5cm]{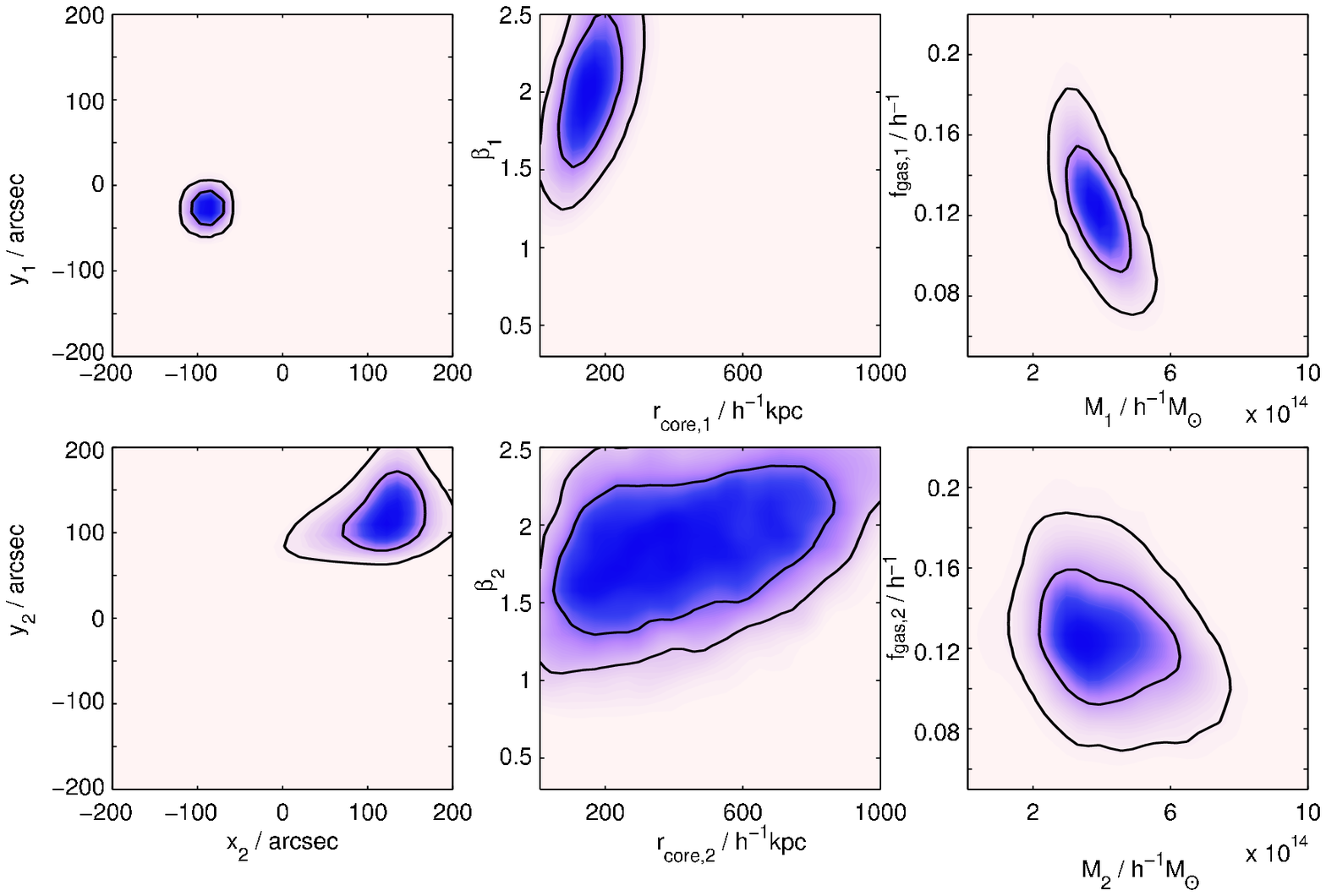}}\quad\quad\qquad\qquad\qquad
\subfloat[Weak lensing 2D posteriors.\label{fig:A851-lensing}]{\includegraphics[width=6.5cm,clip=]{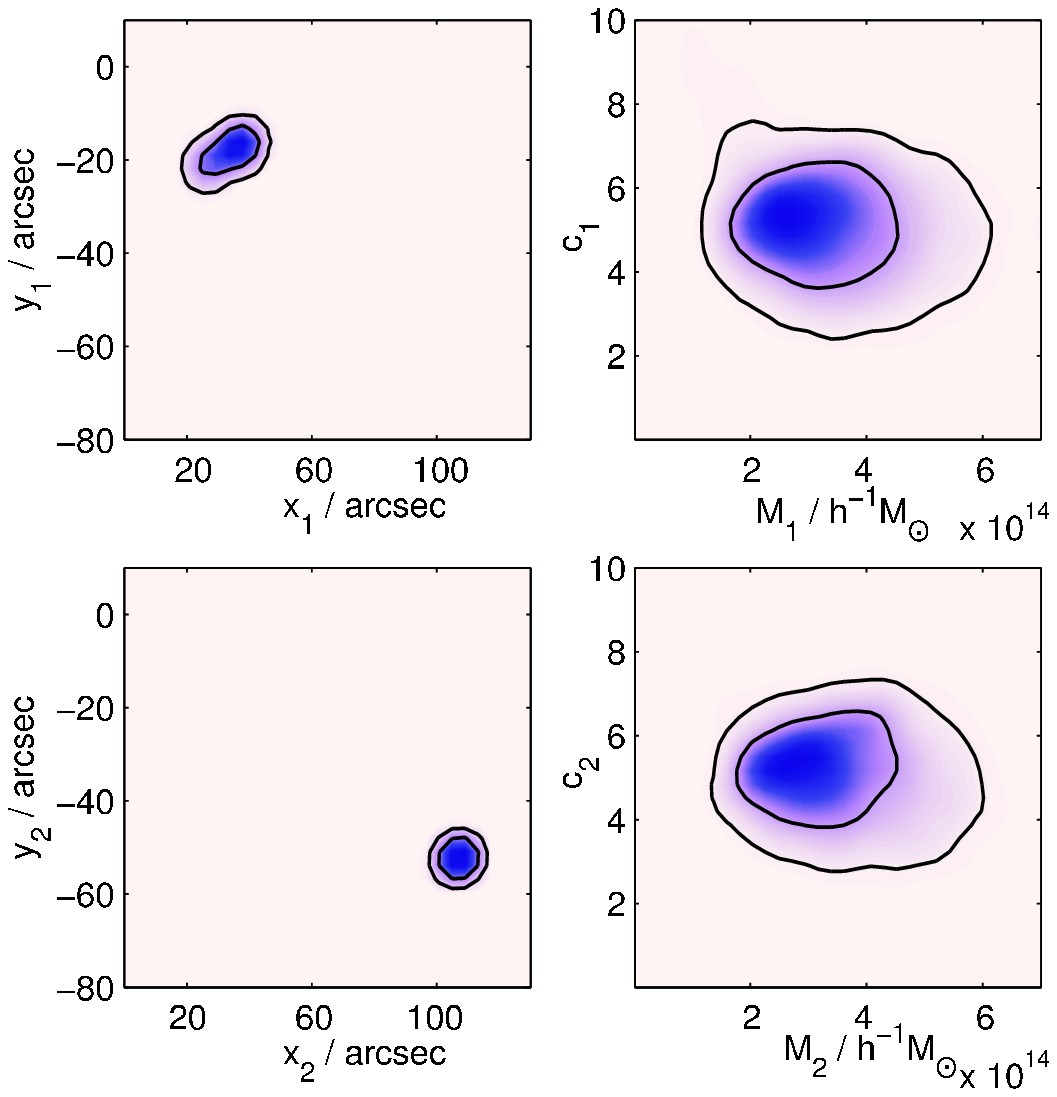}}\quad
\subfloat[X-ray, SZ and weak-lensing composite image.\label{fig:A851}]{\includegraphics[width=8cm,clip=]{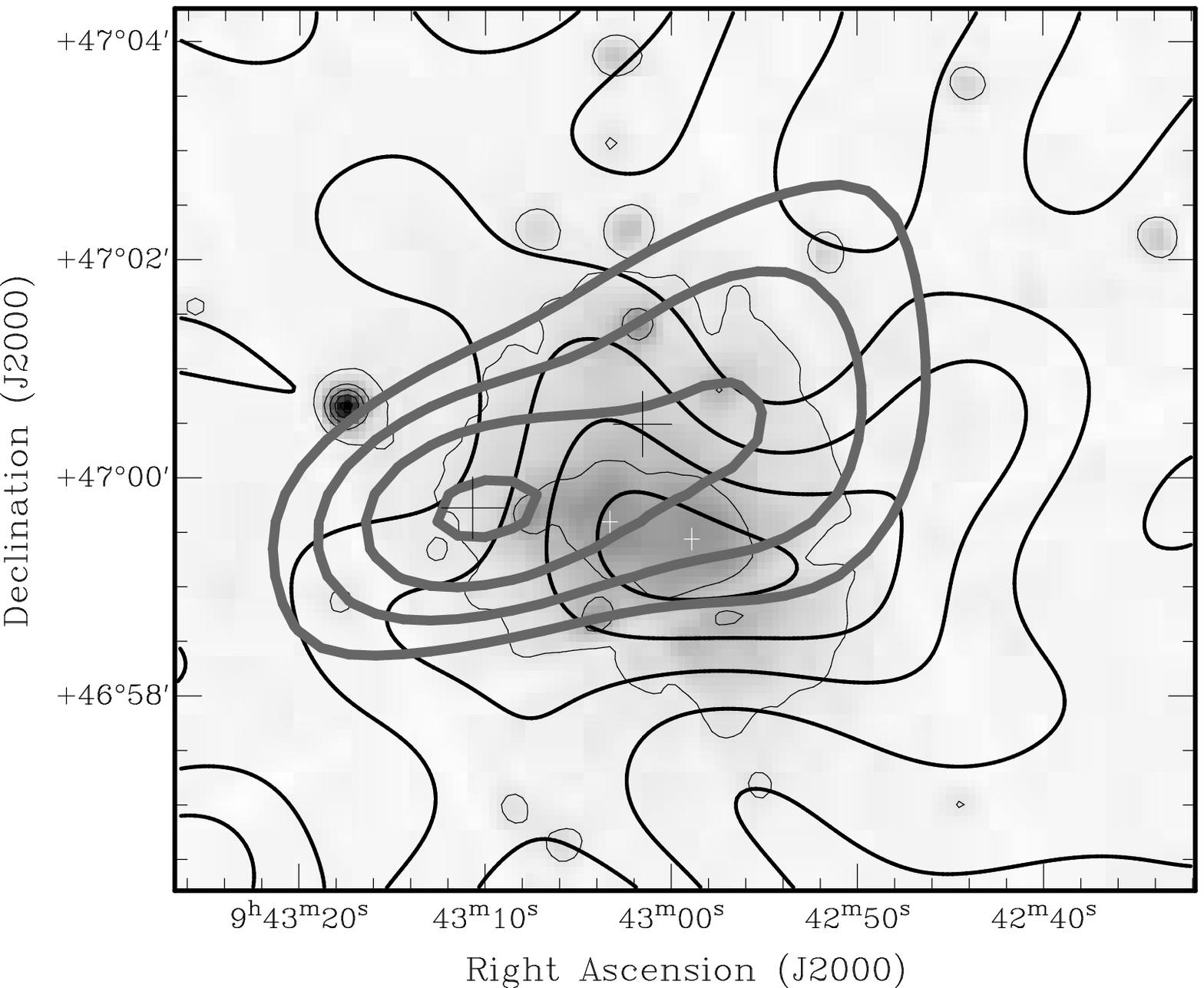}}
\caption{A851.
For the maps and posterior distributions, annotations and axis ranges and labels are as in Fig.~\ref{fig:A1914-all}.
Posterior distributions for two different components fit to each dataset are shown, indicated by the subscripts
on the parameter values.
The parameters of the labelled sources in the SA maps can be found in Table~\ref{tab:A851-sources}.
In the composite image, the greyscale shows \textit{X-ray Multi-Mirror Mission\,-,Newton} X-ray data,
smoothed by a Gaussian of 5\,arcsec, with 20\% contours. Thicker black contours
are 20\% levels of a \textsc{LensEnt} reconstruction using a resolution of
100\,arcsec. Thick grey contours show the SZ decrement in $1\sigma$
intervals, where the map noise $\sigma=75\,\mu$Jy. The black crosses mark the positions
of the two isothermal beta profiles fitted to the SZ data, and the white crosses show the positions
of the two NFW profiles fitted to the lensing data; the arm lengths show $1\sigma$ error bars in each
case.\label{fig:A851-all}}
\end{center}
\end{figure*}

A851 is the highest-redshift cluster in the sample, and is known to
have a very irregular gas distribution, indicating that it is dynamically
young: \citet{2003A+A...404...63D} find that A851 is composed of two subclusters
in the process of merging, the angular
separation of which is 1.5\,arcmin.
The SA map before source subtraction (Fig.~\ref{fig:A851-SA-noss})
shows a busy source environment, with several sources directly over
the cluster. Despite this, the edges of the SZ decrement are still
visible to the east and north-west of the largest conglomeration of
sources. Fitting 11 of the 15 sources (Table~\ref{tab:A851-sources})
%in \textsc{McAdam}
and subtracting them from the SA data results in the
source-subtracted map shown in Fig.~\ref{fig:A851-SA-ss}. The remaining
unsubtracted sources are of flux densities comparable to the SA map
noise. The SZ decrement is clearly visible and there are no significant
positive residuals remaining on the map. The gas distribution is fairly
distorted and ellipsoidal, with a possible extended spur to the north-west.

A $\beta$-model was fitted to these data:
%using \textsc{McAdam}
an elliptical model was preferred
over a spherical model, but the model with the highest evidence was two
spherical $\beta$-models at two separate positions: the resulting parameters
from these fits are listed in Table~\ref{tab:cluster-parameters}.
and the 2D posteriors are
shown in Fig.~\ref{fig:A851-SZ}.

The $z=2$ cluster candidate detected by \citet{1993ApJ...404L..45D} might
produce some SZ signal. Unfortunately its distance from the centre of the A851
decrement is less than 1\,arcmin, so this cannot be resolved by the SA.

Using \textsc{LensEnt} to reconstruct
the mass distribution of A851, at a higher resolution than shown in Fig.~\ref{fig:A851-all},
the two most significant features
are two ellipsoidal subclumps separated by around 1\,arcmin.
These agree fairly well with the positions of the X-ray subclumps
detected by \citeauthor{2003A+A...404...63D} They were also detected as mass
overdensities in lensing analysis carried out by \citet{2000PASJ...52....9I}.

Since two distinct components are clearly identified,% \textsc{McAdam} was used
we fitted two circularly-symmetric NFW profiles to the data. Elliptical models were also
attempted but the evidence did not justify the extra parameters.
Table~\ref{tab:cluster-parameters} shows the resulting parameters, which indicate
that the two subclusters are of roughly equal mass and concentration,
and Fig.~\ref{fig:A851-lensing} shows the 2D lensing posteriors.
The summed mass of these clumps,
$M=6.6\pm1.4\times10^{14}h^{-1}\mathrm{M}_{\odot}$,
 is consistent with the mass limits
determined by \citet{GL/S+K+S96}:
$2.6\times10^{14}h^{-1}\mathrm{M}_{\odot}< M<7.1\times10^{14}h^{-1}\mathrm{M}_{\odot}$.
\citet{2000PASJ...52....9I} do not attempt to make an estimate of the total mass
of the cluster.
Given that the redshift of this cluster is so high, we could be including
more foreground (unlensed) galaxies in the field galaxy selection, reducing
the signal and thus the detected mass, which could explain the 2$\sigma$ difference between
our SZ and lensing mass measurements for this cluster.

Fig.~\ref{fig:A851} shows a composite image of the SZ
decrement, mass reconstruction from lensing and X-ray data. The north-east
mass subclump appears coincident with the X-ray subclump, while the
south-west mass concentration is a little south of the other X-ray subclump.
The main body of the
SZ decrement covers the X-ray-bright area and the two mass clumps,
but there is also an extension to the south-east, which is clearly
visible even before source-subtraction. This might imply that gas
is being `squirted' out of the sides of the cluster, in directions perpendicular
to that of the motion of the subclumps. \citeauthor{2003A+A...404...63D} find
that the compressed gas which also follows this NW--SE extension has
a higher temperature of 6\,keV; if the high temperature extends to the
outskirts of the gas, this will also boost the observed SZ signal.
\section{Discussion}\label{sec:discussion}
\subsection{Mass comparison and modelling}
\begin{figure}
\begin{center}
\includegraphics[angle=-90,width=7cm]{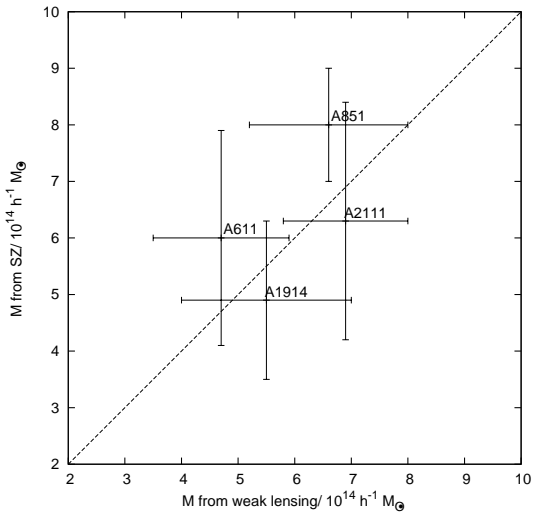}\quad
\caption{Comparison of the different values of $M$ obtained by the lensing
and SZ analyses. The dashed line shows a 1:1 ratio.\label{fig:mass-plot}}
\end{center}
\end{figure}

Fig.~\ref{fig:mass-plot} shows a comparison of the different measured
values of $M$ for the clusters for which we were able to perform both the
SZ and lensing analyses. Measured masses for A611 and A1914 are in very good agreement.
Interestingly, we find better agreement between mass measurements for A1914
than \cite{2010ApJ...721..875O}, who find that including this cluster significantly
distorts their otherwise well-correlated $M$-$T$ relationship. As we use SZ measurements
at $r_{200}$ rather than X-ray at $r_{500}$, this may be indicative of SZ's advantage
in measuring the mass of merging clusters, which are likely to be more common at
the higher redshifts now being probed by new galaxy cluster surveys.

We measure the mass of A851 as larger in the SZ compared to the lensing, although the difference
is not a significant deviation given the errors. However it is the highest
redshift cluster of the sample, it is possible that we are underestimating its mass via lensing,
due to poor field galaxy selection. This may put limits on our ability to measure
weak lensing at higher redshifts, at least with the two optical bands used in this
analysis.

For A2111, the temperature may be somewhat higher than our $M$-$T$
relation predicts, as this cluster is an ongoing merger and has a hot 8\,keV
central component; difficulties in reconciling disturbed clusters with theoretical 
relations have been seen in other studies (e.g. \citealt{2004ApJ...613...95C}).
 Given the small sample discussed here,
it is difficult to draw general conclusions, but the technique would scale
well to larger samples, from which it may be possible to correlate the state of
merger with discrepancy from this $M$-$T$ relation.
\subsection{Gas distribution on large scales}\label{sub:large-scale-gas}
The investigation of \citet{NFW95} into the suitability of the $\beta$-model
for X-ray clusters showed that the measured value of $\beta$ increases
with increasing radius of fit. For relaxed clusters, typical values
of $\beta$ from X-ray and high-resolution SZ, both of which resolve
out large angular scales, are in the range $0.5<\beta<0.8$. In the
context of these core gas measurements, larger values of $\beta$
usually indicate merging processes.

\citet{laroque06} provide the most recent combined SZ-X-ray analysis
that covers the most relaxed clusters in our sample, A2111, A2259 and A611.
We detect the gas masses in the
similar proportions to \citeauthor{laroque06}: A611 and A2111 are of similar mass, and more massive than A2259. This is reassuring,
especially given that the hot ($\simeq8$\,keV) gas in the centre of
A2111 was ignored for our isothermal analysis. It is difficult
to compare our values for $r_{\mathrm{c}}$ and $\beta$ since we are looking
at a much more extended areas of the gas and one would expect the profiles
to be different for different fitting areas.

Using a $\beta$-model presents a problem if extrapolated to high
radii: the density does not steepen quickly enough and large, non-physical
gas masses are predicted. Therefore both high- and low- angular scale
measurements specify a cut-off radius at which to measure the gas
mass, often $r_{500}$, while in this analysis we use $r_{200}$.
It can be difficult to extrapolate
models produced from our data to smaller radii such as $r_{2500}(\approx0.3r_{200})$,
since with a high value of $\beta$, our profiles may not actually reach that
density before the mass goes to zero.
\subsection{Gas fraction}
\begin{figure}
\begin{center}
\includegraphics[angle=0,width=4cm]{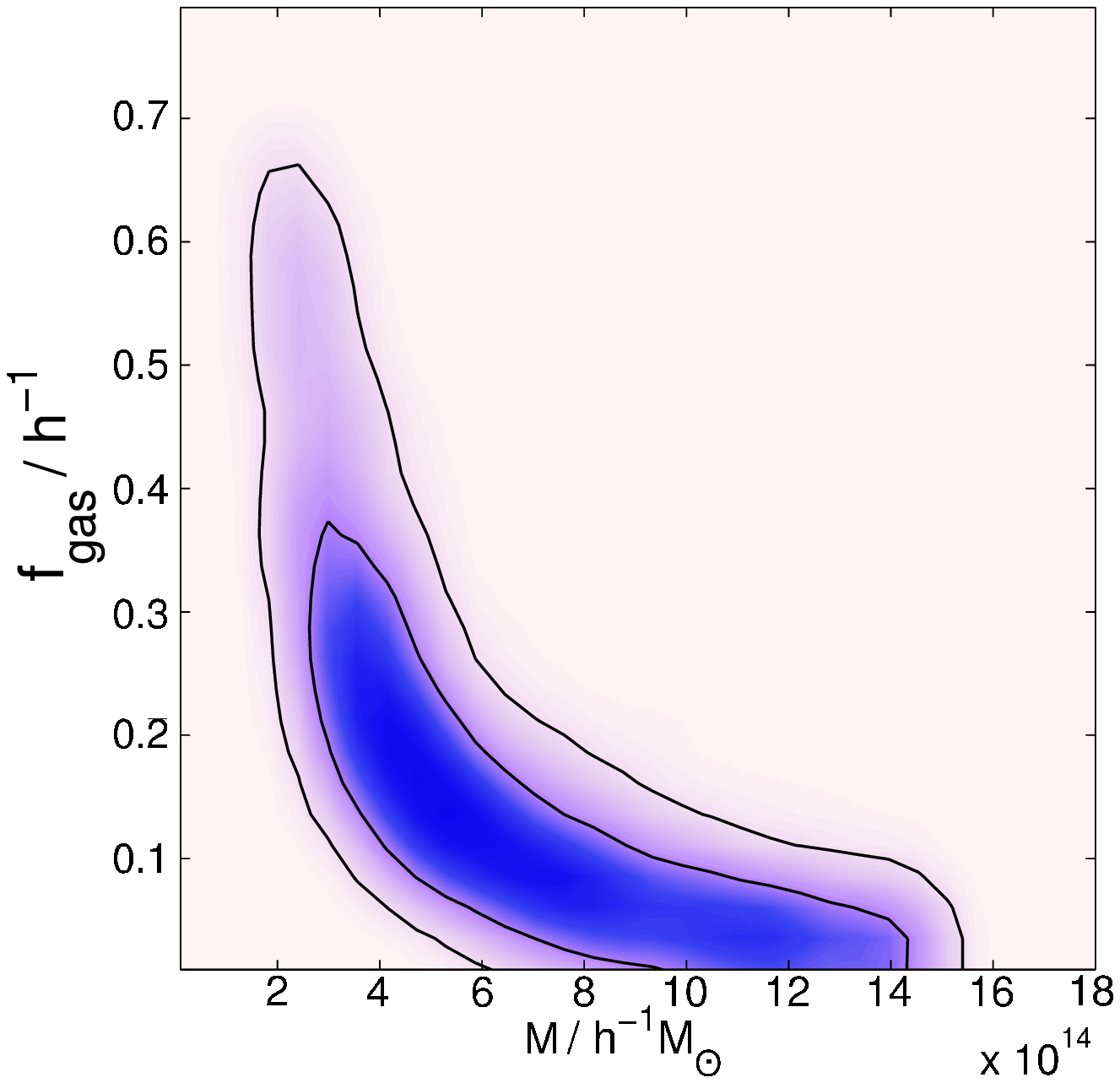}\quad
\includegraphics[angle=0,width=4cm]{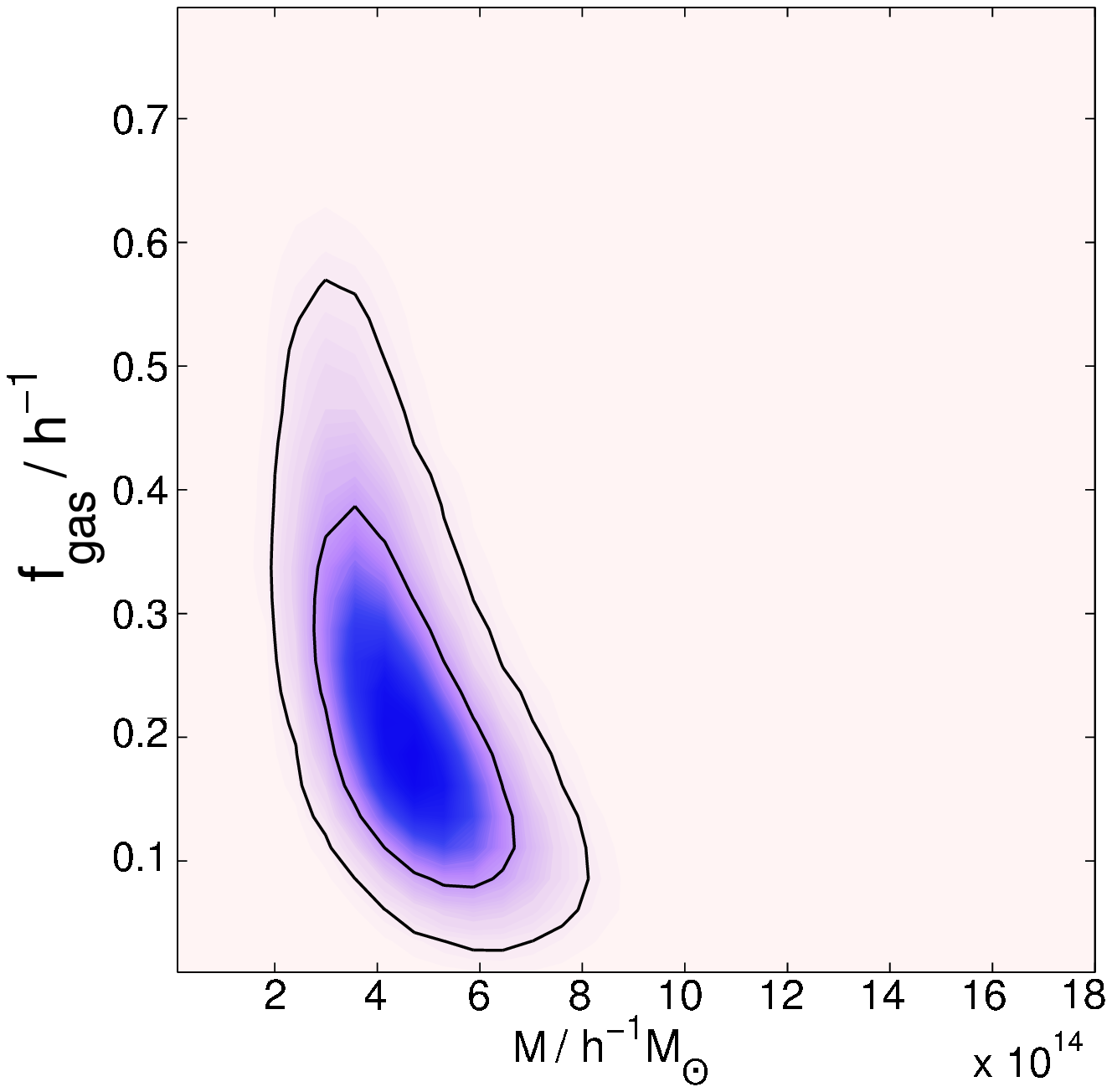}
\caption{$f_{\mathrm{gas}}$--$M$ degeneracy for SZ data alone (left) and SZ \& GL data combined (right).
\label{fig:A611-SZGL2D}}
\end{center}
\end{figure}

For the most relaxed cluster in our sample, A611, we further investigated the gas fraction
using a joint SZ and lensing analysis. We removed the WMAP prior on gas fraction, giving it instead
a uniform prior between 0.01 and 1$h^{-1}$. The posterior plots for the model with and without
the lensing data are shown in Fig.~\ref{fig:A611-SZGL2D}. Together, the datasets constrain the
gas fraction, which cannot be done with either individually. It is noticeable that the SZ data has
no bias in its measurement of the $M-f_{\mathrm{gas}}$ degeneracy; the combined data posterior lies
along the posterior from the SZ data alone. This implies our parameterization is good, and that
with a sensible prior on $f_{\mathrm{gas}}$, SZ data allow us to make good estimates of cluster masses.

The combined data favour a higher value of gas fraction for this cluster than the
WMAP prior normally used: $0.23\pm0.1h^{-1}$ compared to $0.123\pm0.02h^{-1}$. However
the errors are large so it is difficult to tell whether this is significant. It would be illuminating
to extend this analysis to a large dataset of relaxed clusters, as general conclusions about
cluster morphology cannot be made from this single result, and the other lensing-detected clusters
in this paper are complex mergers.

\subsection{Data issues and cluster selection}

This pilot study has been helpful in identifying some areas of difficulty
in joint SZ--weak-gravitational-lensing analysis.
The high density of foreground optical sources in the field of A2259 made measurement
of its low mass difficult via gravitational lensing, while it was well-measured via the SZ. On the other
hand, the SZ analysis of A115 suffered from the effects of contaminating radio sources, and this effect is
also seen in other AMI cluster observations, e.g. by \cite{AMIlocuss1}.

For those clusters observed outside the Galactic plane, with unresolved (easily-modelled) radio sources, the
analysis showed remarkable agreement between measured $M_{200}$ regardless of dynamical state, X-ray luminosity,
temperature, redshift and number of detected sub-structures. Currently the AMI Consortium has over 50 further detections of SZ clusters in similar environments
covering a redshift range of $0.0894\leq z\leq0.686$ and an X-ray luminosity
range of $3.0\times10^{37}\leq L_{\textrm{X}} \leq 28\times10^{37}$\,W, from X-ray-selected samples. Of these, 11 
have associated archival CFHT data, on which may be performed a similar analysis.
Also, the clusters being discovered by \textit{Planck} \citep{PlanckESZ} are highly amenable
to observations by AMI \citep{Tash2PlanckBlind}, so there is potential to expand the sample size by another order
of magnitude over coming years.

\section{Conclusions}\label{sec:conclusions}
\begin{enumerate}
\item{Using AMI SZ data that measure out to $r_{200}$, and thus probe scales
comparable to those of weak-lensing observations, we have used a
fast, Bayesian analysis
%with \textsc{McAdam}
to produce posteriors
for useful parameters for six clusters, on the assumptions of an isothermal $\beta$-model
to SZ data and an NFW model for the lensing data;}
\item{Of the four clusters for which we have both weak lensing and SZ
data, we find that the mass estimates are in very good agreement, and that we
may need to improve our field-galaxy selection for clusters at $z>0.4$
for a larger sample;}
\item{We perform the first multiple-component weak-lensing analysis of A115 and
discover significant substructure;}
\item{We confirm the unusual separation between the gas and mass components in A1914;}
\item{For A611, the most relaxed cluster in the sample, we have carried
out a joint weak-lensing and SZ analysis with the gas fraction
as a free parameter, and find that $f_{\mathrm{gas}}=0.23\pm0.10h^{-1}$.}
\end{enumerate}
\section{ACKNOWLEDGMENTS}

We thank the staff of the Mullard Radio Astronomy Observatory for their invaluable
assistance in the commissioning and operation of AMI, which
is supported by Cambridge University and the STFC.
Computational results were obtained using the COSMOS supercomputer
(DiRAC STFC HPC Facility).
This research used the facilities of the Canadian Astronomy Data
Centre operated by the National Research Council of Canada with the
support of the Canadian Space Agency.
MLD, TMOF, CRG, MO, MPS and TWS acknowledge the
support of STFC studentships.

\begin{table*}
\begin{center}
\begin{tabular}{ccccc|ccccc}
\hline 
ID & RA & Dec &  $S_{\rm{LA}}$/mJy & $\alpha_{\rm{LA}}$ & ID & RA & Dec & $S_{\rm{LA}}$/mJy & $\alpha_{\rm{LA}}$\tabularnewline
\hline
A & 00 56 55.18 & +26 31 24.48 & $52.0\pm 2.8$ & $-0.5\pm 0.3$ & O  & 00 56 01.38 & +26 28 43.78 &  $1.7\pm 0.3$ & $ 1.1\pm 1.4$ \tabularnewline
B & 00 56 02.85 & +26 27 20.98 & $27.6\pm 1.4$ & $ 0.8\pm 0.1$ & P  & 00 56 07.16 & +26 29 46.05 &  $1.4\pm 0.3$ & $ 0.5\pm 1.5$ \tabularnewline
C & 00 57 21.47 & +26 17 27.90 &  $7.1\pm 0.7$ & $ 0.9\pm 1.1$ & Q  & 00 56 06.97 & +26 30 51.52 &  $1.3\pm 0.3$ & $ 0.4\pm 1.8$ \tabularnewline
D & 00 55 50.46 & +26 24 39.02 &  $5.4\pm 0.3$ & $ 2.5\pm 0.4$ & R  & 00 56 00.94 & +26 25 03.79 &  $1.1\pm 0.2$ & $ 1.3\pm 1.3$ \tabularnewline
E & 00 56 20.93 & +26 30 55.24 &  $4.1\pm 0.3$ & $ 1.1\pm 1.0$ & S  & 00 55 59.02 & +26 17 49.46 &  $1.0\pm 0.1$ & $ 1.3\pm 1.3$ \tabularnewline
F & 00 57 06.53 & +26 25 27.03 &  $4.0\pm 0.5$ & $ 1.5\pm 1.4$ & T  & 00 56 10.64 & +26 12 03.40 &  $1.0\pm 0.1$ & $ 0.5\pm 1.4$ \tabularnewline
G & 00 56 08.49 & +26 25 13.46 &  $3.6\pm 0.2$ & $-0.5\pm 0.6$ & U  & 00 56 28.46 & +26 24 03.69 &  $0.9\pm 0.1$ & $ 1.3\pm 1.5$ \tabularnewline
H & 00 56 22.47 & +26 23 03.43 &  $2.8\pm 0.2$ & $ 0.5\pm 0.6$ & V  & 00 55 58.37 & +26 25 33.95 &  $0.9\pm 0.2$ & $ 0.8\pm 1.7$ \tabularnewline
I & 00 56 55.38 & +26 29 12.30 &  $2.7\pm 0.6$ & $ 0.1\pm 1.6$ & W  & 00 55 48.04 & +26 25 56.77 &  $0.7\pm 0.1$ & $ 2.3\pm 1.5$ \tabularnewline
J & 00 56 18.52 & +26 05 00.41 &  $2.3\pm 0.3$ & $-0.9\pm 1.3$ & X  & 00 56 04.35 & +26 25 00.45 &  $0.7\pm 0.2$ & $-0.9\pm 1.5$  \tabularnewline
K & 00 54 55.80 & +26 19 13.30 &  $2.2\pm 0.3$ & $-0.3\pm 1.4$ & Y  & 00 56 05.32 & +26 23 44.74 &  $0.6\pm 0.1$ & $ 0.1\pm 1.6$  \tabularnewline
L & 00 56 50.45 & +26 29 38.97 &  $2.1\pm 0.5$ & $ 0.4\pm 1.6$ & Z  & 00 56 10.69 & +26 24 15.30 &  $0.6\pm 0.1$ & $ 2.2\pm 1.5$  \tabularnewline
M & 00 56 43.35 & +26 29 22.20 &  $2.0\pm 0.4$ & $ 0.2\pm 1.5$ & AA & 00 55 49.68 & +26 22 20.83 &  $0.5\pm 0.1$ & $ 0.2\pm 1.6$ \tabularnewline
N & 00 56 40.69 & +26 30 10.21 &  $1.8\pm 0.4$ & $ 2.2\pm 1.5$ &  & & & & 						 \tabularnewline
\hline
\end{tabular}

\caption{Sources found in the LA image of A115, in order of LA flux density. In this and
all subsequent tables, the flux densities given are measured from the combined-channel 15.7-GHz LA
maps. The spectral index is calculated by measuring the flux densities from the separate
channels and fitting a power-law spectrum. 1$\sigma$ errors are shown for both measurements.
\label{tab:A115-sources}}
\end{center}

\end{table*}

\begin{table*}
\begin{center}
\begin{tabular}{ccccccc}
\hline 
ID & RA & Dec & $S_{\rm{LA}}$/mJy & $\alpha_{\rm{LA}}$ &
$S_{\textrm{SA}}$/mJy & $\alpha_{\textrm{SA}}$\tabularnewline
\hline
A & 14 27 25.06 & +37 46 33.19 & $10.7\pm 0.5$ & $ 0.1\pm 0.6$ & $8.3\pm0.7$ & $0.2 \pm0.5$\tabularnewline
B & 14 25 08.41 & +37 52 42.58 & $ 4.3\pm 0.2$ & $ 1.7\pm 0.7$ & $4.4\pm0.3$ & $1.3 \pm0.5$\tabularnewline
C & 14 25 52.42 & +38 03 04.66 & $ 2.9\pm 0.3$ & $-0.5\pm 1.2$ & $3.8\pm0.3$ & $-0.1\pm1.0$\tabularnewline
D & 14 25 40.83 & +37 45 47.93 & $ 2.7\pm 0.1$ & $ 1.2\pm 0.5$ & $3.1\pm0.2$ & $1.0 \pm0.4$\tabularnewline
E & 14 25 05.12 & +37 55 15.84 & $ 2.7\pm 0.3$ & $ 0.8\pm 1.3$ & $2.5\pm0.3$ & $1.0 \pm1.1$\tabularnewline
F & 14 27 10.92 & +37 55 13.89 & $ 2.7\pm 0.3$ & $ 0.9\pm 1.3$ & $2.5\pm0.4$ & $1.6 \pm1.2$\tabularnewline
G & 14 25 57.37 & +38 01 10.48 & $ 1.4\pm 0.3$ & $ 0.3\pm 1.6$ & $1.2\pm0.2$ & $0.2 \pm1.4$\tabularnewline
H & 14 25 47.75 & +37 47 48.32 & $ 1.2\pm 0.1$ & $-0.4\pm 1.2$ & $1.2\pm0.1$ & $0.0 \pm1.0$\tabularnewline
I & 14 25 56.73 & +37 55 10.47 & $ 0.8\pm 0.1$ & $ 0.7\pm 1.7$& $0.5\pm0.1$ & $0.1 \pm1.4$ \tabularnewline
J & 14 25 50.36 & +37 45 07.81 & $ 0.7\pm 0.1$ & $ 0.8\pm 1.5$& $0.7\pm0.1$ & $1.1 \pm1.5$ \tabularnewline
\hline
K & 14 25 05.78 & +37 47 02.26 & $ 1.2\pm 0.2$ & $-0.5\pm 1.7$  & -- & -- \tabularnewline
L & 14 25 43.15 & +37 40 06.15 & $ 0.8\pm 0.1$ & $-0.8\pm 1.7$ & -- & -- \tabularnewline
M & 14 25 58.12 & +37 43 58.58 & $ 0.6\pm 0.1$ & $ 1.0\pm 1.5$ & -- & -- \tabularnewline
N & 14 25 38.95 & +37 57 32.50 & $ 0.6\pm 0.1$ & $ 1.6\pm 1.7$ & -- & -- \tabularnewline
\hline
\end{tabular}

\caption{Sources found in the LA image of A1914, and fitted to the SA data, in order of LA flux density.
In this and all subsequent tables,
the fitted SA flux densities and spectral indices with $1\sigma$ errors
are shown in the
final two columns; sources without entries in these columns were directly subtracted from the data to reduce the
parameter space to a more manageable size. 
\label{tab:A1914-sources}}

\end{center}
\end{table*}

\begin{table*}
\begin{center}
\begin{tabular}{ccccccc}
\hline
ID & RA & Dec &  $S_{\rm{LA}}$/mJy & $\alpha_{\rm{LA}}$ &
$S_{\textrm{SA}}$/mJy & $\alpha_{\textrm{SA}}$\tabularnewline
\hline
A & 15 40 55.25 & +34 30 17.50 & $17.4\pm 0.7$ & $-0.7\pm 0.4$ & $14.6\pm 0.5$& $-0.1 \pm 0.2$\tabularnewline
B & 15 38 46.64 & +34 18 58.04 & $ 6.3\pm 0.3$ & $ 1.6\pm 0.6$ & $5.5 \pm 0.3$& $1.5  \pm 0.5$\tabularnewline
C & 15 40 31.18 & +34 30 09.72 & $ 2.9\pm 0.2$ & $ 0.2\pm 1.1$ & $2.9 \pm 0.2$& $1.0  \pm 0.7$\tabularnewline
D & 15 39 11.87 & +34 29 33.07 & $ 2.5\pm 0.1$ & $ 0.7\pm 0.7$ & $2.6 \pm 0.1$& $1.5  \pm 0.4$\tabularnewline
E & 15 39 55.06 & +34 20 11.40 & $ 2.3\pm 0.1$ & $ 1.2\pm 0.6$ & $1.9 \pm 0.1$& $1.2  \pm 0.5$\tabularnewline
F & 15 40 41.89 & +34 18 35.72 & $ 2.1\pm 0.4$ & $-0.3\pm 1.3$ & $2.8 \pm 0.2$& $0.7  \pm 1.0$\tabularnewline
G & 15 40 49.55 & +34 30 35.66 & $ 1.9\pm 0.6$ & $ 0.8\pm 1.6$ & $1.9 \pm 0.3$& $0.7  \pm 1.5$\tabularnewline
H & 15 40 31.90 & +34 28 16.52 & $ 1.5\pm 0.3$ & $ 0.2\pm 1.4$ & $2.3 \pm 0.2$& $0.3  \pm 1.0$\tabularnewline
I & 15 38 49.51 & +34 26 56.26 & $ 1.3\pm 0.2$ & $ 1.7\pm 1.5$ & $1.8 \pm 0.2$& $2.3  \pm 1.1$\tabularnewline
J & 15 39 08.11 & +34 21 09.02 & $ 0.8\pm 0.1$ & $ 0.4\pm 1.6$ & $0.7 \pm 0.1$& $0.2  \pm 1.4$\tabularnewline
K & 15 39 56.78 & +34 29 33.38 & $ 0.8\pm 0.1$ & $ 0.8\pm 1.4$ & $0.9 \pm 0.1$& $-0.5 \pm 1.0$\tabularnewline
L & 15 39 30.11 & +34 29 05.46 & $ 0.6\pm 0.1$ & $ 1.2\pm 1.6$ & $0.5 \pm 0.1$& $1.0  \pm 1.4$\tabularnewline
\hline
\end{tabular}

\caption{Sources found in the LA image of A2111, and fitted to the SA data, in order of LA flux density.\label{tab:A2111-sources}}

\end{center}
\end{table*}

\begin{table*}
\begin{center}
\begin{tabular}{ccccccc}
\hline 
ID & RA & Dec & $S_{\mathrm{LA}}$/mJy & $\alpha_{\mathrm{LA}}$ &
$S_{\textrm{SA}}$/mJy & $\alpha_{\textrm{SA}}$ \tabularnewline
\hline
A & 17 19 13.69 & +27 44 49.47 & $ 2.2\pm 0.3$ & $ 2.3\pm 1.3$ & $1.9\pm0.3$& $2.4\pm1.1$ \tabularnewline
B & 17 19 30.18 & +27 46 16.12 & $ 1.1\pm 0.2$ & $ 0.5\pm 1.6$ & $0.9\pm0.2$& $0.5\pm1.4$ \tabularnewline
C & 17 20 51.25 & +27 45 08.05 & $ 1.4\pm 0.1$ & $-0.2\pm 1.5$ & $1.6\pm0.2$& $0.7\pm1.2$ \tabularnewline
D & 17 19 27.06 & +27 32 51.05 & $ 1.0\pm 0.2$ & $ 0.1\pm 1.7$ & $0.6\pm0.2$& $0.0\pm1.6$ \tabularnewline
E & 17 20 20.26 & +27 28 33.96 & $ 1.0\pm 0.2$ & $ 1.3\pm 1.8$ & $1.2\pm0.2$& $2.0\pm1.5$ \tabularnewline
F & 17 20 07.54 & +27 44 35.03 & $ 0.8\pm 0.1$ & $ 0.1\pm 1.3$ & $0.6\pm0.1$& $0.2\pm1.2$ \tabularnewline
G & 17 19 44.57 & +27 32 37.99 & $ 0.6\pm 0.1$ & $ 1.7\pm 1.6$ & $0.4\pm0.1$& $1.4\pm1.5$ \tabularnewline
H & 17 19 41.33 & +27 42 10.07 & $ 0.4\pm 0.1$ & $ 0.4\pm 1.7$ & $0.4\pm0.1$& $0.6\pm1.7$ \tabularnewline
I & 17 20 16.68 & +27 37 50.27 & $ 0.3\pm 0.1$ & $ 1.0\pm 1.7$ & $0.5\pm0.1$& $0.4\pm1.4$ \tabularnewline
\hline
\end{tabular}

\caption{Sources found in the LA image of A2259, and fitted to the SA data, in order of LA flux density.
\label{tab:A2259-sources}}
\end{center}
\end{table*}

\begin{table*}
\begin{center}
\begin{tabular}{cccccccc}
\hline
ID & RA & Dec & $S_{\rm{LA}}$/mJy & $\alpha_{\rm{LA}}$ & $S_{\textrm{SA}}$/mJy & $\alpha_{\textrm{SA}}$\tabularnewline
\hline
A & 08 00 07.83 & +36 04 08.56 & $ 5.9\pm 0.2$ & $ 0.5\pm 0.4$ & $6.1\pm0.2$& $0.4 \pm0.3$\tabularnewline
B & 08 02 00.53 & +36 08 54.64 & $ 3.3\pm 0.3$ & $-0.2\pm 1.0$ & $2.6\pm0.2$& $0.3 \pm0.9$\tabularnewline
C & 08 00 40.55 & +36 14 23.78 & $ 3.1\pm 0.3$ & $ 0.3\pm 1.1$ & $3.6\pm0.3$& $1.2 \pm0.7$\tabularnewline
D & 07 59 51.35 & +36 11 05.49 & $ 3.0\pm 0.4$ & $ 0.2\pm 1.2$ & $2.5\pm0.4$& $0.8 \pm1.1$\tabularnewline
E & 08 00 43.25 & +36 14 03.44 & $ 3.0\pm 0.3$ & $ 0.3\pm 1.1$ & $2.8\pm0.3$& $0.4 \pm0.8$\tabularnewline
F & 07 59 55.92 & +35 58 35.84 & $ 2.8\pm 0.2$ & $ 1.0\pm 1.3$ & $1.5\pm0.2$& $0.6 \pm1.1$\tabularnewline
G & 08 02 12.27 & +36 03 48.26 & $ 2.2\pm 0.2$ & $ 1.0\pm 1.3$ & $1.4\pm0.3$& $1.2 \pm1.2$\tabularnewline
H & 07 59 48.28 & +36 06 41.42 & $ 2.0\pm 0.2$ & $ 0.2\pm 1.5$ & $2.2\pm0.3$& $-0.4\pm1.3$\tabularnewline
I & 08 00 03.25 & +36 00 51.74 & $ 1.0\pm 0.2$ & $-0.5\pm 1.6$ & $0.2\pm0.1$& $-0.3\pm1.6$\tabularnewline
J & 08 00 59.26 & +35 55 51.49 & $ 0.9\pm 0.1$ & $ 0.3\pm 1.3$ & $0.9\pm0.1$& $0.6 \pm1.1$\tabularnewline
K & 08 01 17.11 & +36 04 31.27 & $ 0.6\pm 0.1$ & $ 1.7\pm 1.4$ & $0.5\pm0.1$& $0.7 \pm1.1$\tabularnewline
L & 08 00 30.27 & +36 00 41.98 & $ 0.6\pm 0.1$ & $ 1.1\pm 1.6$ & $0.7\pm0.1$& $0.4 \pm1.4$\tabularnewline
M & 08 01 24.68 & +36 05 37.48 & $ 0.6\pm 0.1$ & $-0.0\pm 1.5$ & $0.5\pm0.1$& $-0.1\pm1.3$\tabularnewline
N & 08 00 52.72 & +36 06 13.52 & $ 0.4\pm 0.1$ & $ 0.7\pm 1.8$ & $0.3\pm0.1$& $0.7 \pm1.6$\tabularnewline
O & 08 00 40.14 & +35 59 51.55 & $ 0.4\pm 0.1$ & $ 1.3\pm 1.6$ & $0.3\pm0.1$& $1.7 \pm1.5$\tabularnewline
\hline
P & 08 00 11.88 & +35 50 15.31 & $ 1.6\pm 0.3$ & $-0.9\pm 1.6$ & --- & --- \tabularnewline
Q & 08 00 54.95 & +36 17 43.76 & $ 1.4\pm 0.4$ & $-0.8\pm 1.7$ & --- & --- \tabularnewline
R & 08 00 38.30 & +36 10 56.43 & $ 0.6\pm 0.1$ & $-0.6\pm 1.6$ & --- & --- \tabularnewline
\hline
\end{tabular}

\caption{Sources found in the LA image of A611, and fitted to the SA data, in order of LA flux density.\label{tab:A611-sources}}

\end{center}
\end{table*}

\begin{table*}
\begin{center}
\begin{tabular}{ccccccc}
\hline 
ID & RA & Dec & $S_{\rm{LA}}$/mJy & $\alpha_{\rm{LA}}$  & $S_{\textrm{SA}}$/mJy & $\alpha_{\textrm{SA}}$\tabularnewline
\hline 
A & 09 44 48.36 & +47 00 02.10 & $ 6.0\pm 0.3$ & $ 0.4\pm 0.6$ & $1.7 \pm 0.8$ & $0.0 \pm1.7$\tabularnewline
B & 09 42 57.44 & +46 58 49.93 & $ 2.2\pm 0.1$ & $ 0.9\pm 0.5$ & $2.9 \pm 0.5$ & $0.5 \pm0.6$\tabularnewline
C & 09 42 34.86 & +47 18 23.55 & $ 2.1\pm 0.5$ & $-0.1\pm 1.7$ & $2.2 \pm 0.1$ & $0.7 \pm0.3$\tabularnewline
D & 09 43 32.44 & +46 47 19.16 & $ 1.4\pm 0.3$ & $-0.7\pm 1.5$ & $1.2 \pm 0.2$ & $-0.5\pm1.3$\tabularnewline
E & 09 43 19.30 & +46 47 32.98 & $ 1.2\pm 0.3$ & $ 0.1\pm 1.6$ & $1.6 \pm 0.2$ & $-0.1\pm1.2$\tabularnewline
F & 09 43 12.10 & +47 04 22.51 & $ 1.2\pm 0.1$ & $ 1.9\pm 0.8$ & $1.2 \pm 0.1$ & $0.9 \pm0.8$\tabularnewline
G & 09 42 24.10 & +47 02 48.19 & $ 1.1\pm 0.0$ & $ 0.9\pm 0.9$ & $1.3 \pm 0.1$ & $1.2 \pm0.5$\tabularnewline
H & 09 42 23.62 & +46 50 07.13 & $ 1.0\pm 0.2$ & $ 0.3\pm 1.4$ & $1.4 \pm 0.2$ & $-0.2\pm1.2$\tabularnewline
I & 09 42 06.14 & +46 57 05.32 & $ 0.9\pm 0.1$ & $ 0.9\pm 1.7$ & $0.6 \pm 0.1$ & $1.1 \pm1.5$\tabularnewline
J & 09 42 49.52 & +46 57 07.06 & $ 0.6\pm 0.1$ & $ 2.1\pm 1.4$ & $0.3 \pm 0.1$ & $1.6 \pm1.2$\tabularnewline
K & 09 43 00.70 & +47 01 27.71 & $ 0.6\pm 0.1$ & $ 1.1\pm 1.4$ & $0.6 \pm 0.1$ & $1.5 \pm1.0$\tabularnewline
\hline
L & 09 44 05.15 & +46 44 02.43 & $ 4.5\pm 0.5$ & $ 0.3\pm 1.4$ & --- & --- \tabularnewline
M & 09 42 19.30 & +46 56 55.84 & $ 0.5\pm 0.1$ & $-0.7\pm 1.7$ & --- & --- \tabularnewline
N & 09 43 44.18 & +47 02 54.42 & $ 0.4\pm 0.1$ & $-0.3\pm 1.7$ & --- & --- \tabularnewline
O & 09 42 31.73 & +46 55 07.53 & $ 0.4\pm 0.1$ & $-0.3\pm 1.6$ & --- & --- \tabularnewline

\hline
\end{tabular}

\caption{Sources found in the LA image of A851, and fitted to the SA data, in order of LA flux density.\label{tab:A851-sources}}

\end{center}
\end{table*}

\bibliographystyle{philthesis}
\bibliography{references}
\bsp 

\label{lastpage}
\end{document}